\documentclass[
 reprint,
nofootinbib,
 amsmath,amssymb,
 aps,
]{revtex4-2}

\usepackage{graphicx}% Include figure files
\usepackage{dcolumn}% Align table columns on decimal point
\usepackage{bm}% bold math
\usepackage[normalem]{ulem}
\usepackage{mathtools}
\usepackage{array,tabularx}
\usepackage{adjustbox}
\usepackage{tikz}
\usepackage{esint}
\usepackage{makecell}

\newcommand{\f}{\frac}

\newcommand{\D}{\mathrm{d}}

\begin{document}

\preprint{APS/123-QED}

\title{Cosmological Dynamics of a Non-Canonical Generalised Brans-Dicke Theory}% Force line breaks with \\

\author{Matthew Debono}
 \email{matthew.debono.20@um.edu.mt}
 \affiliation{%
 Institute of Space Sciences and Astronomy, University of Malta, Msida, Malta
 }%
\author{Gabriel Farrugia}
 \email{gfarr02@um.edu.mt}
 \email{gabriel.farrugia@mcast.edu.mt}
 \affiliation{%
 Department of Physics, University of Malta, Msida, Malta \\
 Institute of Applied Sciences, MCAST, Paola, Malta
 }%
\author{Jackson Levi Said}%
 \email{jackson.said@um.edu.mt}
\affiliation{%
 Institute of Space Sciences and Astronomy, University of Malta, Msida, Malta
}%

\date{\today}% It is always \today, today,

\begin{abstract}
The $\Lambda$CDM model has been presented with a number of cosmic tensions in the face of precision cosmological data, suggesting the presence of a dynamical dark energy component. In this context, we investigate the cosmology arising from a generalisation of Brans-Dicke theory, with a non-minimally coupled scalar field characterising deviations from standard general relativity, and having a non-canonical kinetic term. By reformulating the field equations into an autonomous set of dynamical equations, we use the methods of dynamical systems to investigate the equilibrium states of the system and their stability for a set of widely-used potentials, namely the constant, power-law, and exponential potentials, with the flow visualized using bounded phase portraits. Furthermore, we investigate the physical meaning of the critical points, and we find viable solutions that can reproduce the characteristics of the $\Lambda$CDM model at background level for each of the three potentials. Furthermore, in each case, we observe that the dynamical behaviour differs noticeably from that observed in other scalar-tensor models due to the non-minimal coupling and non-canonical field, despite using similarly defined dynamical variables.
\end{abstract}

\maketitle

%\tableofcontents

\section{Introduction}

The level of statistical consistency between $\Lambda$CDM cosmology and individual survey data sets is high at all levels of cosmology where measurements have been taken \cite{CosmoVerseNetwork:2025alb,Clifton:2011jh}. For this cosmological scenario, cold dark matter (CDM) dominates the physics of large scale structure formation leading to galactic structures \cite{Peebles:2002gy,Baudis:2016qwx,Bertone:2004pz}, and dark energy is expressed as a cosmological constant $\Lambda$ \cite{Copeland:2006wr} which drives late time accelerated cosmic expansion. Moreover, the contribution of cosmic inflation provides a suitable initial point from which this dynamics evolves from the early primordial Universe to its present state \cite{Bahamonde:2017ize}. Despite the attractive feature of this dynamical background, the cosmological constant is well known to have internal consistency problems \cite{Weinberg:1988cp}, and the potential direct detection of CDM particles remains elusive \cite{Gaitskell:2004gd} despite great efforts. The robustness in the measurement of cosmological parameter has led to the appearance of cosmological tensions in recent years \cite{CosmoVerseNetwork:2025alb,DiValentino:2020vhf,DiValentino:2020zio,DiValentino:2020srs} which has led to a plethora of potential new physics settings beyond $\Lambda$CDM \cite{Clifton:2011jh,CANTATA:2021ktz}. On the other hand, these different cosmologies produce different dynamical scenarios which may alter the standard phase portrait scenario. The strongest cosmic tension appears here in the value of cosmic expansion at present time $H_0$, but this is also complemented with a possible tension in the growth of large scale structures \cite{Abdalla:2022yfr,DiValentino:2020vvd}. Together, these tensions and anomalies pose a key question for $\Lambda$CDM cosmology and motivate potential new physics beyond this concordance model.

Confronting these tensions has been a challenge and has led to a wide spectrum of potential solutions in the literature \cite{CANTATA:2021ktz,Capozziello:2002rd,Capozziello:2011et,Clifton:2011jh,Nojiri:2010wj,Nojiri:2017ncd}. This includes new forms of modified CDM \cite{Feng:2010gw,Dodelson:1993je,Joyce:2014kja,Abazajian:2012ys}, dynamical aspects of dark energy \cite{Copeland:2006wr,Benisty:2021gde,Benisty:2020otr,Bamba:2012cp}, and new descriptions of gravitation \cite{Clifton:2011jh,CANTATA:2021ktz,Bahamonde:2021gfp,AlvesBatista:2021gzc,Addazi:2021xuf,Capozziello:2011et} among others. The simplest and most widely acceptable setting is that of a single scalar field that drives dynamical deviations from concordance cosmology.  This is the simplest scenario  where the standard dynamical phase portrait is extended. Generally, scalar fields appear as second order components in the background equations of motion, which means that they are generally free from ghosts and most instabilities \cite{Deffayet:2009mn,Deffayet:2009wt}. Moreover, they have been extensively studied in the literature in frameworks such as Horndeski gravity \cite{Horndeski:1974wa,Kobayashi:2019hrl}.

Cosmological scalar fields have been applied to a variety of settings to produce distinct scenarios. The best known setting is cosmic inflation \cite{1982PhLB..108..389L,1982PhRvL..49.1110G} where a single scalar field drives primordial accelerated expansion in order to resolve the horizon and flatness problems, among others. In the context of cosmic tensions, scalar fields have also been used to produce early dark energy (EDE) \cite{Poulin:2023lkg} where a single scalar field peaks prior to recombination, effectively reducing the sound horizon and increasing the Hubble constant. A variety of potentials, and other settings, have been proposed to describe EDE scenarios. Having the new physics prior to recombination means that the remainder of the concordance model remains intact. On the other hand, having a larger expansion profile prior to recombination means that there will be less time for the seeds of structure formation to form.

The most systematic approach for scalar-tensor extensions of $\Lambda$CDM cosmology has been in the form of Horndeski gravity \cite{Horndeski:1974wa} where the most general formulation of a single scalar field is expressed for second order equations of motion. This has also been extended to other gravitational frameworks \cite{Bahamonde:2019shr,Bahamonde:2019ipm} as well as beyond second order \cite{Gleyzes:2014qga,Kobayashi:2019hrl}, and also to multifield scenarios \cite{Ohashi:2015fma}. Moreover, Horndeski gravity encapsulates many common features of cosmologies beyond $\Lambda$CDM, and can be mapped to these different theoretical formulations through appropriate transformations. On the other hand, the severe constraints imposed by multimessenger signals GW170817 \cite{LIGOScientific:2017vwq} and GRB170817A \cite{Goldstein:2017mmi} which appear in the form of limits on the excess speed of gravitational waves as a ratio of the speed of light. This limit renders only simpler forms of Horndeski gravity models as being physically viable. On the other hand, this still leaves a wide array of possible realisations for possible Horndeski settings. These include EDE, as well as numerous other possible expressions of scalar-tensor scenarios.

The spectrum of lower order Horndeski models has been less studied but remains viable in the context of the recent multimessenger signals. In this work we explore the interesting class of models where an extended non-canonical scalar field is coupled with the Einstein-Hilbert action in three distinct settings. The aim of the work is to explore the dynamical systems of these scenarios and to understand the nature of the parameter models in terms of their potential in producing viable cosmologies. In Sec.~\ref{sec:background}, we cover the background theory and the motivation for this general class of models. The dynamical variables are set up in Sec.~\ref{sec:dynam_sys} where the dynamical variables are established together with the autonomous set of differential equations. The ensuing models are also discussed in this part of the work. The main work is performed in Sec.~\ref{sec:dyn_ana} where the dynamical systems of the three models is explored. Following the established approach, we find the critical points and study their nature. Additionally, we show representative evolutions of the density parameters in each case. This is important for cementing the viability of each model. Finally, we close with a summary of the main results in Sec.~\ref{sec:conc} where we also discuss possible future work.

\section{Action and Field Equations} \label{sec:background}
We consider a class of theories which introduce a scalar field with a non-minimal coupling to gravity, but with a minimal coupling to matter. These theories are described by the action
\begin{equation}
\mathcal{S} = \int \D^4 x~\sqrt{-g}\left\{\f{1}{2}f(\phi)R + K(\phi, X) + \mathcal{L}_m\right\},
\label{genaction}
\end{equation}
working in a choice of units such that $\kappa^2 = 1$. The field equations corresponding to this action are
\begin{align}
f(\phi) G_{\mu\nu} &= T_{\mu\nu}^{(m)} + g_{\mu\nu }K(\phi, X) + (K_X + f_{\phi\phi}) \nabla_{\mu} \phi \nabla_{\nu} \phi \nonumber \\ & - g_{\mu\nu}(f_{\phi} \Box \phi - 2 X f_{\phi\phi}) + f_{\phi}\nabla_{\mu}\nabla_{\nu} \phi,
\label{fieldeq}
\end{align}
and the equation of motion of the scalar field $\phi$ is
\begin{equation}
\nabla^{\mu}[K_X \nabla_{\mu} \phi] + \f{1}{2}f_{\phi} R + K_{\phi} = 0.
\label{eom}
\end{equation}
Here, $X = -(\nabla^{\mu} \phi \nabla_{\mu} \phi)/2$ is the kinetic term of the field, $K(\phi, X)$ and $f(\phi)$ are arbitrary functions of $\phi$ and $X$, $R$ is the Ricci scalar, and $G_{\mu\nu}$ is the Einstein tensor. In addition, $f_{\phi}$ and $K_X$ denote derivatives of $f$ and $K$ with respect to $\phi$ and $X$ respectively; similar notation follows for higher-order derivatives.

Sub-classes of this general scalar-tensor theory can be chosen by specifying $f(\phi)$ and $K(\phi,X)$. For example, the minimally-coupled sector of scalar-tensor theories is given by $f(\phi) = 1$; this includes the $\Lambda$CDM model with $K(\phi,X) = -\Lambda$, and the quintessence models with $K(\phi,X) = X - V(\phi)$. This general model also includes non-minimally coupled theories such as standard Brans-Dicke theory with $f(\phi) = \phi$ and $K(\phi,X) = \omega_{BD}X$, where $\omega_{BD}$ is the Brans-Dicke parameter. Inspired by solutions from Noether symmetries \cite{Capozziello:2018gms,Dialektopoulos:2021ryi,Miranda:2024hhe}, we shall restrict to the class of models with\footnote{Another choice is $f(\phi) = \xi_0 + \xi \phi$ which introduces an additional parameter, $\xi_0$. However, this effectively rescales the Lagrangian, leading to no further change in dynamics under appropriate rescaling of the remaining coupling constants. As such, the choice $\xi_0 = 1$ was considered maintaining $\xi\phi$ as the deviation from GR.} 
\begin{subequations}
\begin{alignat}{2}
f(\phi) &= 1 + \xi \phi, \\
K(\phi, X) &= \f{g(X)}{f(\phi)} - V(\phi),
\end{alignat}
\end{subequations}
where $g(X)$ is an arbitrary function of $X$, and $\xi$ is a coupling constant. These models are a generalisation of Brans-Dicke theory and different choices of $\xi$, $g(X)$, and $V(\phi)$ will lead to different classes of models within this theory, as outlined in Table \ref{tab:submodels}. For example, the choice $\xi = 0$ with $g(X) = 0$ and $V(\phi) = \Lambda$ gives the standard $\Lambda$CDM model, while choosing $g(X) = X$ and keeping $V(\phi)$ arbitrary gives the quintessence class of models.

\begin{table}[!htb]
\centering
\begin{tabular}{c c c c}
\hline
$\xi$ & $g(X)$ & $V(\phi)$ & Theory\\
\hline
0 & 0 & $\Lambda$ & $\Lambda$CDM \\
0 & $X$ & Arbitrary & Quintessence \cite{RatraPeebles_1988} \\
0 & $-X$ & Arbitrary & Phantom dark energy \cite{Roy_2018}\\
1 & $\omega_{BD}X$ & Arbitrary & Generalised Brans-Dicke \cite{Hrycyna_2013a}\\
\hline
\end{tabular}
\caption{Examples of models included in the general scalar-tensor action Eq.~\eqref{genaction} with $f(\phi) = 1 + \xi \phi$ and $K(\phi, X) = \frac{g(X)}{f(\phi)} - V(\phi)$.}
\label{tab:submodels}
\end{table}
While leaving $\xi$ arbitrary, the $f(\phi)$ coupling acts as an effective gravitational coupling, $G_{\text{eff}}^{-1}$. Solar system constraints as well as structure growth restrict $f(\phi) \simeq 1$ \cite{Avilez:2013dxa,Faraoni:2019sxw,Amirhashchi:2019jpf,DeFelice:2011hq}, which can be achieved for $|\xi| \ll 1$. Furthermore, we restrict the model to the branch that does not produce ghosts and Laplacian instabilities which, for FLRW cosmology, is achieved provided \cite{Garriga:1999vw,Kobayashi:2011nu,DeFelice:2011hq,Nojiri:2019dqc}
\begin{equation}
    \frac{K_X + 2X K_{XX} + \frac{3\dot{f}^2}{2fX}}{\left(H + \frac{\dot{f}}{2f}\right)^2} X > 0.
\end{equation}
This sets $K_X + 2X K_{XX} + \frac{3\dot{f}^2}{2fX} > 0$. For the chosen ansatz, a healthy theory requires $\omega + \frac{3\xi^2}{1+\xi\phi} > 0$. For simplicity, the condition $\omega > 0$ will be considered, which satisfies this constraint but also compactifies the dynamical phase space as outlined in the forthcoming section.

\section{Background Cosmological Dynamics} \label{sec:dynam_sys}

Inspired by Brans-Dicke cosmology, whose dynamical behaviour has already been investigated in previous works \cite{Hrycyna_2013a, Hrycyna_2013b, Hrycyna_2014, Hrycyna_2015}, we aim to investigate potentials using this generalised model with $g(X) = \omega X$ and an arbitrary $\xi$, for a coupling constant $\omega$. In other words, we shall be investigating the cosmological dynamics of the scalar-tensor action
\begin{equation}
\mathcal{S} = \int \D^4 x~\sqrt{-g}\left\{\f{1}{2}(1+\xi\phi)R + \f{\omega X}{1+\xi\phi} - V(\phi) + \mathcal{L}_m\right\}.
\end{equation}
Since we are interested in the background cosmological dynamics, we shall treat the universe as homogeneous, isotropic and flat, filled with ideal fluids of pressureless matter and radiation. Equipped with the flat FLRW metric $\D s^2 = -\D t^2 + a^2(t) \delta_{ij}\D x^i \D x^j$\footnote{Here, we assume the lapse function $N(t) = 1$ which is suitable for background behaviour but should not be assumed \textit{a priori} in the study of perturbations \cite{Kobayashi:2011nu,Gleyzes:2014qga,Ahmedov:2024aez}. The implications for numerical relativity of not choosing this gauge choice are discussed in \cite{Aurrekoetxea:2024ypv}.}, where $a(t)$ is the scale factor, the corresponding modified Friedmann equations and Klein-Gordon equation are
\begin{subequations}
\begin{align}
3(1+\xi \phi)H^2 + 3H\xi \dot{\phi} &= \rho_m + \rho_r + \f{1}{2} \f{\omega\dot{\phi}^2}{1+\xi \phi} + V(\phi), \label{eq:00-Friedmann} \\
-(2\dot{H}+3H^2)(1+\xi \phi) &= \f{1}{3}\rho_r + \f{1}{2} \f{\omega\dot{\phi}^2}{1+\xi \phi} - V(\phi) \nonumber \\
&\quad + \xi \ddot{\phi} + 2H\xi\dot{\phi}, \\
\f{\omega}{1+\xi\phi}(\ddot{\phi}+3H\dot{\phi}) &- \f{1}{2}\omega\xi \f{\dot{\phi}^2}{(1+\xi \phi)^2} - 3\xi(\dot{H}+2H^2) \nonumber \\
&\quad + V_{\phi} = 0.
\end{align}
\end{subequations}
where $H(t) \equiv \dot{a}/a$ is the Hubble parameter, and $\rho_m$ and $\rho_r$ are the energy densities of matter and radiation respectively. Unless otherwise specified, overdots represent derivatives with coordinate time, $t$. 

Following the conventions of \cite{Hrycyna_2013a, Hrycyna_2013b}, dividing Eq.~\eqref{eq:00-Friedmann} by $3(1+\xi\phi)H^2$, we define the following dimensionless dynamical variables,
\begin{align}
x &= \f{\dot{\phi}}{\sqrt{6}(1+\xi \phi)H}, &\tilde{\Omega}_m &= \f{\rho_m}{3(1+\xi \phi)H^2}, \nonumber \\
\tilde{\Omega}_r &= \f{\rho_r}{3(1+\xi \phi)H^2}, &y &= \f{\sqrt{V}}{\sqrt{3(1+\xi \phi)H}}.
\label{variables}
\end{align}
Because of the coupling $f(\phi)$, $\tilde{\Omega}_m$ and $\tilde{\Omega}_r$ are not the physical $\Omega_m$ and $\Omega_r$ as defined in \cite{Copeland_1998}. However, they are related to the physical variables by
\begin{equation}
\tilde{\Omega}_m = \f{\Omega_m}{1+\xi \phi}, \quad \tilde{\Omega}_r = \f{\Omega_r}{1+\xi \phi}.    
\end{equation}
The Friedmann constraint Eq.~\eqref{eq:00-Friedmann} is expressed in terms of the dynamical variables as
\begin{equation}\label{eq:friedmann_constraint}
1 = \tilde{\Omega}_m + \tilde{\Omega}_r + \omega x^2 + y^2 - \xi\sqrt{6}x.
\end{equation}
Since we restrict to $f(\phi) > 0$, $\tilde{\Omega}_m$ and $\tilde{\Omega}_r$ are both bounded below by zero, but, by using Eq. \eqref{eq:friedmann_constraint} and performing the change of variables $x \mapsto  x - \frac{\sqrt{6}\xi}{2\omega}$, we see that the maximum value they can take is now $1 + \f{3}{2}\f{\xi^2}{\omega}$, rather than 1. To obtain the autonomous dynamical system of equations, we differentiate the variables defined in Eq.~\eqref{variables} with respect to conformal time $\eta = \log a(t)$, and denote derivatives with respect to $\eta$ by $'$. Additionally, we must also define the variables,
\begin{equation}
\lambda_V = -(1+\xi \phi)\f{V_{\phi}}{V}, \quad \Gamma_V = \f{V_{\phi\phi}V}{V_{\phi}^2}.
\label{lambdaGamma}
\end{equation}
This approach was originally introduced in the context of quintessence models in \cite{Steinhardt_1999} and \cite{Macorra_2000}, where $\lambda_V$ was defined with $\xi = 0$. The variables as defined here are modelled on the approach taken in \cite{Hrycyna_2014} for a model with a non-minimal coupling with gravity. $\lambda_V$ naturally arises when deriving the dynamical equations for the variables in Eq.~\eqref{variables}, and, in order to close the system, we must treat $\lambda_V$ as an additional dynamical variable, which is why $\Gamma_V$ must be introduced. $\lambda_V$ and $\Gamma_V$ will, in general, be functions of $\phi$. For the variables in Eqs.~\eqref{variables} and \eqref{lambdaGamma} to form an autonomous dynamical system of equations, we must assume that we can invert $\lambda_V(\phi)$ to write $\Gamma_V(\phi(\lambda_V)) = \Gamma_V(\lambda_V)$.

This results in the following autonomous system:
\begin{widetext}
\begin{subequations}
\begin{align}
x' & = \f{3}{2\omega+3\xi^2} \biggl\{2\omega^2 x^3 + \xi x^2 (\omega + \sqrt{6}\xi)(\xi - \lambda_V) + \omega x\left(\tilde{\Omega}_m + \f{4}{3}\tilde{\Omega}_r - 2\right) + \xi x \left[3\xi + \lambda_V(3-\tilde{\Omega}_m - \tilde{\Omega}_r)\right] \notag \\ &\qquad\qquad\qquad + \sqrt{6} \left[\f{1}{3}(\lambda_V + 2\xi)(1-\omega x^2 - \tilde{\Omega}_r) - 2\omega \xi x^2\right] - \f{\sqrt{6}}{6}(2\lambda_V + 3\xi)\tilde{\Omega}_m\biggr\} \label{wxeqn} \\
\tilde{\Omega}_m' & = -\f{6\Omega_m}{2\omega+3\xi^2} \biggl\{\omega\left(1 - \tilde{\Omega}_m - \f{4}{3}\tilde{\Omega}_r - 2\omega x^2 \right) - \omega\xi(\xi - \lambda_V)x^2 - \xi\lambda_V(1-\tilde{\Omega}_m - \tilde{\Omega}_r) \notag \\ &\qquad\qquad\qquad + \f{\sqrt{6}}{6}\xi x(3\xi^2 - 6\lambda_V\xi + 10\omega) - \f{1}{2}\xi^2\biggr\} \label{wmeqn} \\
\tilde{\Omega}_r' & = -\f{6\Omega_r}{2\omega+3\xi^2} \biggl\{\omega\left(\f{4}{3} - \tilde{\Omega}_m - \f{4}{3}\tilde{\Omega}_r - 2\omega x^2\right) - \omega\xi(\xi - \lambda_V)x^2 - \xi\lambda_V(1-\tilde{\Omega}_m - \tilde{\Omega}_r) \notag \\ &\qquad\qquad\qquad + \f{\sqrt{6}}{6}\xi x(3\xi^2 - 6\lambda_V\xi + 10\omega) \biggr\}, \label{wreqn} \\
\lambda_V' & = \sqrt{6}[\xi\lambda_V - \lambda_V^2(\Gamma_V - 1)]x. \label{wleqn}
\end{align}
\end{subequations}
\end{widetext}
In the system above, we have eliminated the variable $y$ using the Friedmann constraint Eq.~\eqref{eq:friedmann_constraint}.

Prior to obtaining the respective critical points, it is crucial to define the compact phase space. In particular, Eq. Eq.~\eqref{eq:friedmann_constraint} only defines a compact space for $\omega > 0$, in which case the phase space takes the shape of an ellipse. For $\omega < 0$, the phase space is always non-compact; to see this, consider a projection of the 4-D space defined by the Friedmann constraint with fixed $\tilde{\Omega}_m$ and $\tilde{\Omega}_r$. If $\omega$ is negative, this subspace, given by
\begin{equation}
|\omega|\left(x-\f{\sqrt{6}}{2}\f{\xi}{|\omega|}\right)^2 - y^2 = \f{3\xi^2}{2|\omega|}-1,
\end{equation}
which takes the form of a hyperbola, making it non-compact. Thus, we shall restrict to the $\omega > 0$ case. The case of $\omega < 0$, which would give a phantom scalar field, has been investigated for standard Brans-Dicke theory in \cite{Hrycyna_2013b}; a full dynamical analysis requires the use of the Poincar\'{e} construction to compactify the phase space. However, because we can obtain a bound of $\omega_{BD} \gtrsim 10^4$ from solar system tests of standard Brans-Dicke theory (see \cite{Bertotti:2003rm}), it is reasonable to assume that, in this non-minimally-coupled generalisation, a similar lower bound restricting $\omega$ to be positive may also be obtained. The phase space $\mathcal{P}_{\omega}$ is defined as
\begin{align}
\mathcal{P}_{\omega} &\coloneqq \left\{(x,y,\tilde{\Omega}_m,\tilde{\Omega}_r,\lambda_V): \lambda_V \in (-\infty,\infty), \right.\nonumber \\
&\left.\omega\left(x - \f{\sqrt{6}}{2\omega}\xi\right)^2 + y^2 + \tilde{\Omega}_m + \tilde{\Omega}_r \leq 1 + \f{3 \xi^2}{2\omega}\right\}.
\end{align}
In models where $\lambda_V$ is not constant, the phase space is not compact as $\lambda_V$ is now unbounded, so we must introduce the auxiliary variable $u \in \left[-\f{\pi}{2},\f{\pi}{2}\right]$, defined by $\lambda_V = \tan{u}$, to include the infinite boundary. 

For a fixed value of $\xi$, the phase space will grow as $\omega \rightarrow 0$, and shrink as $\omega \rightarrow \infty$. This means that we should expect the critical points of the dynamical system to have a dependence on $\omega$ that reflects this behaviour; for example, the variables $(x ,\tilde{\Omega}_m, \tilde{\Omega}_r)$ evaluated at critical points on the boundary of the phase space should tend to infinity as $\omega \rightarrow \infty$.

In this work, we shall consider the three models listed in Table~\ref{tab:potentials}. The constant and power-law potentials are derived by solving Eq.~\eqref{lambdaGamma} assuming that $\lambda_V$ is constant; we obtain the power-law for a non-zero $\lambda_V$, while the constant potential emerges as a special case with $\lambda_V = 0$. These three models have all been extensively studied in literature, with and without couplings \cite{Poulin:2023lkg,Schoneberg:2021qvd,Gomez-Valent:2022bku,DAgostino:2022fcx,Bhattacharya:2024hep,Arbey:2021gdg}. The aim is to explore how the chosen ansatz for $f(\phi)$ and $K(\phi, X)$ affects the dynamics for such potentials.

\begin{table}[!htb]
\centering
\begin{tabular}{cccc}
\hline 
\textbf{Model} & $V(\phi)$ & $\Gamma_V$ & $\lambda_V$ \\
\hline
Constant Potential & $V_0$ & 0 & 0 \\
Power-Law Potential & $V_0(1+\xi\phi)^{-n}$ & $1+\f{\lambda_V}{\xi}$ & $n\xi$ \\
Exponential Potential & $V_0 e^{-\lambda \phi}$ & 1 & $(1+\xi \phi)\lambda$ \\
\hline
\end{tabular}
\caption{A table of the models to be considered with the respective forms of $\lambda_V$ and $\Gamma_V$.}
\label{tab:potentials}
\end{table}

\section{Critical Points and Phase Space Portraits} \label{sec:dyn_ana}

An autonomous system of equations such as that in Eqs.~\eqref{wxeqn}--\eqref{wleqn} can be written in the form $\mathbf{x}' = \mathbf{F}(\mathbf{x})$, where $\mathbf{x} = (x, \tilde{\Omega}_m,\tilde{\Omega}_r,\lambda_V)$ is an element of the phase space $\mathcal{P}_{\omega} \subset \mathbb{R}^4$, and $\mathbf{F}(\mathbf{x})$ is viewed as a vector field on $\mathbb{R}^4$ with components $F_i(\mathbf{x})$ corresponding to the right-hand side of Eqs.~\eqref{wxeqn}--\eqref{wleqn}. A dynamical analysis of such systems involves two steps. First, we must find the critical/fixed points of the system; these are the points $\mathbf{x}^*$ such that $\mathbf{F}(\mathbf{x}^*) = \mathbf{0}$. This means that the system is at rest at these fixed points, and could, in principle, remain in this state indefinitely. The second step of a phase space analysis is required to determine if these equilibrium states are stable under small perturbations. For most cosmological applications, and in this work, we shall work in linear stability theory, where we expand $F_i(\mathbf{x})$ around the critical point $\mathbf{x}^*$ and consider only the linear order terms. The object of importance here is the Jacobian matrix $\mathbf{J}(\mathbf{x^*})$, with components
\[J_{ij}(\mathbf{x^*}) = \left.\f{\partial F_i(\mathbf{x})}{\partial x^j}\right|_{\mathbf{x} = \mathbf{x}^*}\]
as it is the eigenvalues of $\mathbf{J}(\mathbf{x}^*)$ which determine the stability of the point. The Hartman-Grobman theorem guarantees that, for hyperbolic fixed points, i.e. those points such that the eigenvalues of $\mathbf{J}(\mathbf{x}^*)$ all have non-zero real part, the stability properties are entirely captured by the linearised system \cite{Strogatz_2018}. Other techniques, such as the method of Lyapunov functions and the analysis of dynamics along the centre manifold, have been developed for non-hyperbolic fixed points (for applications in cosmology, see, for example, \cite{boehmer2010, Charters_2001, leon2013, Rendall_2002, Bhattacharya:2024hep}).

Hyperbolic fixed points can be classified as one of three types: (i) unstable nodes/sources (if all eigenvalues have positive real parts), which repel trajectories in all directions, (ii) saddles (if at least two eigenvalues have a negative real part), which attract trajectories in some directions but repel them in others, and (iii) stable nodes/sinks (if all eigenvalues have a negative real part), which attract trajectories in all directions. This is sufficient for cosmological applications; the interested reader is directed to standard dynamical systems literature such as \cite{Wiggins_1990} for a more detailed discussion of stability theory.

The aim of applying dynamical systems to cosmological models beyond $\Lambda$CDM is to determine whether the model under test can reproduce the broad characteristics of the concordance model. In dynamical systems terminology, these correspond to a sequence of three critical points, two of which (corresponding to radiation and matter domination) are saddles, with the other (corresponding to dark energy domination) being a late-time attractor. To determine the physical properties of the universe at these critical points, we must define a set of cosmographic parameters, which, when evaluated at the critical points, give us the required information. The relevant parameters are the effective equation of state, $w_{\text{eff}}$, which gives information on the dominating fluid(s), and the deceleration parameter $q$, which tells us whether the expansion is accelerating ($q < 0$) or decelerating ($q > 0$). Moreover, because we are working with a scalar degree of freedom as a dark energy source, we also require the scalar equation of state parameter $w_{\phi}$ to understand the behaviour of the dark energy fluid at the critical points. These are defined below, with $\rho_{\phi}$ and $p_{\phi}$ the energy density and pressure of the scalar field $\phi$ respectively:
\begin{widetext}
\begin{subequations}
\begin{alignat}{4}
q &:= -1 - \f{\dot{H}}{H^2} = -1 + \f{3}{2\omega + 3\xi^2}\biggl\{\omega\left[\tilde{\Omega}_m + \f{4}{3}\tilde{\Omega}_r + (2\omega + \xi^2)x^2\right] + 3\xi\lambda_V(1-\omega x^2 - \tilde{\Omega}_m - \tilde{\Omega}_r) \notag \\ &\qquad\qquad\qquad\quad + 3\sqrt{6}\xi x\left(\lambda_V \xi - \f{4}{3}\omega\right) + 6\xi^2\biggr\}, \\
% q &:= -1 - \f{\dot{H}}{H^2} = \f{\omega(3\tilde{\Omega}_m + 4\tilde{\Omega}_r + 6\omega x^2 - 2) + 3\xi \lambda_V(1+\sqrt{6}\xi x - \omega x^2 - \tilde{\Omega}_m - \tilde{\Omega}_r) + 3\xi^2(1+\omega x^2) - 4\sqrt{6} \omega \xi x}{2\omega + 3\xi^2}, \\
w_{\phi} &:= \f{p_{\phi}}{\rho_{\phi}} = \f{2\omega(\tilde{\Omega}_m + \tilde{\Omega}_r + 2\omega x^2 -1) + 2\xi\lambda_V(1-\omega x^2 - \tilde{\Omega}_m - \tilde{\Omega}_r) + 2\sqrt{6}\xi x\left(\xi\lambda_V - \f{4}{3}\omega\right) + \xi^2(1 + 2\omega x^2 - \tilde{\Omega}_r)}{(2\omega + 3\xi^2)(\omega x^2 + y^2 - \xi\sqrt{6}x)}, \\
% w_{\phi} &:= \f{p_{\phi}}{\rho_{\phi}} = \f{8\sqrt{6}\xi \omega x - 6\xi \lambda_V(1+\sqrt{6}\xi x - \omega x^2 - \tilde{\Omega}_m - \tilde{\Omega}_r) - 6 \omega(2\omega x^2 + \tilde{\Omega}_m + \tilde{\Omega}_r - 1) + 3\xi^2(\tilde{\Omega}_r - 2\omega x^2 -1)}{3(2\omega + 3\xi^2)(\xi\sqrt{6}x - \omega x^2 - y^2)}, \\
w_{\text{eff}} &:= \f{p_{\text{tot}}}{\rho_{\text{tot}}} = \f{p_m + p_r + p_{\phi}}{\rho_m + \rho_r + \rho_{\phi}} = \f{1}{3}\Omega_r + w_{\phi}\Omega_{\phi},\\
\Omega_{\phi} &:= \omega x^2 + y^2 - \xi\sqrt{6}x.
\end{alignat}
\end{subequations}
\end{widetext}

These parameters become very important when determining which of the critical points corresponds to a dark energy-dominated solution, with the scalar field acting like a cosmological constant, i.e. $q \approx -1$, $w_{\phi} = w_{\text{eff}} \approx -1$. For the other critical points, they will serve to check that we recover the expected behaviour ($w_{\text{eff}} = \f{1}{3}$ and $q > 0$ for a radiation-dominated epoch, for example).

\subsection{The Constant Potential: $V(\phi) = V_0$}
The coordinates of the critical points for Eqs.~\eqref{wxeqn}--\eqref{wreqn} with the constant potential are found in Table \ref{tab:wcase1points}, and their respective physical properties in Table \ref{tab:wcase1params}. In this case, the critical points represent the following physical behaviours:

\begin{table*}[!htb]
\centering
\begin{tabular}{ccccc}
\hline
Point & $x$ & $\tilde{\Omega}_m$ & $\tilde{\Omega}_r$ & Stability \\
\hline
$P_1$ & 0 & 0 & 1 & Saddle \\
$P_2$ & $\f{\xi}{\sqrt{6}(\xi^2 + \omega)}$ & $1+\f{\xi^2(5\omega + 6\xi^2)}{6(\xi^2 + \omega)^2}$ & 0 & Saddle \\
$P_3$ & $\sqrt{\f{2}{3}}\f{2\xi}{\xi^2 + 2\omega}$ & 0 & 0 & Stable node \\
$P_4^{\pm}$ & $\f{\sqrt{6}}{2\omega}\left(\xi \pm \sqrt{\xi^2 + \f{2}{3}\omega}\right)$ & 0 & 0 & Unstable node \\
\hline
\end{tabular}
\caption{Critical points for the constant potential $V(\phi) = V_0$. Note that as $\omega > 0$, all critical points exist.} 
\label{tab:wcase1points}
\end{table*}

\begin{table*}[!htb]
\centering
\begin{tabular}{cccccc}
\hline
Point & $\tilde{\Omega}_{\phi}$ & $w_{\phi}$ & $w_{\text{eff}}$ & $q$ & Accel. \\
\hline
$P_1$ & $0$ & - & $\f{1}{3}$ & 1 & No \\
$P_2$ & $-\f{\xi^2(5\omega + 6\xi^2)}{6(\xi^2 + \omega)^2}$ & $-\f{2(\xi^2 + \omega)}{6\xi^2 + 5\omega}$ & $\f{\xi^2}{3(\xi^2 + \omega)}$ & $1-\f{\omega}{2(\xi^2 + \omega)}$ & No \\
$P_3$ & 1 & $-1 + \f{4\xi^2}{3(\xi^2 + 2\omega)}$ & $-1 + \f{4\xi^2}{3(\xi^2 + 2\omega)}$ & $-1 + \f{2\xi^2}{\xi^2 + 2\omega}$ & $\omega > \f{\xi^2}{2}$ \\
$P_4^{\pm}$ & 1 & $\f{6\xi^2 + 3\omega \pm 2\xi\sqrt{9\xi^2 + 6\omega}}{3\omega}$ & $\f{6\xi^2 + 3\omega \pm 2\xi\sqrt{9\xi^2 + 6\omega}}{3\omega}$ & $\f{3\xi^2 + 2\omega \pm \xi\sqrt{9\xi^2 + 6\omega}}{\omega}$ & No \\
\hline
\end{tabular}
\caption{Physical properties of the critical points for the constant potential $V(\phi) = V_0$. For the critical point $P_1$, $w_\phi$ is undefined. Nonetheless, it does not contribute to the dynamics as the effective scalar-field density is zero.}
\label{tab:wcase1params}
\end{table*}

\begin{itemize}
    \item $P_1$: this is a radiation-dominated epoch, similar to that appearing in standard quintessence, and is a saddle as expected.
    \item $P_2$: both pressureless matter and the scalar field contribute to the expansion at this point; such behaviour does not appear in simpler scalar-tensor models such as quintessence. For $\omega > 0$, we will always have $\tilde{\Omega}_m \geq 1$, but, by tuning the values of $\omega$ and $\xi$, we can minimize the contribution from the scalar field. Consider the ratio of $\tilde{\Omega}_{\phi}$ to $\tilde{\Omega}_m$:
    \begin{equation}
        \f{\tilde{\Omega}_{\phi}}{\tilde{\Omega}_m} = -\f{\xi^2(5\omega + 6\xi^2)}{(2\omega+3\xi^2)(3\omega + 4\xi^2)}.
    \end{equation}
    $P_2$ is only entirely matter-dominated if $\xi = 0$ and $\omega \neq 0$ or if $\xi \neq 0$ and $5\omega = -6\xi^2$. The first case eliminates the non-minimal coupling to gravity and reduces this point to the standard matter-dominated $\Lambda$CDM fixed point, and the latter case cannot be achieved due to the $\omega > 0$ restriction. However, we can reduce the fractional contribution of the scalar (but not eliminate it completely) by choosing small values of $\xi$, which is in line with having $f(\phi) \simeq 1$.
    \item $P_3$: here, only the scalar field contributes. For $\omega > \f{\xi^2}{6}$, the equation of state becomes dark, and the universe undergoes an accelerating expansion for $\omega > \f{\xi^2}{2}$. Moreover, for $\omega > 0$, all the eigenvalues are negative, making this an attractor for all values of $\xi$. Therefore, $P_3$ can represent a universe experiencing a late-time accelerated expansion due to a scalar field behaving like a cosmological constant.
    \item $P_4^{\pm}$: the dominant fluid here is the scalar field, but $w_{\phi}$, $w_{\text{eff}}$, and $q$ evaluated at this point will always be positive for $\omega > 0$, so the scalar field does not act as a cosmological constant. Moreover, investigating the eigenvalues of the Jacobian evaluated at this point, as given in Appendix~\ref{app:eigenvalues} Table \ref{tab:wcase1eigen}, shows that these points are sources for all values of $\xi$ and $\omega > 0$.
\end{itemize}

\begin{figure*}[!htb]
\minipage{0.4\textwidth}
\includegraphics[width=\linewidth]{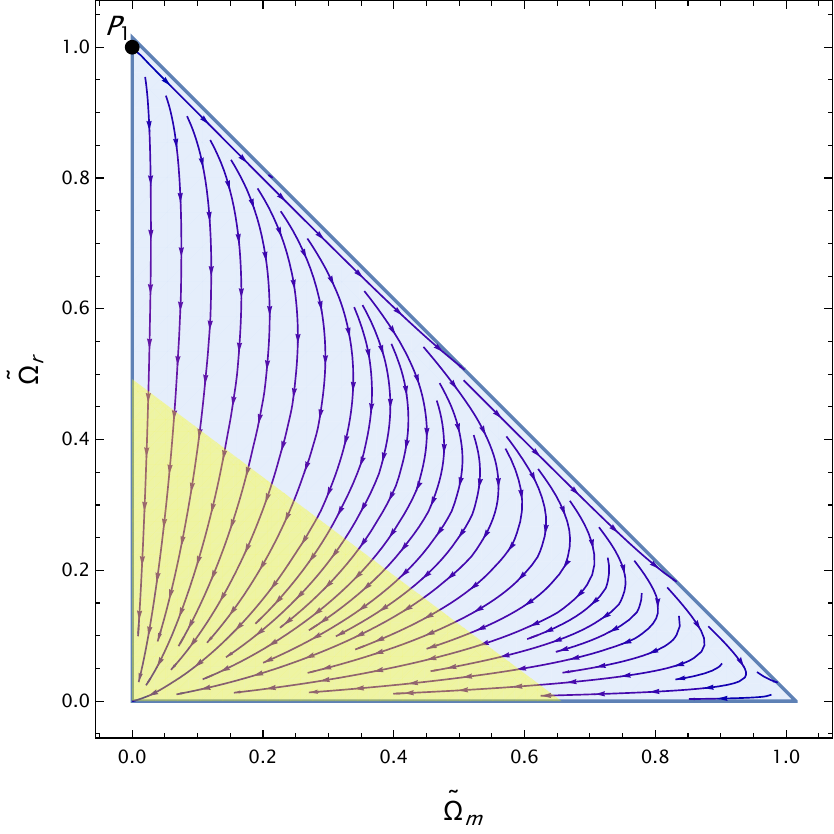}
\endminipage\hspace{0.1\textwidth}
\minipage{0.4\textwidth}
\includegraphics[width=\linewidth]{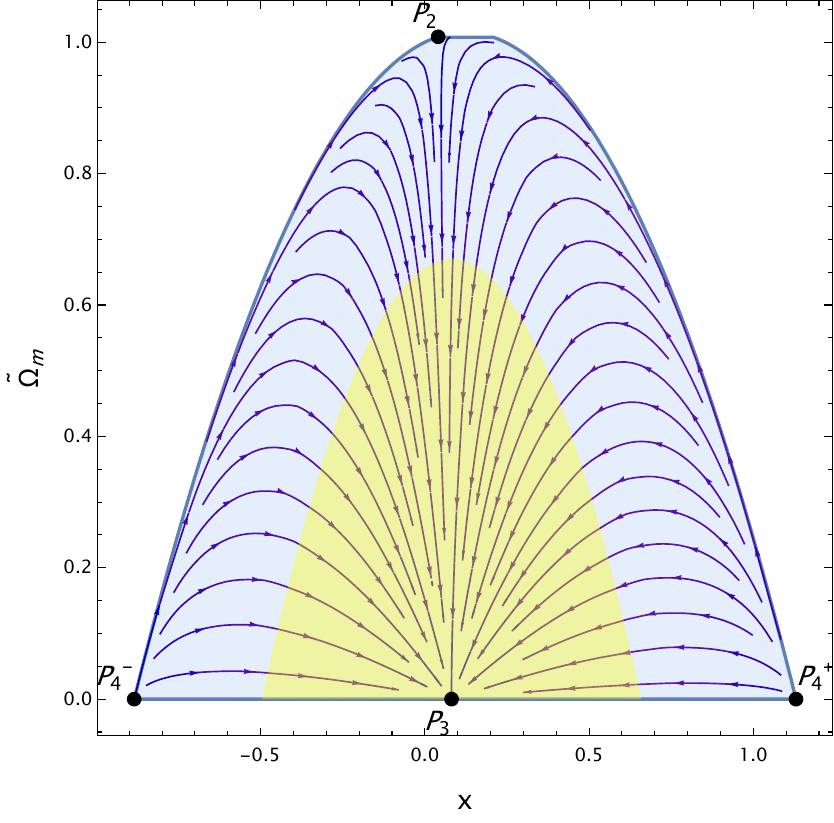}
\endminipage
\caption{Phase portraits for the constant potential $V(\phi) = V_0$ in the $x = 0$ plane (top) and the $\Tilde{\Omega}_r = 0$ plane (bottom) with $\omega = 1$ and $\xi = 0.1$. The region highlighted in yellow denotes the subset of the phase space where the universe is accelerating.}
\label{fig:model1_0.1_pp}
\end{figure*}

\begin{figure*}[!htb]
\minipage{0.4\textwidth}
\includegraphics[width=\linewidth]{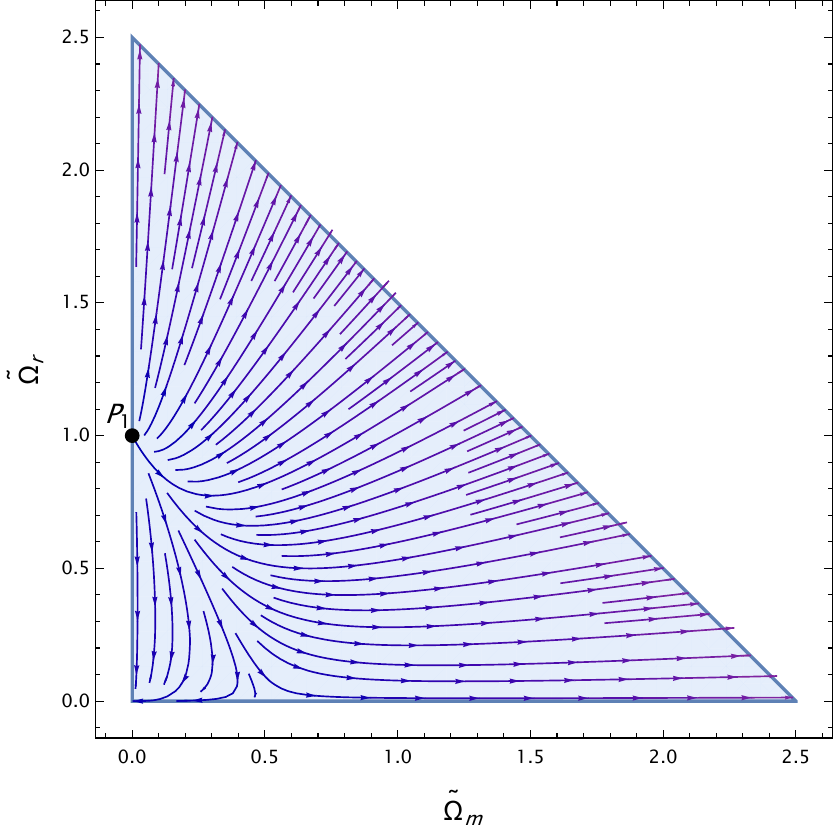}
\endminipage\hspace{0.1\textwidth}
\minipage{0.4\textwidth}
\includegraphics[width=\linewidth]{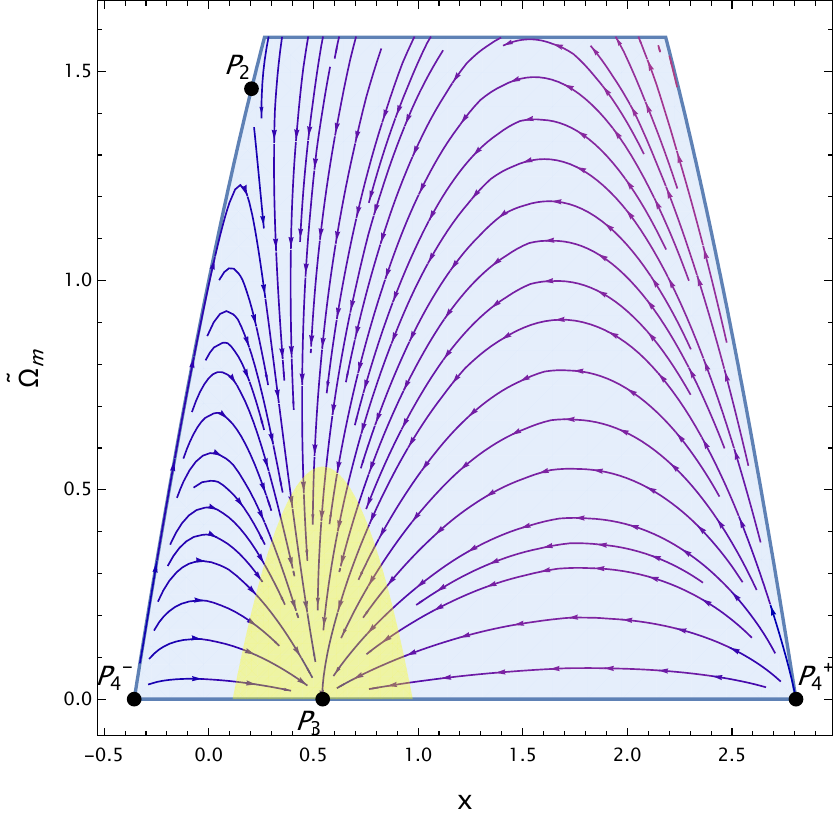}
\endminipage
\caption{Phase portraits for the constant potential $V(\phi) = V_0$ in the $x = 0$ plane (top) and the $\Tilde{\Omega}_r = 0$ plane (bottom) with $\omega = 1$ and $\xi = 1$. The region highlighted in yellow denotes the subset of the phase space where the universe is accelerating. Observe that the $\Tilde{\Omega}_r$--$\Tilde{\Omega}_m$ phase space no longer has a region with acceleration.}
\label{fig:model1_1_pp}
\end{figure*}

\begin{figure}[!htb]
\centering
\includegraphics[width=0.9\linewidth]{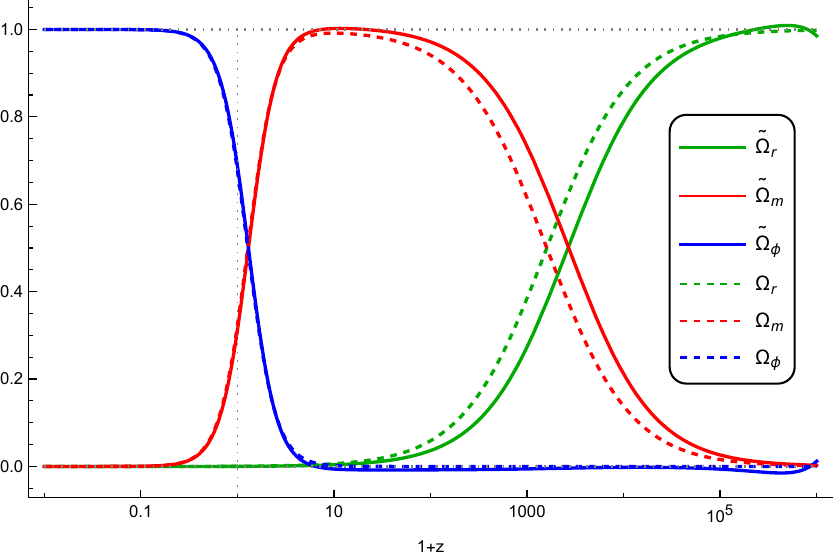}
\caption{The evolution of the cosmographic parameters $\tilde{\Omega}_m$, $\tilde{\Omega}_r$, and $\tilde{\Omega}_{\phi}$ as a function of redshift (solid lines) for the constant potential $V(\phi) = V_0$ with $\omega = 1$ and $\xi = 0.1$. The evolution for the $\Lambda$CDM cosmographic parameters (dashed lines) is also plotted for comparison.}
\label{fig:model1_0.1_evol}
\end{figure}

The phase portraits of the system for $\omega = 1$ can be found in Fig.~\ref{fig:model1_0.1_pp} for $\xi = 0.1$ and Fig.~\ref{fig:model1_1_pp} for $\xi = 1$. For $\xi = 0.1$, the flow in the $x=0$ projection of the phase space, as seen in Fig. \ref{fig:model1_0.1_pp}, is similar to that obtained for the $\Lambda$CDM model \cite{Bahamonde_2018}. Therefore, it is perhaps unsurprising that a $\Lambda$CDM-like evolution, as seen in Fig. \ref{fig:model1_0.1_evol}, is possible in this case. Here, we can see that the effect of the coupling is to shift matter-radiation equality to a slightly earlier redshift, thereby producing a longer matter-dominated epoch. As $\xi$ increases, however, the region of acceleration disappears in the $x = 0$ projection, and is visibly smaller in the $\tilde{\Omega}_r = 0$ projection as seen in Fig.~\ref{fig:model1_1_pp}. While a physically viable trajectory may be possible in this case, the dynamical behaviour suggests that deviations from GR, characterised by $\xi$, should be small.

\begin{figure}[!htb]
\centering
\includegraphics[width=0.9\linewidth]{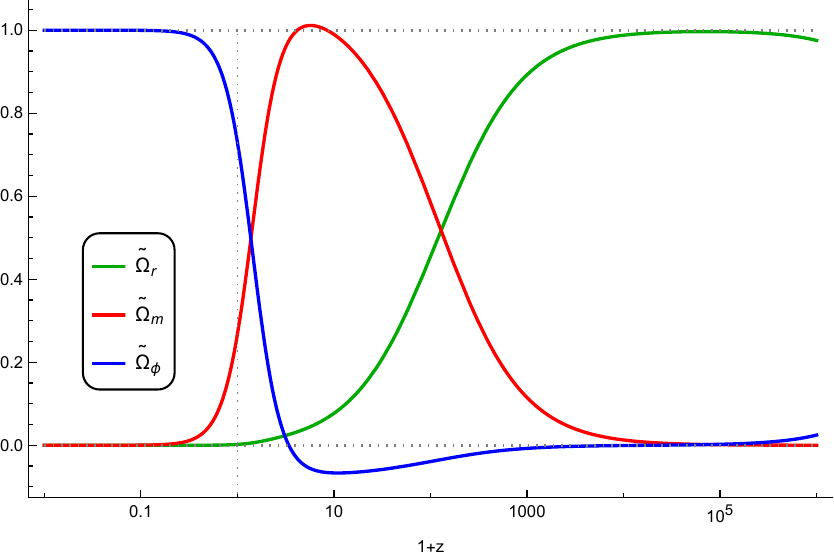}
\caption{The evolution of the cosmographic parameters $\tilde{\Omega}_m$, $\tilde{\Omega}_r$, and $\tilde{\Omega}_{\phi}$ as a function of redshift for the constant potential $V(\phi) = V_0$ with $\omega = 0.1$ and $\xi = 0.1$. In this case, the negative values of $\tilde{\Omega}_\phi$ become more apparent compared to the $\omega = 1$ case, with a notable shift in the matter-radiation equality occurring at $z \sim 100$.}
\label{fig:model1_w_0.1_evol}
\end{figure}

Because we are restricting $\omega$ to be positive, the properties of the critical points are entirely unchanged from the earlier analysis with $\omega = 1$. To investigate the $\omega$-dependence, consider, for example, the points $P_4^{\pm}$, which lie on the boundary of the phase space, and the point $P_3$. Assuming we are only considering small deviations from GR, i.e. $|\xi| \leq 1$, $P_4^{\pm}$ approach $P_3$ as $\omega \rightarrow \infty$. In finding an evolution mimicking that of $\Lambda$CDM, we would like to avoid this behaviour, as $P_4^{\pm}$ correspond to kination solutions due to the kinetic energy of the scalar degree of freedom dominating. In principle, this issue can be somewhat alleviated by choosing $\xi$ to have the same order of magnitude as $\omega$, but this would contradict the assumption of having only small deviations from GR if $|\xi| \sim \omega \gg 1$. In the case. The effect of having $|\xi| \sim \omega \ll 1$ can be seen in Fig. \ref{fig:model1_w_0.1_evol}, where matter-radiation equality at $1 + z_{\text{eq}} = \mathcal{O}(10^2)$, contrary to observational constraints.

\begin{figure}[!htb]
\centering
\includegraphics[width=0.9\linewidth]{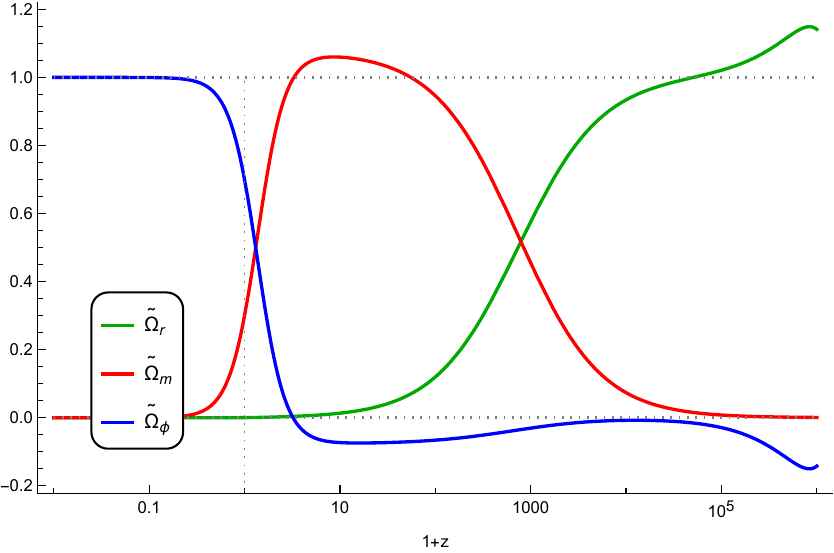}
\caption{The evolution of the cosmographic parameters $\tilde{\Omega}_m$, $\tilde{\Omega}_r$, and $\tilde{\Omega}_{\phi}$ as a function of redshift for the constant potential $V(\phi) = V_0$ with $\omega = 10$ and $\xi = 1$. The negative values of $\tilde{\Omega}_\phi$ become more pronounced especially towards high redshifts, with larger values for $\tilde{\Omega}_m$ and $\tilde{\Omega}_r$ during their respective domination epochs.}
\label{fig:model1_w_10_evol}
\end{figure}

To understand the effect of not restricting $\omega$ to 1, we shall examine two cases: $\omega = 0.1$, and $\omega = 10$. In this case, as can be seen by comparing Figs. \ref{fig:model1_w_0.1_evol} and \ref{fig:model1_w_10_evol}, having $\omega < 1$ means that $\tilde{\Omega}_m$, $\tilde{\Omega}_r$, and $\tilde{\Omega}_{\phi}$ are all approximately bounded above by 1 throughout the evolution, but $\tilde{\Omega}_{\phi}$ takes negative values; compare this to the evolution obtained with $\xi = 0.1$ and $\omega = 1$ in Figure \ref{fig:model1_0.1_evol}. This does not contradict compactness; rather, it merely means that the trajectory in phase space is orbiting $P_2$, which has $\tilde{\Omega}_{\phi} < 0$, as can be seen in Table \ref{tab:wcase1params}. An important feature of Fig.~\ref{fig:model1_w_0.1_evol} is the redshift of matter-radiation equality, which happens at $1+z_{eq} \sim \mathcal{O}(10^2)$ for $\omega = 0.1$. However, \cite{Planck:2018vyg} $z_{\text{eq}}$ gives strong constraints on $z_{\text{eq}}$ of $\mathcal{O}(10^3)$. Related to this is the length of time the trajectory spends orbiting $P_2$, which represents a matter-dominated universe; a longer matter-dominated era means more time available for the growth of large-scale structure. In this regard, the model with $\omega = 0.1$ can be discarded, and no phase portraits have been presented for this reason.

\begin{figure}[!htb]
\centering
\includegraphics[width=0.8\columnwidth]{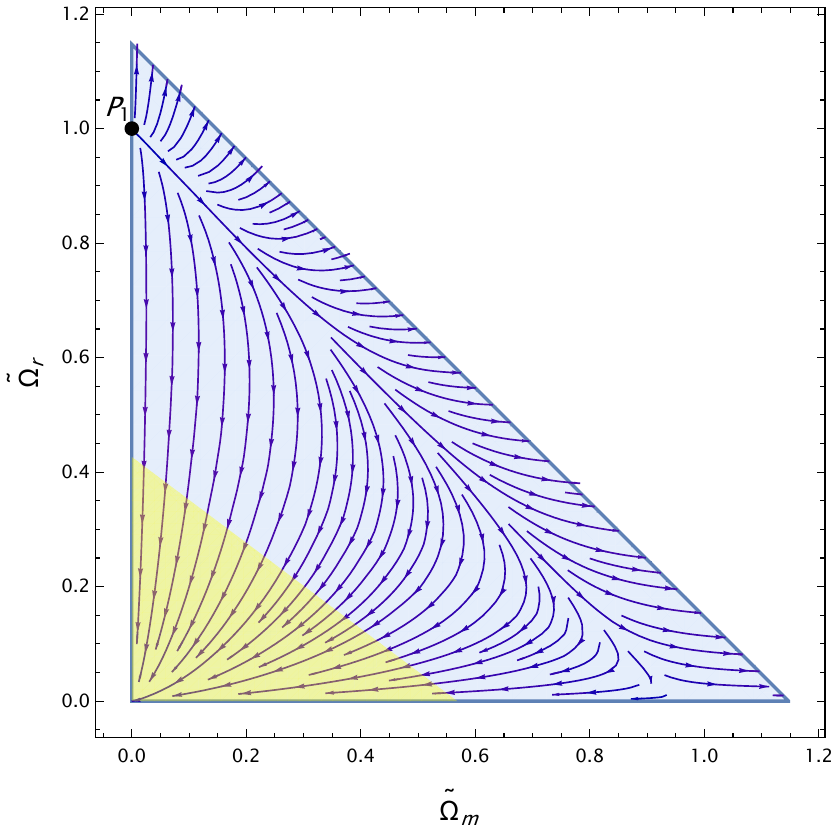}

\vspace{1em}

\includegraphics[width=0.8\columnwidth]{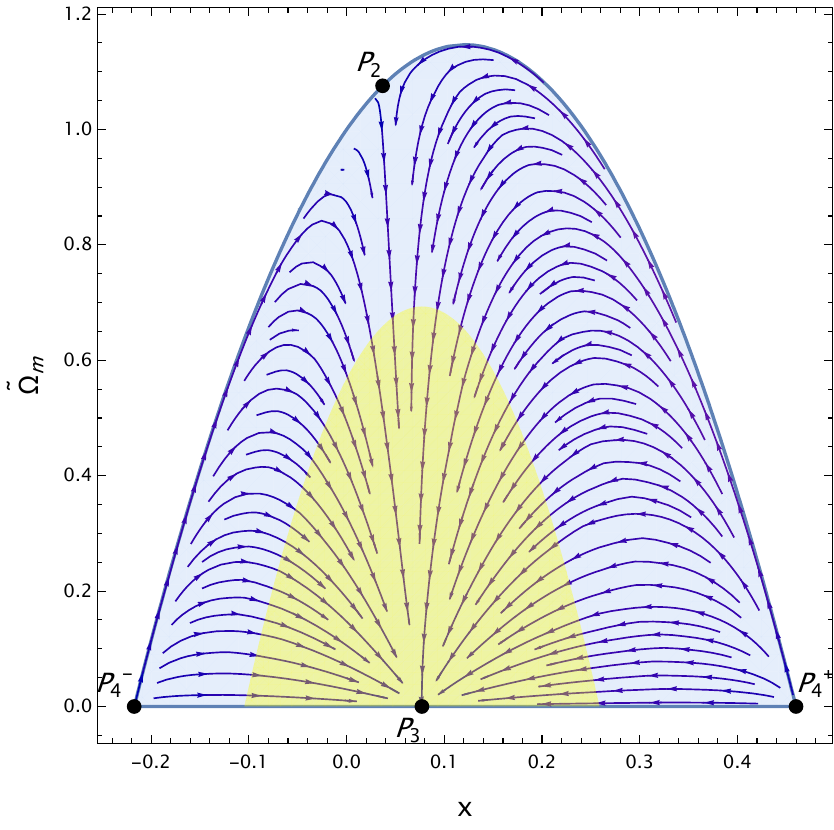}
\caption{Phase portraits for the constant potential $V(\phi) = V_0$ in the $x = 0$ plane (top) and the $\Tilde{\Omega}_r = 0$ plane (bottom) with $\omega = 10$ and $\xi = 1$. The region highlighted in yellow denotes the subset of the phase space where the universe is accelerating. Compared to the $\omega = 1$ and $\xi = 1$ case, the $\Tilde{\Omega}_r$--$\Tilde{\Omega}_m$ phase space exhibits an accelerating region.}
\label{fig:model1_w_10_pp}
\end{figure}

For $\omega = 10$, by comparing the phase portraits in Fig.~\ref{fig:model1_w_10_pp} to those in Fig. \ref{fig:model1_1_pp}, we notice that, for $\xi = 1$, the choice of $\omega = 10$ gives phase portraits which, besides the boundary of the phase space, are closer to those obtained with quintessence models than those produced with $\omega = 1$. For example, one obvious feature that appears in Fig. \ref{fig:model1_w_10_pp} and does not appear in Fig. \ref{fig:model1_1_pp} is the accelerating region in the $x=0$ projection. This suggests that having $\omega > 1$ can allow $\xi \sim \mathcal{O}(1)$. Moreover, $\omega = 10$ gives $1 + z_{\text{eq}} \sim \mathcal{O}(10^3)$, matching observations.

% \begin{figure*}[!htb]
% \centering
% \minipage{0.49\textwidth}
% \includegraphics[width=\linewidth]{Model 1_ω/ωξ=0.1,0.1_MR.pdf}
% \endminipage
% \hfill
% \minipage{0.49\textwidth}
% \includegraphics[width=\linewidth]{Model 1_ω/ωξ=0.1,0.1_XM.pdf}
% \endminipage
% \caption{Phase portraits for the constant potential $V(\phi) = V_0$ in the $x = 0$ plane on the left and the $\Tilde{\Omega}_r = 0$ plane on the right with $\omega = 0.1$ and $\xi = 0.1$. The region highlighted in yellow denotes the subset of the phase space where the universe is accelerating. Observe that compared to $\omega = 1$ and $\xi = 0.1$, there is a shift in the positions of the critical points $P_2$, $P_3$ and $P^{\pm}_4$ due to their $\omega$-dependence. Furthermore, the parameters' maxima and minima change due to changes in the phase-space $\mathcal{P}_\omega$.}
% \label{fig:model1_w_0.1_pp}
% \end{figure*}

\subsection{The Power-Law Potential: $V(\phi) = V_0(1+\xi \phi)^{-n}$}

The power-law potential is usually taken to be of the form $V_0\phi^n$. The advantage of using the form given here is that $\lambda_V$ is constant, thus eliminating the dynamical equation for $\lambda_V$ and the need to compactify the phase space. Nonetheless, requiring $f(\phi) \simeq 1$ reduces this to the standard power-law form, allowing the choice of $\xi$ to clearly distinguish between small and larger deviations from GR. In this case, the critical points are outlined in Table \ref{tab:wcase2points} with their respective physical properties in Table \ref{tab:wcase2params}. Here, the critical points represent the following cosmological behaviours:
\begin{table*}[!htb]
\centering
\begin{tabular}{ccccc}
\hline
$P$ & $x$ & $\tilde{\Omega}_m$ & $\tilde{\Omega}_r$ & Stability \\ 
\hline
$P_1$ & 0 & 0 & 1 & Saddle \\
$P_2$ & $\f{\xi}{\sqrt{6}(\xi^2 + \omega)}$ & $1+\f{6 + 5\varsigma}{6(1 + \varsigma)^2}$ & 0 & Saddle for $n \leq 3$ or $(n > 3 \text{ and } 1 < \f{3\varsigma}{n-3})$ \\
$P_3$ & $-\f{\sqrt{6}(n+2)\xi}{3(n-1)\xi^2 - 6\omega}$ & 0 & 0 & \makecell{Stable node for $\xi = 0$ or \\ $\xi \neq 0$ and $\left[(4n+7)^2< 73 + 48\varsigma \text{ or } n > 1 + 2\varsigma\right]$}\\ 
$P_4^{\pm}$ & $\f{\sqrt{6}}{2\omega}\left(\xi \pm \sqrt{\xi^2 + \f{2}{3}\omega}\right)$ & 0 & 0 & Unstable node \\
$P_5$ & $\sqrt{\f{2}{3}}\f{2}{n\xi}$ & 0 & $1+\f{2(2n-1 - 2\varsigma)}{n^2}$ & Saddle \\
$P_6$ & $\sqrt{\f{3}{2}}\f{1}{n\xi}$ & $1+\f{7n-3 - 6\varsigma}{2n^2}$ & 0 & Saddle \\
\hline
\end{tabular}
\caption{Critical points for the power-law potential $V(\phi) = V_0(1+\xi \phi)^{-n}$, $n \neq 0$. Note that $P_3$ exists provided that $\omega \neq \f{1}{2}(n-1)\xi^2$. Here, $\varsigma \coloneqq \omega/\xi^2$.}
\label{tab:wcase2points}
\end{table*}

\begin{table*}[!htb]
\centering
\begin{tabular}{cccccc}
\hline
Point & $\tilde{\Omega}_{\phi}$ & $w_{\phi}$ & $w_{\text{eff}}$ & $q$ & Acc. \\
\hline
$P_1$ & 0 & - & $\f{1}{3}$ & 1 & No \\
$P_2$ & $-\f{(6 + 5\varsigma)}{6(1 + \varsigma)^2}$ & $-\f{2(1+\varsigma)}{6 + 5\varsigma}$ & $\f{1}{3(1 + \varsigma)}$ & $\f{1}{2} + \f{1}{2(1 + \varsigma)}$ & No \\
$P_3$ & 1 & $-1 -\f{2(n+1)(n+2)}{3(n-1) - 6\varsigma}$ & $-1 -\f{2(n+1)(n+2)}{3(n-1) - 6\varsigma}$ & $-1 -\f{(n+1)(n+2)}{(n-1) - 2\varsigma}$ & \makecell{$\xi = 0$. If $\xi \neq 0$: $[\varsigma > \f{1}{2}(n^2 + 4n + 1)$ \\
and $(n < -2-\sqrt{3}$ or $n > -2 + \sqrt{3})]$; \\
or $-2-\sqrt{3} \leq n < -2 + \sqrt{3}$; \\
or $[1 < n \leq 3 \text{ and } \varsigma < \f{1}{2}(n-1)]$; \\
or $[n > 3 \text{ and } \f{1}{3}(n-3) < \varsigma < \f{1}{2}(n-1)]$} \\
$P_4^{\pm}$ & 1 & $\f{6\xi^2 + 3\omega \pm 2\xi\sqrt{9\xi^2 + 6\omega}}{3\omega}$ & $\f{6\xi^2 + 3\omega \pm 2\xi\sqrt{9\xi^2 + 6\omega}}{3\omega}$ & $\f{3\xi^2 + 2\omega \pm \xi\sqrt{9\xi^2 + 6\omega}}{\omega}$ & No \\
$P_5$ & $-\f{2(2n-1 - 2\varsigma)}{n^2}$ & $\f{1}{3} + \f{2}{3}\f{3n-1}{3(1-2n) + 2\varsigma}$ & $\f{1}{3} + \f{4}{3n}$ & $1 + \f{2}{n}$ & $-2 < n < 0$ \\
$P_6$ & $-\f{7n-3 - 6\varsigma}{2n^2}$ & $\f{2n}{(3-7n) + 6\varsigma}$ & $\f{1}{n}$ & $\f{1}{2} + \f{3}{2n}$ & $-3 < n < 0$ \\
\hline
\end{tabular}
\caption{Physical properties of the critical points for the power-law potential $V(\phi) = V_0(1+\xi \phi)^{-n}$, $n \neq 0$. For the $P_1$ critical point, $\omega_\phi$ is undefined. Nonetheless, it does not contribute to the dynamics as the effective scalar-field density is zero. Here, $\varsigma \coloneqq \omega/\xi^2$.}
\label{tab:wcase2params}
\end{table*}
\begin{itemize}
    \item $P_1$: this is a standard radiation-dominated era, as appearing in quintessence;
    \item $P_2$: this is similar to $P_2$ for the constant potential, in that both the scalar and pressureless matter contribute to the total energy density. As such, the ratio of $\tilde{\Omega}_{\phi}$ to $\tilde{\Omega}_m$ and the related discussion are unchanged. However, the stability now depends on $n$. For this point to correspond to the matter-dominated epoch of the universe, we must choose $n$ and $\xi$ such that $n < 3\left(1+\f{\omega}{\xi^2}\right)$ to make this point a saddle;
    \item $P_3$: this point is dominated by the scalar field. We recognise that this is the only critical point of the system which can correspond to a late-time accelerating universe. As such, we can impose restrictions on $\omega$, $n$, and $\xi$ to ensure that this is the case. Example cases are discussed based on some parameter choices in the forthcoming phase space discussion;
    \item $P_4^{\pm}$: these points correspond to a universe dominated by the kinetic energy of the scalar field, and, as such, cannot serve to represent a dark energy-dominated universe;
    \item $P_5$: this epoch contains contributions from both the scalar and radiation. To minimize the contribution of $\phi$ and have this point serve as the standard radiation era, we must have $n \rightarrow \infty$. In this limit, $\tilde{\Omega}_r\rightarrow 1^-$, and $\tilde{\Omega}_{\phi} \rightarrow 0^+$. For $-2 < n < 0$, the universe can undergo an accelerated expansion; in this case, the effective EoS $w_{\text{eff}}$ is bounded above by $-2$, and the deceleration parameter has no lower bound, but the EoS of the field, $w_{\phi}$, will be positive for $\omega > 0$ and all values of $\xi$, so this does not correspond to a Big Rip scenario; 
    \item $P_6$: this has similar properties as $P_5$, but for matter rather than radiation. It can only serve as a standard matter era for $n \rightarrow \infty$. Similarly to $P_5$, we can have a universe with an accelerating expansion in this case for $-3 < n < 0$, with the effective EoS then being bounded above by $-3$. For $n$ in this range, the scalar EoS, $w_{\phi}$, is bounded above by 0.
\end{itemize}

\begin{figure}[!htb]
\includegraphics[width=0.8\columnwidth]{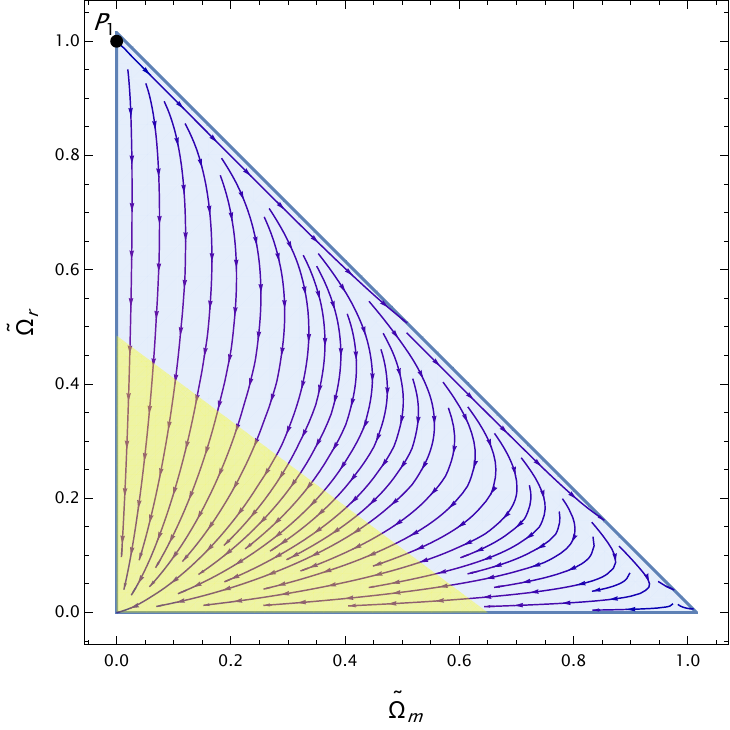}

\vspace{1em}

\includegraphics[width=0.8\columnwidth]{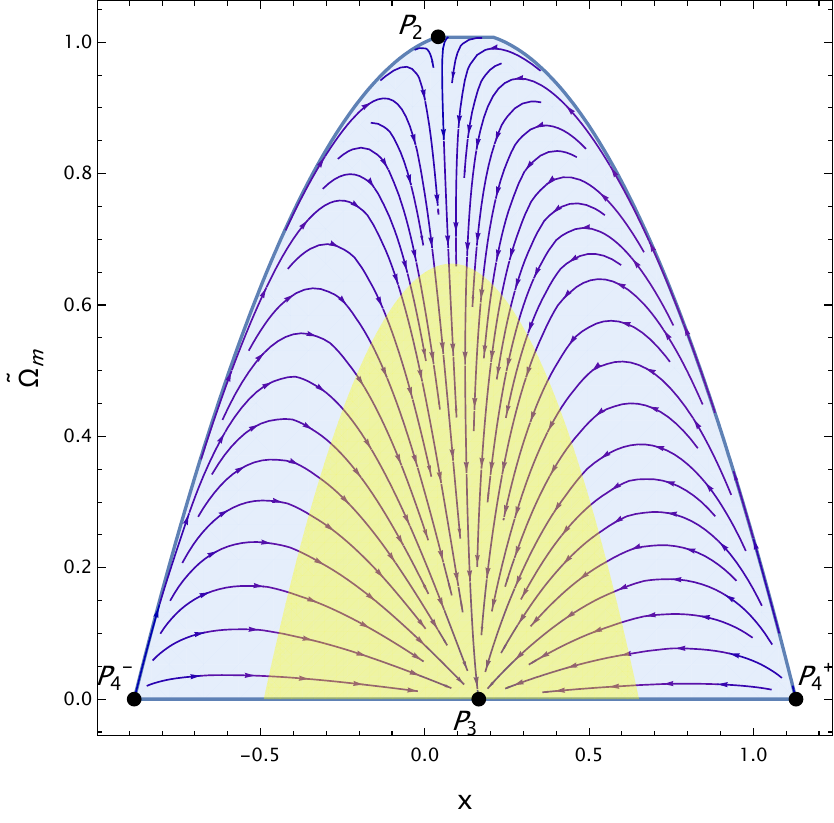}
\caption{Phase portraits for the power-law potential $V(\phi) = V_0(1+\xi \phi)^{-n}$ in the $x = 0$ plane (top) and the $\Tilde{\Omega}_r = 0$ plane (bottom) with $n = 2$, $\omega = 1$, and $\xi = 0.1$. The region highlighted in yellow denotes the subset of the phase space where the universe is accelerating.}
\label{fig:model2_2_0.1_pp}
\end{figure}

\begin{figure}[!htb]
\includegraphics[width=0.8\columnwidth]{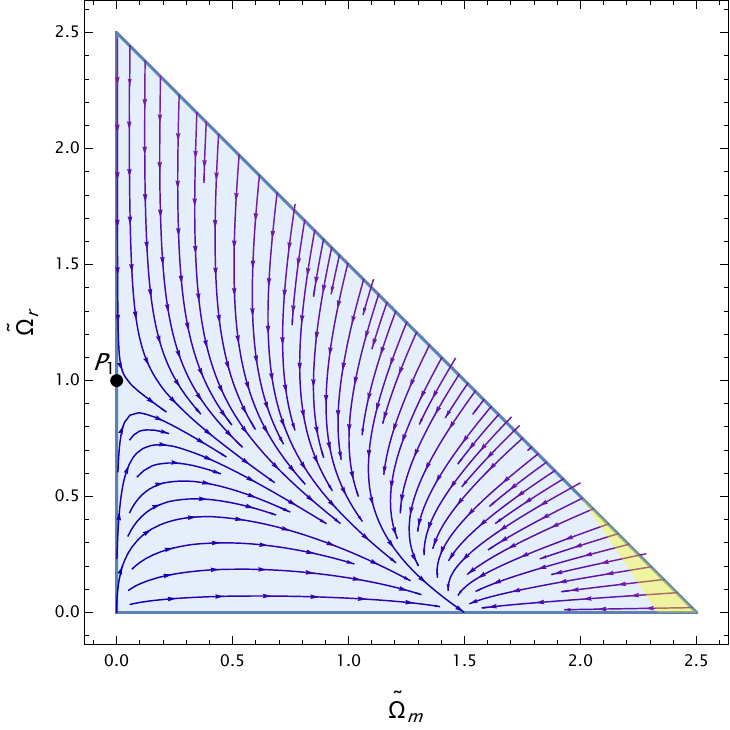}

\includegraphics[width=0.8\columnwidth]{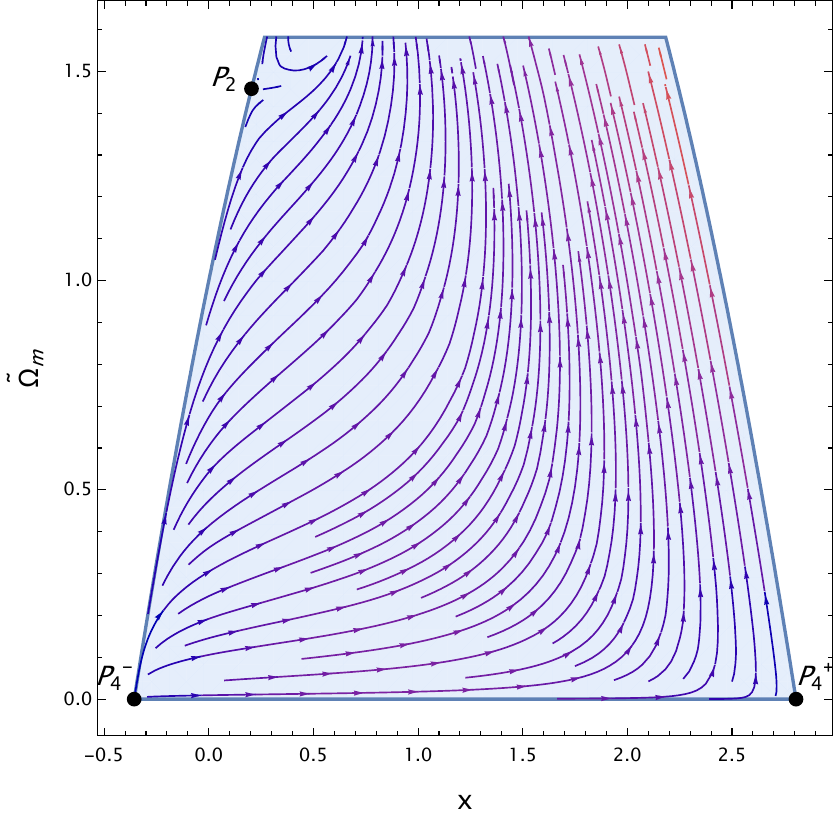}
\caption{Phase portraits for the power-law potential $V(\phi) = V_0(1+\xi \phi)^{-n}$ in the $x = 0$ plane (top) and the $\Tilde{\Omega}_r = 0$ plane (bottom) with $n = 2$, $\omega = 1$, and $\xi = 1$. Observe that acceleration (the region highlighted in yellow) is predominantly absent.}
\label{fig:model2_2_1_pp}
\end{figure}

Phase portraits for $n = 2$ for $\omega = 1$ with $\xi = 0.1$ and $\xi = 1$ can be seen in Figs. \ref{fig:model2_2_0.1_pp} and \ref{fig:model2_2_1_pp} respectively. For a physical evolution, the phase space trajectory should start at $P_1$, pass close to $P_2$, and end at $P_3$. This means that $P_2$ should be a saddle, and $P_3$ should be a stable node. Moreover, $P_3$ should have properties similar to that of a cosmological constant, i.e. the effective equation of state should be negative, and the expansion of the universe should be accelerating. To determine the ranges of values of $\omega$, $\xi$ and $n$ which allow the correct sequence of critical points to exist, we restrict all the eigenvalues of $P_3$ to be negative, $q$ and $w_{\text{eff}}$ evaluated at $P_3$ to be negative, and the eigenvalue $l_1$ for the point $P_2$ to be positive (the other eigenvalues are negative for $\omega > 0$; see Appendix~\ref{app:eigenvalues} Table~\ref{tab:wcase2eigen}). To proceed, we specify the value of $\xi$ and obtain the corresponding bounds on $\omega$ and $n$. For example, $\xi \sim \mathcal{O}(10^{-9})$ gives bounds on $\omega$ and $n$ of the order $\mathcal{O}(10^9)$, meaning that virtually any choice of $(n,\omega)$ will give the correct sequence of critical points. However, as $\xi$ increases, the range of allowed $n$ becomes significantly more restricted. Below, we give the expressions for $\xi = 1$:
\begin{itemize}
    \item $\omega > \f{1}{2}(1+4n+n^2)$ and $(n \leq -2 - \sqrt{3} \text{ or } n < -2 + \sqrt{3})$;
    \item $-2-\sqrt{3} < n \leq -2 + \sqrt{3}$;
    \item $1 < n \leq 3$ and $\omega < \f{1}{2}(n-1)$;
    \item $n > 3$ and $\f{1}{3}(n-3) < \omega < \f{1}{2}(n-1)$.
\end{itemize}
% In principle, we can also specify $\omega$. For example, for $(\omega, \xi) = (1,1)$, we have $3 < n < 6$ or $-2-\sqrt{5} < n < -2 + \sqrt{5}$.

\begin{figure}[!htb]
\centering
\includegraphics[width=0.9\linewidth]{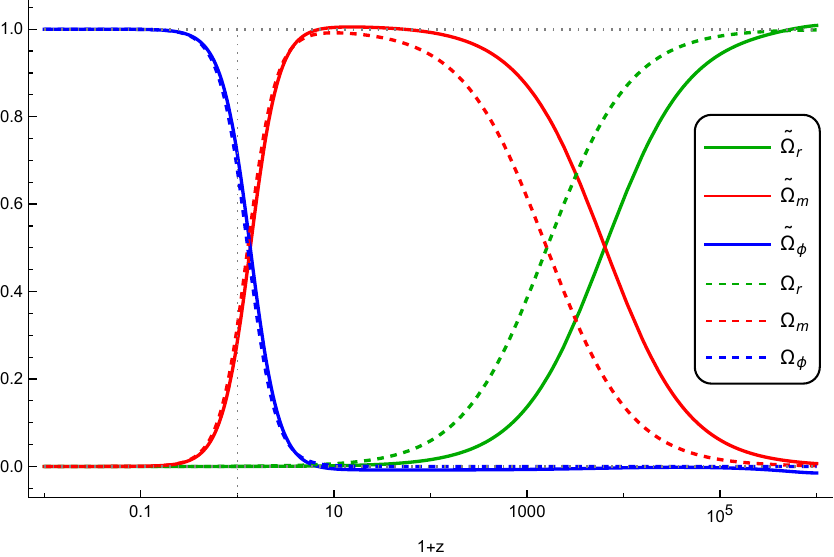}
\caption{The evolution of the cosmographic parameters $\tilde{\Omega}_m$, $\tilde{\Omega}_r$, and $\tilde{\Omega}_{\phi}$ as a function of redshift (solid lines) for the power-law potential $V(\phi) = V_0(1+\xi \phi)^{-n}$ with $\omega = 1$, $n = 2$ and $\xi = 0.1$. The evolution for the $\Lambda$CDM cosmographic parameters (dashed lines) is also plotted for comparison.}
\label{fig:model2evol1}
\end{figure}

Comparing the phase portraits, it is clear that, for $n = 2$, the value of $\xi = 0.1$ leads to a physical evolution. Additionally, the phase space flow in Fig.~\ref{fig:model2_2_1_pp} reflects the constraints on $(n,\xi)$ obtained by investigating the stability properties of the critical points of the system: $n = 2$ is not in the set of values of $n$ which simultaneously make $P_2$ a saddle and $P_3$ a stable node for $\xi = 1$. Fig.~\ref{fig:model2evol1} can be used to compared the evolution in the $\Lambda$CDM model to that for $\xi = 0.1$. We see that the late-time behaviour is relatively unchanged, with the effect of the non-minimal coupling and potential being to shift matter-radiation equality to an earlier redshift and making the matter-dominated era longer as a result. It should be noted that, in minimally coupled theories like quintessence, scalar fields with inverse power-law potentials often exhibit tracking behaviour, which is important in avoiding the coincidence problem \cite{Zlatev_1999, Steinhardt_1999}. However, in the analysis performed here, no tracking behaviour was observed. This is expected as tracking solutions were not initially constructed for theories with a non-minimal coupling. 

\begin{figure}[!htb]
\centering
\includegraphics[width=0.8\columnwidth]{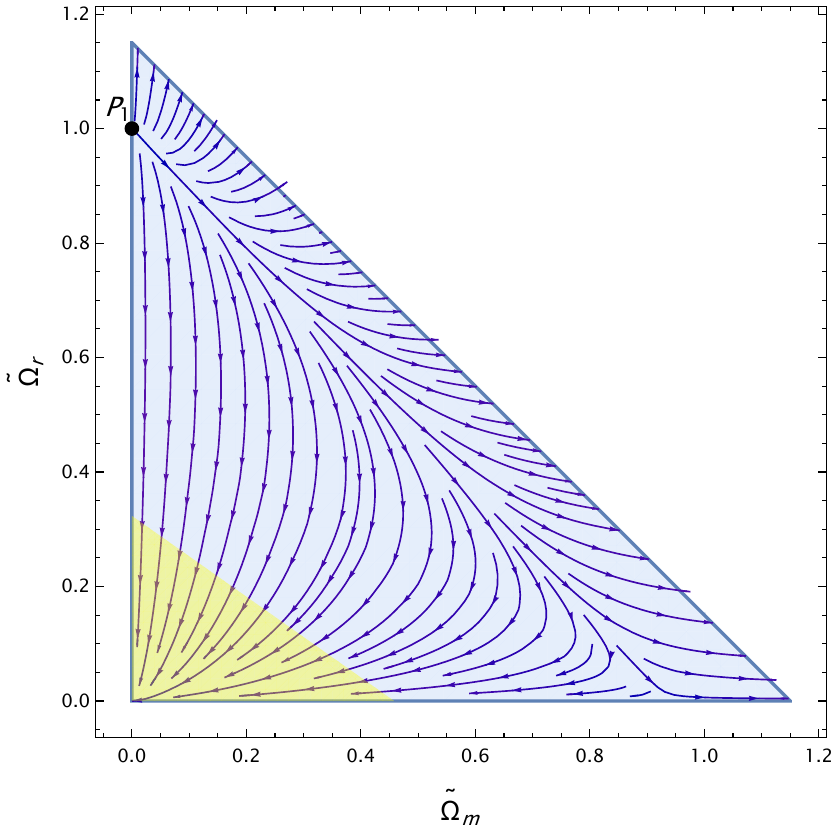}

\vspace{1em}

\includegraphics[width=0.8\columnwidth]{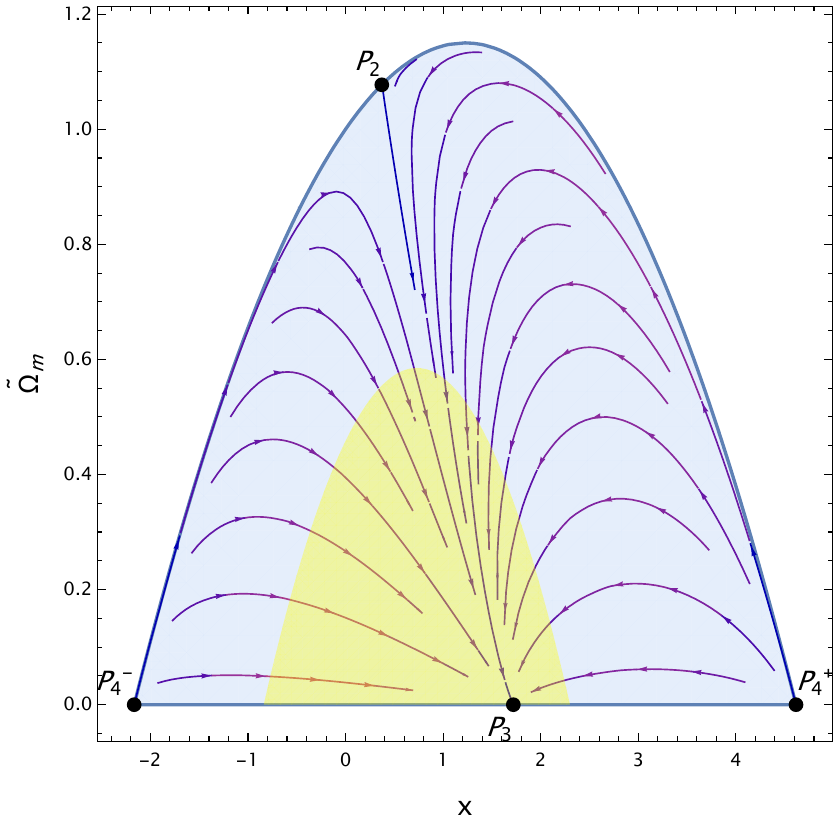}
\caption{Phase portraits for the power-law potential $V(\phi) = V_0(1+\xi \phi)^{-n}$ in the $x = 0$ plane (top) and the $\Tilde{\Omega}_r = 0$ plane (bottom) with $n = 2$, $\omega = 0.1$, and $\xi = 0.1$. For this case, compared to $\omega = 1$, there is a shift in the positions of the critical points $P_2$, $P_3$ and $P^{\pm}_4$ due to their $\omega$-dependence, in addition to a change in the region where there is an accelerated expansion.}
\label{fig:model2_w_0.1_pp}
\end{figure}

\begin{figure*}[!htb]
\centering
\minipage{0.4\textwidth}
\includegraphics[width=\linewidth]{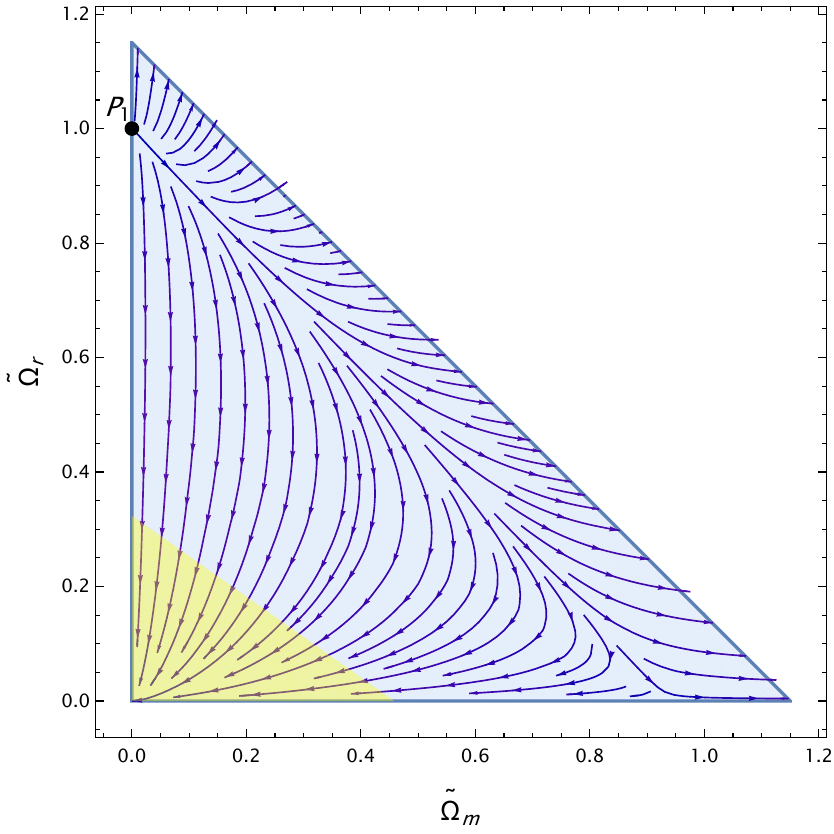}
\endminipage
\hspace{0.1\textwidth}
\minipage{0.4\textwidth}
\includegraphics[width=\linewidth]{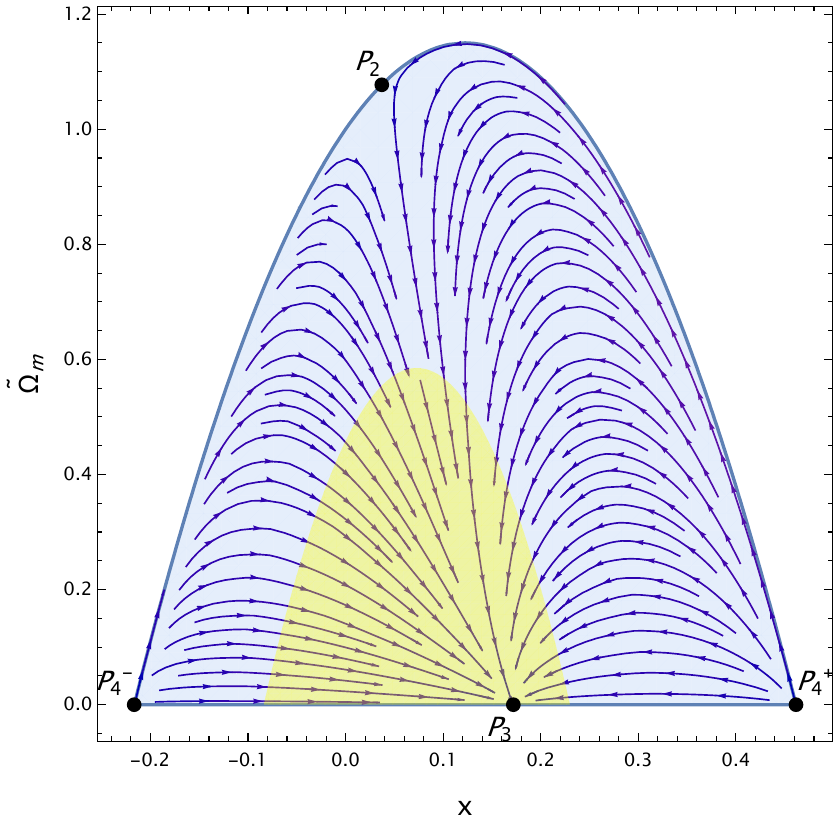}
\endminipage
\caption{Phase portraits for the power-law potential $V(\phi) = V_0(1+\xi \phi)^{-n}$ in the $x = 0$ plane on the left and the $\Tilde{\Omega}_r = 0$ plane on the right with $n = 2$, $\omega = 10$, and $\xi = 1$. Similar to the $\omega = 0.1$ and $\xi = 0.1$ case, a shift in the critical points and acceleration region is observed.}
\label{fig:model2_w_10_pp}
\end{figure*}

When $\omega \neq 1$, as with the constant potential, the critical points have a dependence on $\omega$, but their properties remain unchanged from those for $\omega = 1$. To understand how varying $\omega$ affects the dynamics in this case, we shall examine the $n = 2$ potential for $(\omega,\xi) = (0.1,0.1)$ and $(\omega,\xi) = (10,1)$. Excluding the size of the accelerating region, there is no significant difference between the phase portraits in Fig.~\ref{fig:model2_w_0.1_pp} and those in Fig.~\ref{fig:model2_2_0.1_pp}. However, comparing the phase portraits for $(\omega,\xi) = (10,1)$ in Fig. \ref{fig:model2_w_10_pp} to those for $(\omega,\xi) = (1,1)$ in Fig. \ref{fig:model2_2_1_pp}, it is very clear that allowing $\omega$ to vary affects the dynamics; in the latter case, it follows immediately from the phase portraits that it is not possible to find a physical evolution. 

\begin{figure*}[!htb]
\centering
\minipage{0.4\textwidth}
\includegraphics[width=\linewidth]{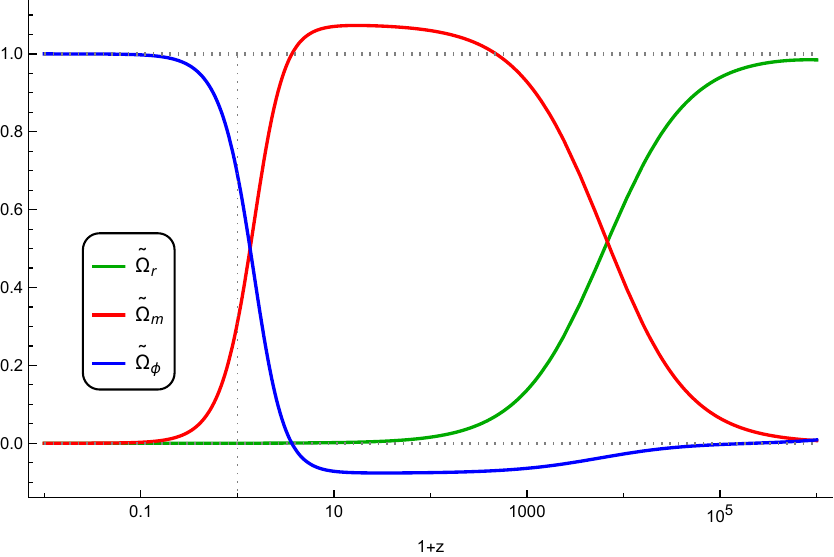}
\endminipage
\hspace{0.1\textwidth}
\minipage{0.4\textwidth}
\includegraphics[width=\linewidth]{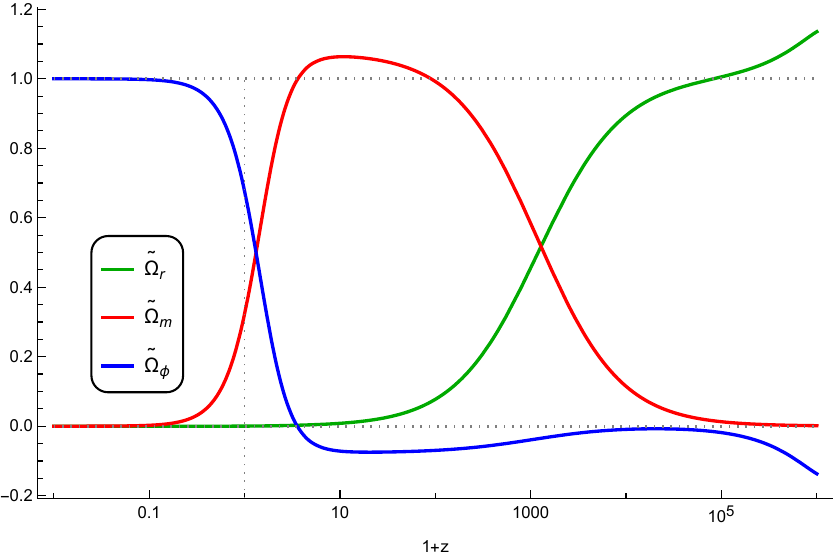}
\endminipage
\caption{The evolution of the cosmographic parameters $\tilde{\Omega}_m$, $\tilde{\Omega}_r$, and $\tilde{\Omega}_{\phi}$ as a function of redshift for the power-law potential $V(\phi) = V_0(1+\xi \phi)^{-n}$ with $n = 2$ for $\omega = 0.1$ and $\xi = 0.1$ for the left figure and $\omega = 10$ and $\xi = 1$ for the right figure. Evidently, there is a difference in the evolutionary growth especially at higher redshift and the time when matter-radiation equality occurs.}
\label{fig:model2_w_evol}
\end{figure*}

This effect is reflected in Fig.~\ref{fig:model2_w_evol}, where we can see that taking $\omega \neq 1$ makes the orbit around the matter-dominated fixed point $P_2$ clear, with $\tilde{\Omega}_{\phi}$ taking negative values during this epoch. The choice of $(\omega,\xi) = (10,1)$ still gives a viable evolution, keeping in mind that $\tilde{\Omega}_r$ and $\tilde{\Omega}_m$ are not the physical densities. However, $(\omega,\xi) = (0.1,0.1)$ now also gives an evolution which mimics that of $\Lambda$CDM. Moreover, the matter-dominated epoch lasts longer in this case, with matter-radiation equality still taking place at $1+z \sim \mathcal{O}(10^3)$. This suggests that, for the power-law potential, having $\omega < 1$ can still lead to a physical evolution for $\xi < 1$, which remains consistent with the assumption of small deviations from GR. 

As a final point, we show that we recover the cosmological dynamics arising from standard Brans-Dicke theory with a power-law potential. Under $n \mapsto 2n$, we find that, up to a redefinition of the dynamical variables which only involves numerical pre-factors, the points $P_2$, $P_3$, $P_4^{\pm}$, and $P_5$ correspond exactly to the critical points as presented in \cite{Bahamonde_2018}, where the authors consider a universe containing only pressureless dust. As such, the critical points $P_1$ and $P_6$ do not appear in their analysis.

\subsection{The Exponential Potential: $V(\phi) = V_0 e^{-\lambda \phi}$} 

The exponential potential is a common test case in scalar-tensor models of dark energy, both due to its mathematical simplicity, and also because it can be easily motivated using high-energy physics (see \cite{baumann2014inflationstringtheory} for example). It has been extensively studied using a dynamical systems approach in sub-classes of the general scalar-tensor theory investigated here (for examples in quintessence, see \cite{Copeland_1998, Urena_Lopez_2012, Tamanini_2014}). However, the analysis presented here is distinct as it involves a non-minimal gravitational coupling, which means that we expect the dynamical behaviour to change. One immediate observation is that, due to the non-minimal coupling, $\lambda_V$ is now dynamical - the equivalent parameter for quintessence would be constant and equal to the slope of the exponential potential. Using the aforementioned change of variables to compactify the phase space, the equation for $u = \arctan \lambda_V$ becomes:
\begin{equation}
u' = \sqrt{6}\xi x \sin u \cos u.
\label{ueqn}
\end{equation}

\begin{table*}[!htb]
\centering
\begin{tabular}{cccccc}
\hline
Point & $x$ & $\Tilde{\Omega}_m$ & $\Tilde{\Omega}_r$ & $u$ & Stability \\ 
\hline
$P_1$ & 0 & 0 & 1 & 0 & Saddle\\
$P_2$ & $\f{\xi}{\sqrt{6}(1+\xi^2)}$ & $1 + \f{\xi^2(5+6\xi^2)}{6(1+\xi^2)^2}$ & 0 & 0 & Saddle\\
$P_3$ & $\f{2\xi}{2+\xi^2}\sqrt{\f{2}{3}}$ & 0 & 0 & 0 & Saddle \\
$P_4^{\pm}$ & $\f{\sqrt{6}}{2}\left(\xi \pm \sqrt{\f{2}{3}+\xi^2}\right)$ & 0 & 0 & 0 & $P_4^+$ is a source and $P_4^-$ is a saddle for $\xi > 0$ (vice-versa for $\xi < 0$) \\
$P_5$ & 0 & 0 & 0 & $-\arctan 2 \xi$ & Stable node\\
\hline
\end{tabular}
\caption{Critical points for the exponential potential $V(\phi) = V_0 e^{-\lambda\phi}$.}
\label{tab:case4points}
\end{table*}

\begin{table*}[!htb]
\centering
\begin{tabular}{cccccc}
\hline
Point & $\tilde{\Omega}_{\phi}$ & $w_{\phi}$ & $w_{\text{eff}}$ & $q$ & Acc.\\ 
\hline
$P_1$ & 0 & - & $\f{1}{3}$ & 1 & No\\
$P_2$ & $\f{\xi^2(5+6\xi^2)}{6(1+\xi^2)^2}$ & $-\f{1}{3} - \f{1}{3(5+6\xi^2)}$ & $\f{\xi^2}{3(1+\xi^2)}$ & $\f{1}{2} + \f{\xi^2}{2(1+\xi^2)}$ & No\\
$P_3$ & 1 & $-1 + \f{4\xi^2}{3(\xi^2 + 2)}$ & $-1 + \f{4\xi^2}{3(\xi^2 + 2)}$ & $-1 + \f{2\xi^2}{\xi^2 + 2}$ & $|\xi| < \sqrt{2}$\\
$P_4^{\pm}$ & 1 & $1 + 2\xi^2 \pm 2\xi\sqrt{\f{2}{3}+\xi^2}$ & $1 + 2\xi^2 \pm 2\xi\sqrt{\f{2}{3}+\xi^2}$ & $2 + 3\xi^2 \pm \xi\sqrt{6+9\xi^2}$ & No\\
$P_5$ & 1 & $-1$ & $-1$ & $-1$ & Always\\
\hline
\end{tabular}
\caption{Physical properties of the critical points for the exponential potential $V(\phi) = V_0 e^{-\lambda\phi}$. For the critical point $P_1$, $w_\phi$ is undefined. Nonetheless, it does not contribute to the dynamics as the effective scalar-field density is zero.}
\label{tab:case4params}
\end{table*}

For simplicity, only the case for $\omega = 1$ has been explored. This is further justified by the forthcoming phase space analysis and discussion, in which we find that the dynamics are independent of the slope of the potential, $\lambda$. This means that, specifically for this scalar-tensor model, $\lambda$ can be fixed, effectively reducing this to the constant potential case. Hence, the analysis for $\omega \neq 1$ would reduce to that discussed previously. The critical points are tabulated in Table~\ref{tab:case4points}, with their physical properties in Table~\ref{tab:case4params}. In this case:

\begin{itemize}
    \item $P_1$: this corresponds to $P_1$ for the constant potential; it has one additional null eigenvalue, making it non-hyperbolic. However, a centre manifold analysis is not necessary here as both the stable and unstable subspaces are non-empty here, making this point a saddle;
    \item $P_2$: both the scalar and matter contribute to this era. It occurs at $\lambda_V = 0$, meaning that it corresponds exactly to $P_2$ for the constant potential. The only difference is the addition of one positive eigenvalue, but the point remains a saddle;
    \item $P_3$: this is similar to the critical point $P_3$ for the constant potential, with the exception of having one additional eigenvalue which is strictly positive, making it a saddle. Thus, it cannot serve as a late-time de Sitter attractor;
    \item $P_4^{\pm}$: this has similar properties to $P_4^{\pm}$ for the constant potential, meaning that the universe is entirely dominated by the kinetic energy of the scalar field, and $w_{\phi}$, $q$ and $w_{\text{eff}}$ are all positive.
    
    For $\xi > 0$, $P_4^+$ is a source, while $P_4^-$ is a saddle, and vice-versa for $\xi < 0$.   
    \item $P_5$: this is fully dominated by the scalar field, and is dark and accelerating. Moreover, it is an attractor for all values of $\xi$, making it a good candidate for a late-time dark energy attractor. This critical point is new in the context of dark energy sourced by a scalar field with an exponential potential due to the presence of the coupling contribution $f(\phi)$ - it cannot appear in standard quintessence models with the exponential potential as these have constant $\lambda_V$.
\end{itemize}

\begin{figure}[!htb]
\centering
\includegraphics[width=0.8\columnwidth]{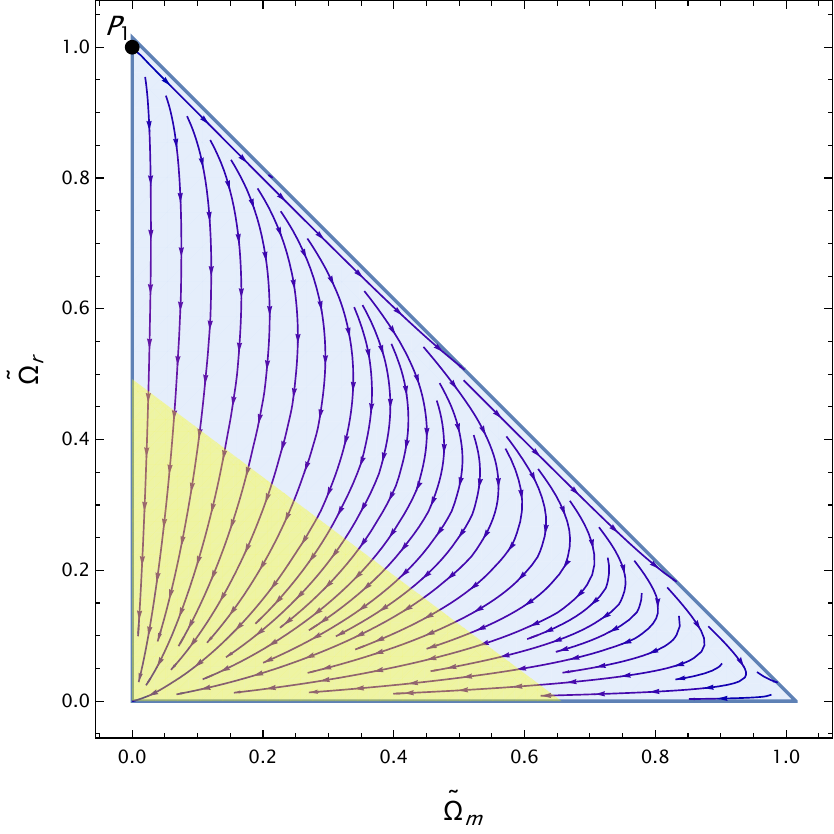}

\vspace{1em}

\includegraphics[width=0.8\columnwidth]{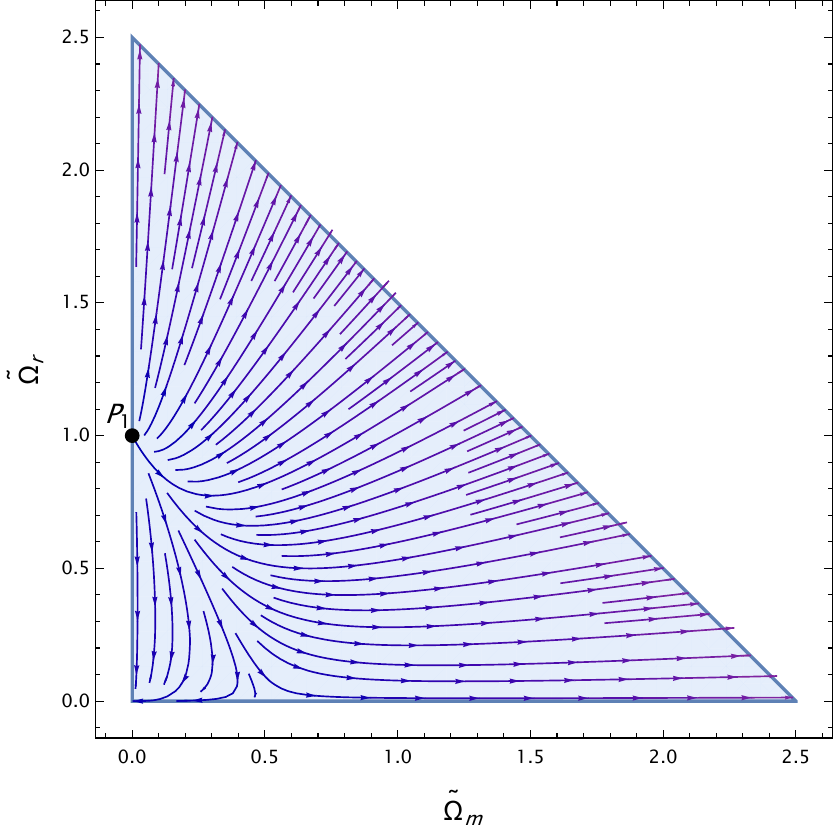}
\caption{$\Tilde{\Omega}_r$--$\Tilde{\Omega}_m$ phase portraits for the exponential potential $V(\phi) = V_0 e^{-\lambda \phi}$ in the $(x,u) = (0,0)$ projection,  for $\omega = 1$ with $\xi = 0.1$ and 1 for the top and bottom figures respectively. Clearly, $\xi = 1$ does not produce a region of universe acceleration even when there is a scalar-field domination. This is due to $P_5$ not being present within this projection and hence the trajectory occurs in a different plane.}
\label{fig:model3_P1}
\end{figure}

\begin{figure}[!htb]
\centering
\includegraphics[width=0.8\columnwidth]{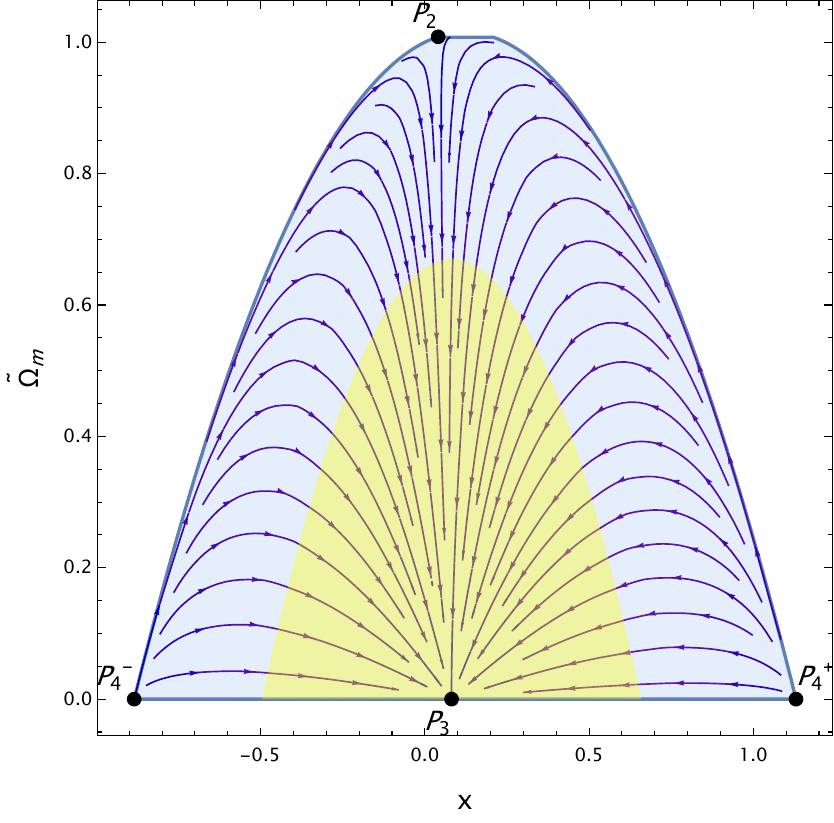}

\vspace{1em}

\includegraphics[width=0.8\columnwidth]{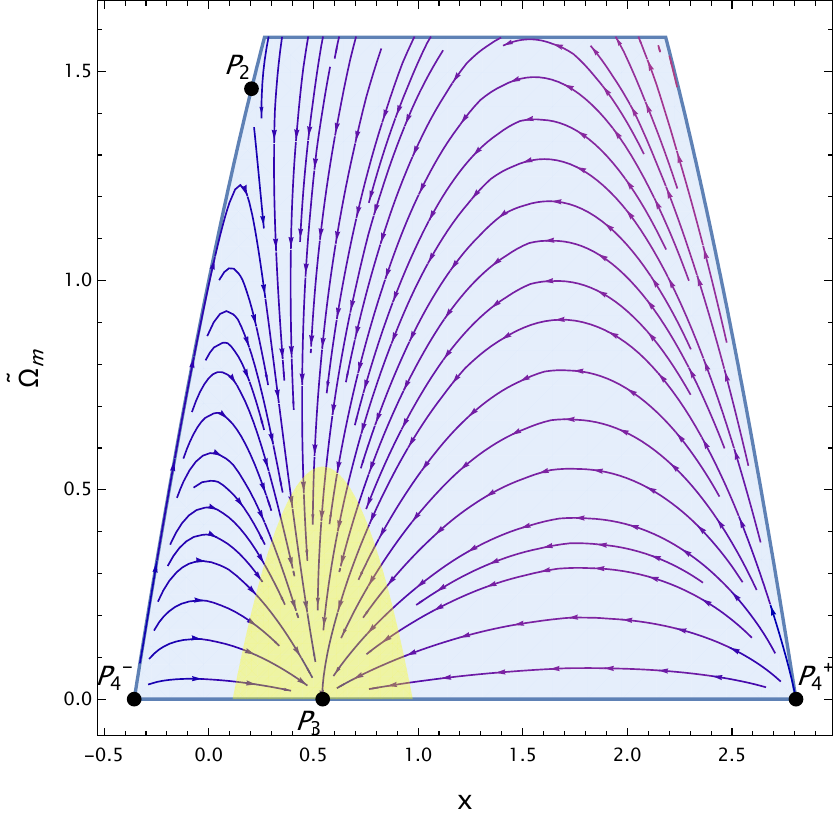}
\caption{$\Tilde{\Omega}_m$--$x$ phase portraits for the exponential potential $V(\phi) = V_0 e^{-\lambda \phi}$ in the $(\tilde{\Omega}_r,u) = (0,0)$ projection, for $\omega = 1$ with $\xi = 0.1$ and 1 for the top and bottom figures respectively. In this case, the accelerating region close to the scalar-field domination is realised through $P_3$, although it is not a late-time attractor, with the saddle behaviour not being realised within this projection.}
\label{fig:model3_P234}
\end{figure}

\begin{figure}[!htb]
\centering
\includegraphics[width=0.8\columnwidth]{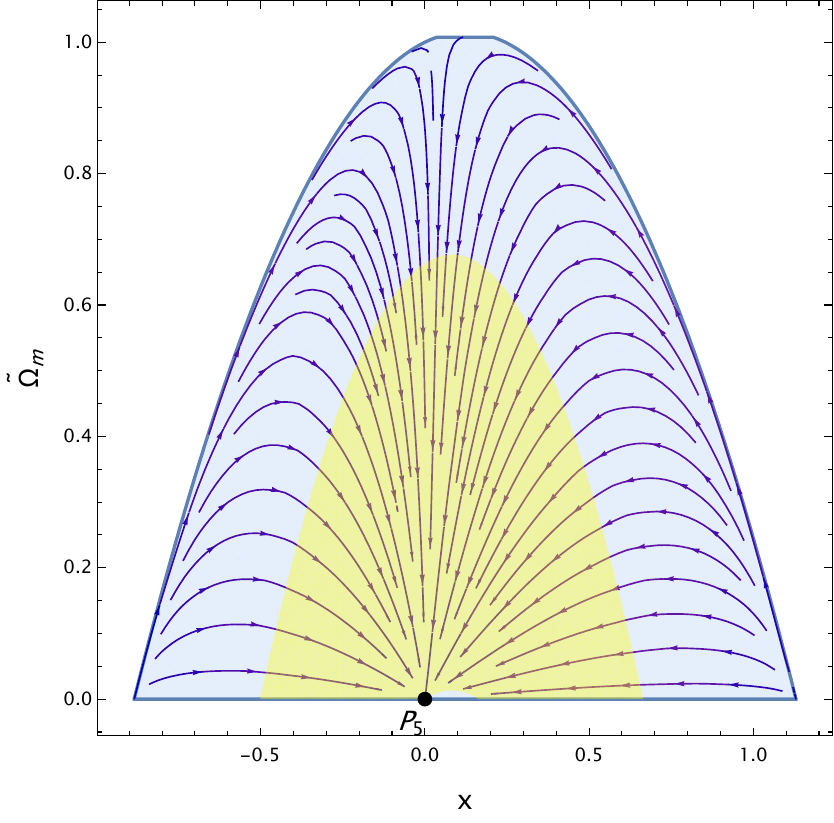}

\vspace{1em}

\includegraphics[width=0.8\columnwidth]{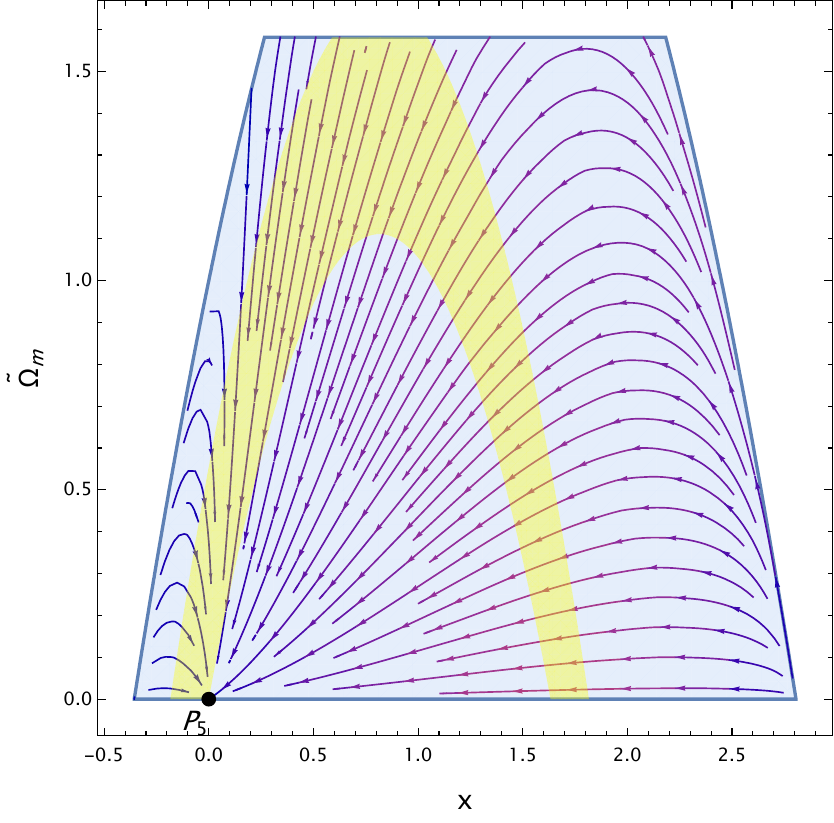}
\caption{$\Tilde{\Omega}_m$--$x$ phase portraits for the exponential potential $V(\phi) = V_0 e^{-\lambda \phi}$ in the $(\tilde{\Omega}_r,u) = (0,-\arctan(2\xi))$ projection, for $\omega = 1$ with $\xi = 0.1$ and 1 for the left and right figures respectively. Since the $u = -\arctan(2\xi)$ projection is considered, the presence of the accelerating scalar field late-time attractor, $P_5$, is now visible.}
\label{fig:model3_P5}
\end{figure}

\begin{figure}[!htb]
\centering
\includegraphics[width=0.8\columnwidth]{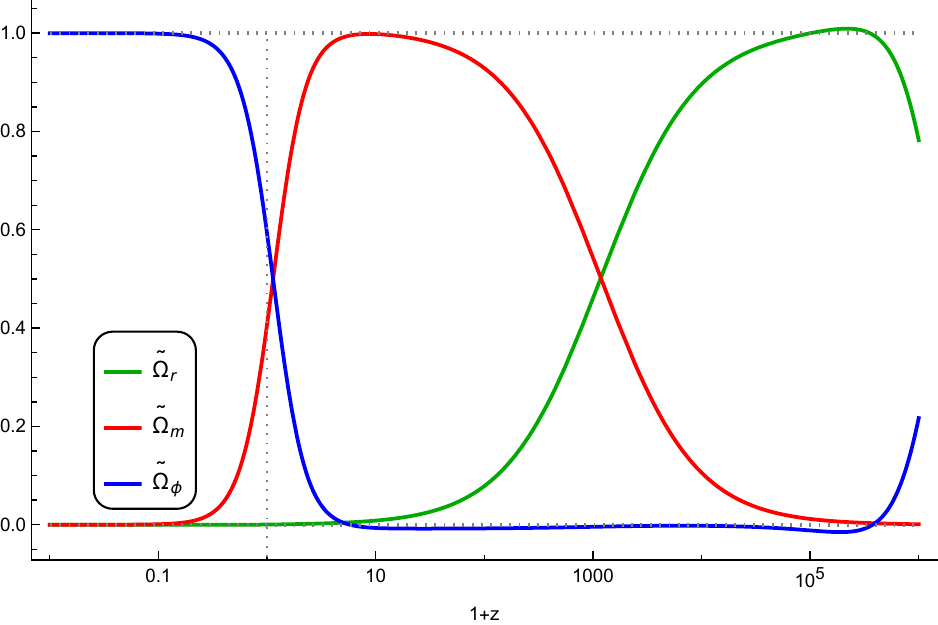}

\vspace{1em}

\includegraphics[width=0.8\columnwidth]{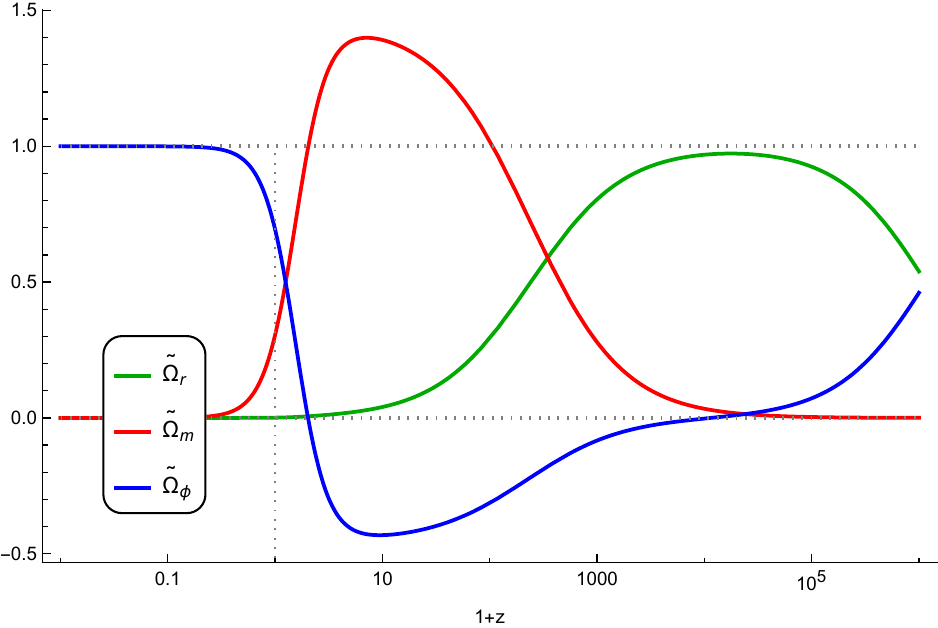}
\caption{The evolution of the cosmographic parameters $\tilde{\Omega}_m$, $\tilde{\Omega}_r$, and $\tilde{\Omega}_{\phi}$ as a function of redshift for the exponential potential $V(\phi) = V_0 e^{-\lambda \phi}$, for $\omega = 1$ with $\xi = 0.1$ and 1 for the left and right figures respectively. $\xi = 0.1$ evolves closer to $\Lambda$CDM behaviour, whereas a larger shift in the modified matter and scalar-field densities is observed for $\xi = 1$ during the matter-dominated era. Yet, both models show a clear shift in early scalar field domination. Additionally, matter-radiation equality occurs later for $\xi = 1$.}
\label{fig:model3_evol}
\end{figure}

By examining the critical points, it is already clear that the behaviour of the system with the exponential potential deviates only slightly from that with the constant potential. This is reflected in the phase portraits of the system, which are found in Figs.~\ref{fig:model3_P1}, \ref{fig:model3_P234}, and \ref{fig:model3_P5}. Two sets of phase portraits are presented, for $\xi = 0.1$ and 1, showing how the choice of $\xi$ affects the dynamics. As has been clear for both the constant potential and the power-law potential, $\xi = 0.1$ is more consistent with $\Lambda$CDM. In this case, this is further reflected in Fig.~\ref{fig:model3_evol}, where the evolution is plotted for both $\xi = 0.1$ and $\xi = 1$. 

Although, with further fine-tuning, it may be possible to find a physically viable evolution with $\xi = 1$, the evidence thus far is consistent with $|\xi| < 1$. Interestingly, in this case, the evolution for $\xi = 0.1$, while still mimicking that of the $\Lambda$CDM model, shows that, for larger redshifts, the scalar field dominates. This is due to the fixed points $P_4^{\pm}$. While these points exist in all the models examined so far, kination has been successfully suppressed in the evolution of the universe. In this case, this may still be achievable with fine-tuning of boundary conditions. However, a caveat of this model is that the phase space is 4-D, since $\lambda_V$ is not constant. Because $\lambda_V$ (or its compact equivalent $u$), is not a physical quantity, it is more difficult to choose appropriate boundary conditions. Moreover, the slope of the potential, $\lambda$, remains a free parameter as it cannot be constrained through the dynamics. This suggests that, if the dynamical behaviour of these models is to be explored further at background level (for example, in a phase space analysis with $\omega < 0$), the exponential potential can be discarded in favour of the simpler constant potential.

\section{Conclusion} \label{sec:conc}

In this work, the cosmological dynamics for a non-canonical generalised Brans-Dicke theory were explored, for the specific couplings $f(\phi) = 1 + \xi\phi$ and $K(\phi,X) = \frac{\omega X}{1 + \xi\phi} - V(\phi)$ for three different potentials. By setting constraints to conform with Solar System tests, growth of structure and avoidance of instabilities, the respective phase space variables and dynamics were explored.

Starting with the constant potential, the critical points showcase a sequence which can start from a non-accelerating scalar-field dominated point, passing through radiation and matter dominated epochs, and ending with another scalar-field dominated attractor point. The final point can represent the late-time accelerated universe provided $\omega > \frac{\xi^2}{2}$, which is easily achieved for virtually any value of $\omega$ when $|\xi| \ll 1$. Yet, the trajectory highlights the absence of an early inflationary point, meaning that constant potential models can best describe the late universe. Furthermore, choosing $\omega$ to be at least an order of magnitude larger than $\xi$ is able to achieve matter-radiation equality at $z_{eq} \sim \mathcal{O}(10^3)$ as desired.

The power-law potential offers richer dynamics with the addition of two further saddle points where both the scalar field and either radiation $(P_5)$ or matter $(P_6)$ contribute. However, such points represent transition periods which only exhibit an acceleration for a small range of negative $n$. Thus, the cosmological trajectory is again similar to the constant potential case provided that such accelerating transient periods are avoided. Otherwise, these points may potentially act as initial inflationary saddle points but this does not ensure a consistent cosmology for later epochs. For the investigated case of the inverse square potential $n = 2$, $\omega$ and $\xi$ have to be chosen such that the matter-dominated fixed point acts as a saddle and the late-time accelerating attractor point exists. This requires either $\f{\omega}{\xi^2} < \f{1}{2}$ or $\f{\omega}{\xi^2} > \f{13}{2}$.

Finally, the exponential potential was explored for the case of $\omega = 1$; having $\omega \neq 1$ would lead to dynamics similar to those obtained for the constant potential. Indeed, the critical points are similar, differing in the nature of the final attractor being strictly a de Sitter, dark energy point. Nonetheless, similar to the constant potential, the trajectory always starts from a non-accelerating, scalar-field dominated source, meaning the cosmology does not originate from an inflationary point. Even so, an inflationary transition point may be achieved for $|\xi| < 2$. Indeed, setting $|\xi| \ll 1$ causes this saddle point to have a dark energy-like equation of state $\omega_\phi \simeq -1$ and $q \simeq -1$. However, this does not guarantee that the inflationary epoch underwent the necessary $e$-folds. Yet, the late-time behaviour is dissimilar from $\Lambda$CDM.

These results show the potential for such Brans-Dicke inspired models to describe a viable cosmological history, provided that appropriate parameter restrictions are considered. Future work aims to analyse the evolution of cosmological perturbations in these models and test their viability against observations with the purpose of addressing cosmic tensions.

\begin{acknowledgments}
This article is based upon work from COST Action CA21136 \emph{Addressing observational tensions in cosmology with systematics and fundamental physics} (CosmoVerse), supported by COST (European Cooperation in Science and Technology). JLS would also like to acknowledge funding from ``Xjenza Malta'' as part of the ``Technology Development Programme'' DTP-2024-014 (CosmicLearning) Project.
\end{acknowledgments}

\begin{table*}[!htb]
\centering
\begin{tabular}{cccc}
\hline
Point & $l_1$ & $l_2$ & $l_3$ \\
\hline
$P_1$ & $-1$ & 1 & 4 \\
$P_2$ & $-1$ & 3 & $-\f{4\xi^2 + 3\omega}{2(\xi^2 + \omega)}$ \\
$P_3$ & $-4$ & $-3$ & $-\f{5\xi^2 + 6\omega}{\xi^2 + 2\omega}$ \\
$P_4^{\pm}$ & $\f{3\xi^2 + 2\omega \pm \xi\sqrt{9\xi^2 + 6\omega}}{\omega}$ & $\f{3\xi^2 + 3\omega \pm \xi\sqrt{9\xi^2 + 6\omega}}{\omega}$ & $\f{3\xi^2 + 6\omega \pm \xi\sqrt{9\xi^2 + 6\omega}}{\omega}$ \\
\hline
\end{tabular}
\caption{Eigenvalues for the critical points of the constant potential $V(\phi) = V_0$. As $\omega > 0$, all eigenvalues are real and all points are hyperbolic.}
\label{tab:wcase1eigen}
\end{table*}

\begin{table*}[!htb]
\centering
\begin{tabular}{cccc}
\hline
Point & $l_1$ & $l_2$ & $l_3$ \\
\hline
$P_1$ & $-1$ & 1 & 4 \\
$P_2$ & $\f{3\omega-(n-3)\xi^2}{\xi^2 + \omega}$ & $-1$ & $-\f{4\xi^2 + 3\omega}{2(\xi^2 + \omega)}$ \\
$P_3$ & $\f{(3-7n-2n^2)\xi^2 + 6\omega}{(n-1)\xi^2 - 2\omega}$ & $\f{8\omega -2(n^2+4n-2)\xi^2}{(n-1)\xi^2 - 2\omega}$ & $\f{6\omega - (n^2+4n-5)\xi^2}{(n-1)\xi^2 - 2\omega}$ \\
$P_4^{\pm}$ & $\f{3\xi^2 + 2\omega \pm \xi \sqrt{9\xi^2 + 6\omega}}{\omega}$ & $\f{3\xi^2 + 3\omega \pm \xi \sqrt{9\xi^2 + 6\omega}}{\omega}$ & $\f{6\omega - 3(n-1)\xi^2 \pm (n-1)\xi \sqrt{9\xi^2+6\omega}}{\omega}$ \\
$P_5$ & 1 & $\Delta_1^+(\omega)$ & $\Delta_1^-(\omega)$ \\
$P_6$ & $-1$ & $\Delta_2^+(\omega)$ & $\Delta_2^-(\omega)$ \\
\hline
\end{tabular}
\caption{Eigenvalues for the critical points of the power-law potential $V(\phi) = V_0 (1+\xi \phi)^{-n}$, $n \neq 0$. Here, we define $\Delta_1^{\pm}(\omega) = -\f{1}{2} - \f{1}{n} \pm \f{1}{2}\sqrt{\f{64\omega}{n^2 \xi^2} + 3\left(\f{12}{n^2}-\f{20}{n}-5\right)}$, and $\Delta_2^{\pm}(\omega) = -\f{3}{4} - \f{3}{4n} \pm \f{\sqrt{3}}{4}\f{1}{n\xi(3\xi^2+2\omega)}\delta(\omega)$, where $\delta(\omega)^2 = (3\xi^2+2\omega)[144\omega^2-6(7n^2+34n-37)\xi^2\omega + (81-174n+17n^2+16n^3)\xi^4]$. Contrary to the constant potential, $P_2$--$P_6$ can be non-hyperbolic depending on the choices of $n$, $\omega$ and $\xi$. For example, for $\omega = \f{1}{3}(n-3)\xi^2$, $P_2$ becomes non-hyperbolic as $l_1$ vanishes. When $P_2$ is non-hyperbolic, it also coincides with $P_6$.}
\label{tab:wcase2eigen}
\end{table*}

\begin{table*}[!htb]
\centering
\begin{tabular}{ccccc}
\hline
Point & $l_1$ & $l_2$ & $l_3$ & $l_4$ \\
\hline
$P_1$ & 0 & $-1$ & 1 & 4 \\
$P_2$ & $\f{\xi^2}{1+\xi^2}$ & $-1$ & 3 & $-\f{3+4\xi^2}{2(1+\xi^2)}$ \\
$P_3$ & $\f{4\xi^2}{\xi^2 + 2}$ & $-4$ & $-3$ & $-\f{6+5\xi^2}{\xi^2 + 2}$ \\
$P_4^{\pm}$ & $\xi(3\xi \pm \sqrt{6+9\xi^2})$ & $2 + 3\xi^2 \pm \xi\sqrt{6+9\xi^2}$ & $3 + 3\xi^2 \pm \xi\sqrt{6+9\xi^2}$ & $6 + 3\xi^2 \pm \xi\sqrt{6+9\xi^2}$ \\
$P_5$ & $-4$ & $-3$ & $\Delta_3^+(\xi)$ & $\Delta_3^-(\xi)$\\
\hline
\end{tabular}
\caption{Eigenvalues for the critical points of the exponential potential $V(\phi) = V_0 e^{-\lambda\phi}$ for $\omega = 1$. Here, $\Delta_3^{\pm}(\xi) = -\f{3}{2} \pm \f{\sqrt{3}}{2}\f{1+4\xi^2}{12\xi^4 + 11\xi^2 + 2}\sqrt{12 + 4\xi^2 - 21\xi^4}$. An additional eigenvalue appears due to having a dynamical $\lambda_V$. In this case, both $P^{\pm}_4$ and $P_5$ are non-hyperbolic for $\xi = 0$.}
\label{tab:case2eigen}
\end{table*}

\appendix

\section{Stability Properties}\label{app:eigenvalues}

The eigenvalues for each critical point for each respective scalar field potential (constant, power-law, and exponential) are presented in Tables~\ref{tab:wcase1eigen}--\ref{tab:case2eigen}, together with the respective special limiting cases when the critical points become non-hyperbolic. One simple scenario requires $\xi = 0$ which reduces to a minimal coupling to curvature. However, as this was not of interest in this work, such cases were not explored. Other special cases were not explored from a practical, cosmological viewpoint. For instance, setting $\omega = \f{1}{3}(n-3)\xi^2$ for the power-law potential would require specific fine-tuning between the parameters. Taking $n = 2$, when $|\xi| \ll 1$, this would set $\omega \ll 1$, which can negatively effect the redshift at which matter-radiation equality occurs as seen for the case $(\omega,\xi)=(0.1,0.1)$ in Fig.~\ref{fig:model2_w_evol}. Hence, such cases were not explored in detail.


\begin{thebibliography}{81}%
\makeatletter
\providecommand \@ifxundefined [1]{%
 \@ifx{#1\undefined}
}%
\providecommand \@ifnum [1]{%
 \ifnum #1\expandafter \@firstoftwo
 \else \expandafter \@secondoftwo
 \fi
}%
\providecommand \@ifx [1]{%
 \ifx #1\expandafter \@firstoftwo
 \else \expandafter \@secondoftwo
 \fi
}%
\providecommand \natexlab [1]{#1}%
\providecommand \enquote  [1]{``#1''}%
\providecommand \bibnamefont  [1]{#1}%
\providecommand \bibfnamefont [1]{#1}%
\providecommand \citenamefont [1]{#1}%
\providecommand \href@noop [0]{\@secondoftwo}%
\providecommand \href [0]{\begingroup \@sanitize@url \@href}%
\providecommand \@href[1]{\@@startlink{#1}\@@href}%
\providecommand \@@href[1]{\endgroup#1\@@endlink}%
\providecommand \@sanitize@url [0]{\catcode `\\12\catcode `\$12\catcode `\&12\catcode `\#12\catcode `\^12\catcode `\_12\catcode `\%12\relax}%
\providecommand \@@startlink[1]{}%
\providecommand \@@endlink[0]{}%
\providecommand \url  [0]{\begingroup\@sanitize@url \@url }%
\providecommand \@url [1]{\endgroup\@href {#1}{\urlprefix }}%
\providecommand \urlprefix  [0]{URL }%
\providecommand \Eprint [0]{\href }%
\providecommand \doibase [0]{https://doi.org/}%
\providecommand \selectlanguage [0]{\@gobble}%
\providecommand \bibinfo  [0]{\@secondoftwo}%
\providecommand \bibfield  [0]{\@secondoftwo}%
\providecommand \translation [1]{[#1]}%
\providecommand \BibitemOpen [0]{}%
\providecommand \bibitemStop [0]{}%
\providecommand \bibitemNoStop [0]{.\EOS\space}%
\providecommand \EOS [0]{\spacefactor3000\relax}%
\providecommand \BibitemShut  [1]{\csname bibitem#1\endcsname}%
\let\auto@bib@innerbib\@empty
%</preamble>
\bibitem [{\citenamefont {Di~Valentino}\ \emph {et~al.}(2025)\citenamefont {Di~Valentino} \emph {et~al.}}]{CosmoVerseNetwork:2025alb}%
  \BibitemOpen
  \bibfield  {author} {\bibinfo {author} {\bibfnamefont {E.}~\bibnamefont {Di~Valentino}} \emph {et~al.} (\bibinfo {collaboration} {CosmoVerse Network}),\ }\bibfield  {title} {\bibinfo {title} {{The CosmoVerse White Paper: Addressing observational tensions in cosmology with systematics and fundamental physics}},\ }\href {https://doi.org/10.1016/j.dark.2025.101965} {\bibfield  {journal} {\bibinfo  {journal} {Phys. Dark Univ.}\ }\textbf {\bibinfo {volume} {49}},\ \bibinfo {pages} {101965} (\bibinfo {year} {2025})},\ \Eprint {https://arxiv.org/abs/2504.01669} {arXiv:2504.01669 [astro-ph.CO]} \BibitemShut {NoStop}%
\bibitem [{\citenamefont {Clifton}\ \emph {et~al.}(2012)\citenamefont {Clifton}, \citenamefont {Ferreira}, \citenamefont {Padilla},\ and\ \citenamefont {Skordis}}]{Clifton:2011jh}%
  \BibitemOpen
  \bibfield  {author} {\bibinfo {author} {\bibfnamefont {T.}~\bibnamefont {Clifton}}, \bibinfo {author} {\bibfnamefont {P.~G.}\ \bibnamefont {Ferreira}}, \bibinfo {author} {\bibfnamefont {A.}~\bibnamefont {Padilla}},\ and\ \bibinfo {author} {\bibfnamefont {C.}~\bibnamefont {Skordis}},\ }\bibfield  {title} {\bibinfo {title} {{Modified Gravity and Cosmology}},\ }\href {https://doi.org/10.1016/j.physrep.2012.01.001} {\bibfield  {journal} {\bibinfo  {journal} {Phys. Rept.}\ }\textbf {\bibinfo {volume} {513}},\ \bibinfo {pages} {1} (\bibinfo {year} {2012})},\ \Eprint {https://arxiv.org/abs/1106.2476} {arXiv:1106.2476 [astro-ph.CO]} \BibitemShut {NoStop}%
\bibitem [{\citenamefont {Peebles}\ and\ \citenamefont {Ratra}(2003)}]{Peebles:2002gy}%
  \BibitemOpen
  \bibfield  {author} {\bibinfo {author} {\bibfnamefont {P.~J.~E.}\ \bibnamefont {Peebles}}\ and\ \bibinfo {author} {\bibfnamefont {B.}~\bibnamefont {Ratra}},\ }\bibfield  {title} {\bibinfo {title} {{The Cosmological Constant and Dark Energy}},\ }\href {https://doi.org/10.1103/RevModPhys.75.559} {\bibfield  {journal} {\bibinfo  {journal} {Rev. Mod. Phys.}\ }\textbf {\bibinfo {volume} {75}},\ \bibinfo {pages} {559} (\bibinfo {year} {2003})},\ \Eprint {https://arxiv.org/abs/astro-ph/0207347} {arXiv:astro-ph/0207347} \BibitemShut {NoStop}%
\bibitem [{\citenamefont {Baudis}(2016)}]{Baudis:2016qwx}%
  \BibitemOpen
  \bibfield  {author} {\bibinfo {author} {\bibfnamefont {L.}~\bibnamefont {Baudis}},\ }\bibfield  {title} {\bibinfo {title} {{Dark matter detection}},\ }\href {https://doi.org/10.1088/0954-3899/43/4/044001} {\bibfield  {journal} {\bibinfo  {journal} {J. Phys. G}\ }\textbf {\bibinfo {volume} {43}},\ \bibinfo {pages} {044001} (\bibinfo {year} {2016})}\BibitemShut {NoStop}%
\bibitem [{\citenamefont {Bertone}\ \emph {et~al.}(2005)\citenamefont {Bertone}, \citenamefont {Hooper},\ and\ \citenamefont {Silk}}]{Bertone:2004pz}%
  \BibitemOpen
  \bibfield  {author} {\bibinfo {author} {\bibfnamefont {G.}~\bibnamefont {Bertone}}, \bibinfo {author} {\bibfnamefont {D.}~\bibnamefont {Hooper}},\ and\ \bibinfo {author} {\bibfnamefont {J.}~\bibnamefont {Silk}},\ }\bibfield  {title} {\bibinfo {title} {{Particle dark matter: Evidence, candidates and constraints}},\ }\href {https://doi.org/10.1016/j.physrep.2004.08.031} {\bibfield  {journal} {\bibinfo  {journal} {Phys. Rept.}\ }\textbf {\bibinfo {volume} {405}},\ \bibinfo {pages} {279} (\bibinfo {year} {2005})},\ \Eprint {https://arxiv.org/abs/hep-ph/0404175} {arXiv:hep-ph/0404175} \BibitemShut {NoStop}%
\bibitem [{\citenamefont {Copeland}\ \emph {et~al.}(2006)\citenamefont {Copeland}, \citenamefont {Sami},\ and\ \citenamefont {Tsujikawa}}]{Copeland:2006wr}%
  \BibitemOpen
  \bibfield  {author} {\bibinfo {author} {\bibfnamefont {E.~J.}\ \bibnamefont {Copeland}}, \bibinfo {author} {\bibfnamefont {M.}~\bibnamefont {Sami}},\ and\ \bibinfo {author} {\bibfnamefont {S.}~\bibnamefont {Tsujikawa}},\ }\bibfield  {title} {\bibinfo {title} {{Dynamics of dark energy}},\ }\href {https://doi.org/10.1142/S021827180600942X} {\bibfield  {journal} {\bibinfo  {journal} {Int. J. Mod. Phys. D}\ }\textbf {\bibinfo {volume} {15}},\ \bibinfo {pages} {1753} (\bibinfo {year} {2006})},\ \Eprint {https://arxiv.org/abs/hep-th/0603057} {arXiv:hep-th/0603057} \BibitemShut {NoStop}%
\bibitem [{\citenamefont {Bahamonde}\ \emph {et~al.}(2018{\natexlab{a}})\citenamefont {Bahamonde}, \citenamefont {B{\"o}hmer}, \citenamefont {Carloni}, \citenamefont {Copeland}, \citenamefont {Fang},\ and\ \citenamefont {Tamanini}}]{Bahamonde:2017ize}%
  \BibitemOpen
  \bibfield  {author} {\bibinfo {author} {\bibfnamefont {S.}~\bibnamefont {Bahamonde}}, \bibinfo {author} {\bibfnamefont {C.~G.}\ \bibnamefont {B{\"o}hmer}}, \bibinfo {author} {\bibfnamefont {S.}~\bibnamefont {Carloni}}, \bibinfo {author} {\bibfnamefont {E.~J.}\ \bibnamefont {Copeland}}, \bibinfo {author} {\bibfnamefont {W.}~\bibnamefont {Fang}},\ and\ \bibinfo {author} {\bibfnamefont {N.}~\bibnamefont {Tamanini}},\ }\bibfield  {title} {\bibinfo {title} {{Dynamical systems applied to cosmology: dark energy and modified gravity}},\ }\href {https://doi.org/10.1016/j.physrep.2018.09.001} {\bibfield  {journal} {\bibinfo  {journal} {Phys. Rept.}\ }\textbf {\bibinfo {volume} {775-777}},\ \bibinfo {pages} {1} (\bibinfo {year} {2018}{\natexlab{a}})},\ \Eprint {https://arxiv.org/abs/1712.03107} {arXiv:1712.03107 [gr-qc]} \BibitemShut {NoStop}%
\bibitem [{\citenamefont {Weinberg}(1989)}]{Weinberg:1988cp}%
  \BibitemOpen
  \bibfield  {author} {\bibinfo {author} {\bibfnamefont {S.}~\bibnamefont {Weinberg}},\ }\bibfield  {title} {\bibinfo {title} {{The Cosmological Constant Problem}},\ }\href {https://doi.org/10.1103/RevModPhys.61.1} {\bibfield  {journal} {\bibinfo  {journal} {Rev. Mod. Phys.}\ }\textbf {\bibinfo {volume} {61}},\ \bibinfo {pages} {1} (\bibinfo {year} {1989})}\BibitemShut {NoStop}%
\bibitem [{\citenamefont {Gaitskell}(2004)}]{Gaitskell:2004gd}%
  \BibitemOpen
  \bibfield  {author} {\bibinfo {author} {\bibfnamefont {R.~J.}\ \bibnamefont {Gaitskell}},\ }\bibfield  {title} {\bibinfo {title} {{Direct detection of dark matter}},\ }\href {https://doi.org/10.1146/annurev.nucl.54.070103.181244} {\bibfield  {journal} {\bibinfo  {journal} {Ann. Rev. Nucl. Part. Sci.}\ }\textbf {\bibinfo {volume} {54}},\ \bibinfo {pages} {315} (\bibinfo {year} {2004})}\BibitemShut {NoStop}%
\bibitem [{\citenamefont {Di~Valentino}\ \emph {et~al.}(2021{\natexlab{a}})\citenamefont {Di~Valentino} \emph {et~al.}}]{DiValentino:2020vhf}%
  \BibitemOpen
  \bibfield  {author} {\bibinfo {author} {\bibfnamefont {E.}~\bibnamefont {Di~Valentino}} \emph {et~al.},\ }\bibfield  {title} {\bibinfo {title} {{Snowmass2021 - Letter of interest cosmology intertwined I: Perspectives for the next decade}},\ }\href {https://doi.org/10.1016/j.astropartphys.2021.102606} {\bibfield  {journal} {\bibinfo  {journal} {Astropart. Phys.}\ }\textbf {\bibinfo {volume} {131}},\ \bibinfo {pages} {102606} (\bibinfo {year} {2021}{\natexlab{a}})},\ \Eprint {https://arxiv.org/abs/2008.11283} {arXiv:2008.11283 [astro-ph.CO]} \BibitemShut {NoStop}%
\bibitem [{\citenamefont {Di~Valentino}\ \emph {et~al.}(2021{\natexlab{b}})\citenamefont {Di~Valentino} \emph {et~al.}}]{DiValentino:2020zio}%
  \BibitemOpen
  \bibfield  {author} {\bibinfo {author} {\bibfnamefont {E.}~\bibnamefont {Di~Valentino}} \emph {et~al.},\ }\bibfield  {title} {\bibinfo {title} {{Snowmass2021 - Letter of interest cosmology intertwined II: The hubble constant tension}},\ }\href {https://doi.org/10.1016/j.astropartphys.2021.102605} {\bibfield  {journal} {\bibinfo  {journal} {Astropart. Phys.}\ }\textbf {\bibinfo {volume} {131}},\ \bibinfo {pages} {102605} (\bibinfo {year} {2021}{\natexlab{b}})},\ \Eprint {https://arxiv.org/abs/2008.11284} {arXiv:2008.11284 [astro-ph.CO]} \BibitemShut {NoStop}%
\bibitem [{\citenamefont {Di~Valentino}\ \emph {et~al.}(2021{\natexlab{c}})\citenamefont {Di~Valentino} \emph {et~al.}}]{DiValentino:2020srs}%
  \BibitemOpen
  \bibfield  {author} {\bibinfo {author} {\bibfnamefont {E.}~\bibnamefont {Di~Valentino}} \emph {et~al.},\ }\bibfield  {title} {\bibinfo {title} {{Snowmass2021 - Letter of interest cosmology intertwined IV: The age of the universe and its curvature}},\ }\href {https://doi.org/10.1016/j.astropartphys.2021.102607} {\bibfield  {journal} {\bibinfo  {journal} {Astropart. Phys.}\ }\textbf {\bibinfo {volume} {131}},\ \bibinfo {pages} {102607} (\bibinfo {year} {2021}{\natexlab{c}})},\ \Eprint {https://arxiv.org/abs/2008.11286} {arXiv:2008.11286 [astro-ph.CO]} \BibitemShut {NoStop}%
\bibitem [{\citenamefont {Saridakis}\ \emph {et~al.}(2021)\citenamefont {Saridakis}, \citenamefont {Lazkoz}, \citenamefont {Salzano}, \citenamefont {Vargas~Moniz}, \citenamefont {Capozziello}, \citenamefont {Beltr{\'a}n~Jim{\'e}nez}, \citenamefont {De~Laurentis},\ and\ \citenamefont {Olmo}}]{CANTATA:2021ktz}%
  \BibitemOpen
  \bibinfo {editor} {\bibfnamefont {E.~N.}\ \bibnamefont {Saridakis}}, \bibinfo {editor} {\bibfnamefont {R.}~\bibnamefont {Lazkoz}}, \bibinfo {editor} {\bibfnamefont {V.}~\bibnamefont {Salzano}}, \bibinfo {editor} {\bibfnamefont {P.}~\bibnamefont {Vargas~Moniz}}, \bibinfo {editor} {\bibfnamefont {S.}~\bibnamefont {Capozziello}}, \bibinfo {editor} {\bibfnamefont {J.}~\bibnamefont {Beltr{\'a}n~Jim{\'e}nez}}, \bibinfo {editor} {\bibfnamefont {M.}~\bibnamefont {De~Laurentis}},\ and\ \bibinfo {editor} {\bibfnamefont {G.~J.}\ \bibnamefont {Olmo}},\ eds.,\ \href {https://doi.org/10.1007/978-3-030-83715-0} {\emph {\bibinfo {title} {{Modified Gravity and Cosmology. An Update by the CANTATA Network}}}}\ (\bibinfo  {publisher} {Springer},\ \bibinfo {year} {2021})\ \Eprint {https://arxiv.org/abs/2105.12582} {arXiv:2105.12582 [gr-qc]} \BibitemShut {NoStop}%
\bibitem [{\citenamefont {Abdalla}\ \emph {et~al.}(2022)\citenamefont {Abdalla} \emph {et~al.}}]{Abdalla:2022yfr}%
  \BibitemOpen
  \bibfield  {author} {\bibinfo {author} {\bibfnamefont {E.}~\bibnamefont {Abdalla}} \emph {et~al.},\ }\bibfield  {title} {\bibinfo {title} {{Cosmology intertwined: A review of the particle physics, astrophysics, and cosmology associated with the cosmological tensions and anomalies}},\ }\href {https://doi.org/10.1016/j.jheap.2022.04.002} {\bibfield  {journal} {\bibinfo  {journal} {JHEAp}\ }\textbf {\bibinfo {volume} {34}},\ \bibinfo {pages} {49} (\bibinfo {year} {2022})},\ \Eprint {https://arxiv.org/abs/2203.06142} {arXiv:2203.06142 [astro-ph.CO]} \BibitemShut {NoStop}%
\bibitem [{\citenamefont {Di~Valentino}\ \emph {et~al.}(2021{\natexlab{d}})\citenamefont {Di~Valentino} \emph {et~al.}}]{DiValentino:2020vvd}%
  \BibitemOpen
  \bibfield  {author} {\bibinfo {author} {\bibfnamefont {E.}~\bibnamefont {Di~Valentino}} \emph {et~al.},\ }\bibfield  {title} {\bibinfo {title} {{Cosmology Intertwined III: $f \sigma_8$ and $S_8$}},\ }\href {https://doi.org/10.1016/j.astropartphys.2021.102604} {\bibfield  {journal} {\bibinfo  {journal} {Astropart. Phys.}\ }\textbf {\bibinfo {volume} {131}},\ \bibinfo {pages} {102604} (\bibinfo {year} {2021}{\natexlab{d}})},\ \Eprint {https://arxiv.org/abs/2008.11285} {arXiv:2008.11285 [astro-ph.CO]} \BibitemShut {NoStop}%
\bibitem [{\citenamefont {Capozziello}(2002)}]{Capozziello:2002rd}%
  \BibitemOpen
  \bibfield  {author} {\bibinfo {author} {\bibfnamefont {S.}~\bibnamefont {Capozziello}},\ }\bibfield  {title} {\bibinfo {title} {{Curvature quintessence}},\ }\href {https://doi.org/10.1142/S0218271802002025} {\bibfield  {journal} {\bibinfo  {journal} {Int. J. Mod. Phys. D}\ }\textbf {\bibinfo {volume} {11}},\ \bibinfo {pages} {483} (\bibinfo {year} {2002})},\ \Eprint {https://arxiv.org/abs/gr-qc/0201033} {arXiv:gr-qc/0201033} \BibitemShut {NoStop}%
\bibitem [{\citenamefont {Capozziello}\ and\ \citenamefont {De~Laurentis}(2011)}]{Capozziello:2011et}%
  \BibitemOpen
  \bibfield  {author} {\bibinfo {author} {\bibfnamefont {S.}~\bibnamefont {Capozziello}}\ and\ \bibinfo {author} {\bibfnamefont {M.}~\bibnamefont {De~Laurentis}},\ }\bibfield  {title} {\bibinfo {title} {{Extended Theories of Gravity}},\ }\href {https://doi.org/10.1016/j.physrep.2011.09.003} {\bibfield  {journal} {\bibinfo  {journal} {Phys. Rept.}\ }\textbf {\bibinfo {volume} {509}},\ \bibinfo {pages} {167} (\bibinfo {year} {2011})},\ \Eprint {https://arxiv.org/abs/1108.6266} {arXiv:1108.6266 [gr-qc]} \BibitemShut {NoStop}%
\bibitem [{\citenamefont {Nojiri}\ and\ \citenamefont {Odintsov}(2011)}]{Nojiri:2010wj}%
  \BibitemOpen
  \bibfield  {author} {\bibinfo {author} {\bibfnamefont {S.}~\bibnamefont {Nojiri}}\ and\ \bibinfo {author} {\bibfnamefont {S.~D.}\ \bibnamefont {Odintsov}},\ }\bibfield  {title} {\bibinfo {title} {{Unified cosmic history in modified gravity: from F(R) theory to Lorentz non-invariant models}},\ }\href {https://doi.org/10.1016/j.physrep.2011.04.001} {\bibfield  {journal} {\bibinfo  {journal} {Phys. Rept.}\ }\textbf {\bibinfo {volume} {505}},\ \bibinfo {pages} {59} (\bibinfo {year} {2011})},\ \Eprint {https://arxiv.org/abs/1011.0544} {arXiv:1011.0544 [gr-qc]} \BibitemShut {NoStop}%
\bibitem [{\citenamefont {Nojiri}\ \emph {et~al.}(2017)\citenamefont {Nojiri}, \citenamefont {Odintsov},\ and\ \citenamefont {Oikonomou}}]{Nojiri:2017ncd}%
  \BibitemOpen
  \bibfield  {author} {\bibinfo {author} {\bibfnamefont {S.}~\bibnamefont {Nojiri}}, \bibinfo {author} {\bibfnamefont {S.~D.}\ \bibnamefont {Odintsov}},\ and\ \bibinfo {author} {\bibfnamefont {V.~K.}\ \bibnamefont {Oikonomou}},\ }\bibfield  {title} {\bibinfo {title} {{Modified Gravity Theories on a Nutshell: Inflation, Bounce and Late-time Evolution}},\ }\href {https://doi.org/10.1016/j.physrep.2017.06.001} {\bibfield  {journal} {\bibinfo  {journal} {Phys. Rept.}\ }\textbf {\bibinfo {volume} {692}},\ \bibinfo {pages} {1} (\bibinfo {year} {2017})},\ \Eprint {https://arxiv.org/abs/1705.11098} {arXiv:1705.11098 [gr-qc]} \BibitemShut {NoStop}%
\bibitem [{\citenamefont {Feng}(2010)}]{Feng:2010gw}%
  \BibitemOpen
  \bibfield  {author} {\bibinfo {author} {\bibfnamefont {J.~L.}\ \bibnamefont {Feng}},\ }\bibfield  {title} {\bibinfo {title} {{Dark Matter Candidates from Particle Physics and Methods of Detection}},\ }\href {https://doi.org/10.1146/annurev-astro-082708-101659} {\bibfield  {journal} {\bibinfo  {journal} {Ann. Rev. Astron. Astrophys.}\ }\textbf {\bibinfo {volume} {48}},\ \bibinfo {pages} {495} (\bibinfo {year} {2010})},\ \Eprint {https://arxiv.org/abs/1003.0904} {arXiv:1003.0904 [astro-ph.CO]} \BibitemShut {NoStop}%
\bibitem [{\citenamefont {Dodelson}\ and\ \citenamefont {Widrow}(1994)}]{Dodelson:1993je}%
  \BibitemOpen
  \bibfield  {author} {\bibinfo {author} {\bibfnamefont {S.}~\bibnamefont {Dodelson}}\ and\ \bibinfo {author} {\bibfnamefont {L.~M.}\ \bibnamefont {Widrow}},\ }\bibfield  {title} {\bibinfo {title} {{Sterile-neutrinos as dark matter}},\ }\href {https://doi.org/10.1103/PhysRevLett.72.17} {\bibfield  {journal} {\bibinfo  {journal} {Phys. Rev. Lett.}\ }\textbf {\bibinfo {volume} {72}},\ \bibinfo {pages} {17} (\bibinfo {year} {1994})},\ \Eprint {https://arxiv.org/abs/hep-ph/9303287} {arXiv:hep-ph/9303287} \BibitemShut {NoStop}%
\bibitem [{\citenamefont {Joyce}\ \emph {et~al.}(2015)\citenamefont {Joyce}, \citenamefont {Jain}, \citenamefont {Khoury},\ and\ \citenamefont {Trodden}}]{Joyce:2014kja}%
  \BibitemOpen
  \bibfield  {author} {\bibinfo {author} {\bibfnamefont {A.}~\bibnamefont {Joyce}}, \bibinfo {author} {\bibfnamefont {B.}~\bibnamefont {Jain}}, \bibinfo {author} {\bibfnamefont {J.}~\bibnamefont {Khoury}},\ and\ \bibinfo {author} {\bibfnamefont {M.}~\bibnamefont {Trodden}},\ }\bibfield  {title} {\bibinfo {title} {{Beyond the Cosmological Standard Model}},\ }\href {https://doi.org/10.1016/j.physrep.2014.12.002} {\bibfield  {journal} {\bibinfo  {journal} {Phys. Rept.}\ }\textbf {\bibinfo {volume} {568}},\ \bibinfo {pages} {1} (\bibinfo {year} {2015})},\ \Eprint {https://arxiv.org/abs/1407.0059} {arXiv:1407.0059 [astro-ph.CO]} \BibitemShut {NoStop}%
\bibitem [{\citenamefont {Abazajian}\ \emph {et~al.}(2012)\citenamefont {Abazajian} \emph {et~al.}}]{Abazajian:2012ys}%
  \BibitemOpen
  \bibfield  {author} {\bibinfo {author} {\bibfnamefont {K.~N.}\ \bibnamefont {Abazajian}} \emph {et~al.},\ }\bibfield  {title} {\bibinfo {title} {{Light Sterile Neutrinos: A White Paper}},\ }\href@noop {} {\  (\bibinfo {year} {2012})},\ \Eprint {https://arxiv.org/abs/1204.5379} {arXiv:1204.5379 [hep-ph]} \BibitemShut {NoStop}%
\bibitem [{\citenamefont {Staicova}\ and\ \citenamefont {Benisty}(2022)}]{Benisty:2021gde}%
  \BibitemOpen
  \bibfield  {author} {\bibinfo {author} {\bibfnamefont {D.}~\bibnamefont {Staicova}}\ and\ \bibinfo {author} {\bibfnamefont {D.}~\bibnamefont {Benisty}},\ }\bibfield  {title} {\bibinfo {title} {{Constraining the dark energy models using baryon acoustic oscillations: An approach independent of H0 {\ensuremath{\cdot}} rd}},\ }\href {https://doi.org/10.1051/0004-6361/202244366} {\bibfield  {journal} {\bibinfo  {journal} {Astron. Astrophys.}\ }\textbf {\bibinfo {volume} {668}},\ \bibinfo {pages} {A135} (\bibinfo {year} {2022})},\ \Eprint {https://arxiv.org/abs/2107.14129} {arXiv:2107.14129 [astro-ph.CO]} \BibitemShut {NoStop}%
\bibitem [{\citenamefont {Benisty}\ and\ \citenamefont {Staicova}(2021)}]{Benisty:2020otr}%
  \BibitemOpen
  \bibfield  {author} {\bibinfo {author} {\bibfnamefont {D.}~\bibnamefont {Benisty}}\ and\ \bibinfo {author} {\bibfnamefont {D.}~\bibnamefont {Staicova}},\ }\bibfield  {title} {\bibinfo {title} {{Testing late-time cosmic acceleration with uncorrelated baryon acoustic oscillation dataset}},\ }\href {https://doi.org/10.1051/0004-6361/202039502} {\bibfield  {journal} {\bibinfo  {journal} {Astron. Astrophys.}\ }\textbf {\bibinfo {volume} {647}},\ \bibinfo {pages} {A38} (\bibinfo {year} {2021})},\ \Eprint {https://arxiv.org/abs/2009.10701} {arXiv:2009.10701 [astro-ph.CO]} \BibitemShut {NoStop}%
\bibitem [{\citenamefont {Bamba}\ \emph {et~al.}(2012)\citenamefont {Bamba}, \citenamefont {Capozziello}, \citenamefont {Nojiri},\ and\ \citenamefont {Odintsov}}]{Bamba:2012cp}%
  \BibitemOpen
  \bibfield  {author} {\bibinfo {author} {\bibfnamefont {K.}~\bibnamefont {Bamba}}, \bibinfo {author} {\bibfnamefont {S.}~\bibnamefont {Capozziello}}, \bibinfo {author} {\bibfnamefont {S.}~\bibnamefont {Nojiri}},\ and\ \bibinfo {author} {\bibfnamefont {S.~D.}\ \bibnamefont {Odintsov}},\ }\bibfield  {title} {\bibinfo {title} {{Dark energy cosmology: the equivalent description via different theoretical models and cosmography tests}},\ }\href {https://doi.org/10.1007/s10509-012-1181-8} {\bibfield  {journal} {\bibinfo  {journal} {Astrophys. Space Sci.}\ }\textbf {\bibinfo {volume} {342}},\ \bibinfo {pages} {155} (\bibinfo {year} {2012})},\ \Eprint {https://arxiv.org/abs/1205.3421} {arXiv:1205.3421 [gr-qc]} \BibitemShut {NoStop}%
\bibitem [{\citenamefont {Bahamonde}\ \emph {et~al.}(2023)\citenamefont {Bahamonde}, \citenamefont {Dialektopoulos}, \citenamefont {Escamilla-Rivera}, \citenamefont {Farrugia}, \citenamefont {Gakis}, \citenamefont {Hendry}, \citenamefont {Hohmann}, \citenamefont {Levi~Said}, \citenamefont {Mifsud},\ and\ \citenamefont {Di~Valentino}}]{Bahamonde:2021gfp}%
  \BibitemOpen
  \bibfield  {author} {\bibinfo {author} {\bibfnamefont {S.}~\bibnamefont {Bahamonde}}, \bibinfo {author} {\bibfnamefont {K.~F.}\ \bibnamefont {Dialektopoulos}}, \bibinfo {author} {\bibfnamefont {C.}~\bibnamefont {Escamilla-Rivera}}, \bibinfo {author} {\bibfnamefont {G.}~\bibnamefont {Farrugia}}, \bibinfo {author} {\bibfnamefont {V.}~\bibnamefont {Gakis}}, \bibinfo {author} {\bibfnamefont {M.}~\bibnamefont {Hendry}}, \bibinfo {author} {\bibfnamefont {M.}~\bibnamefont {Hohmann}}, \bibinfo {author} {\bibfnamefont {J.}~\bibnamefont {Levi~Said}}, \bibinfo {author} {\bibfnamefont {J.}~\bibnamefont {Mifsud}},\ and\ \bibinfo {author} {\bibfnamefont {E.}~\bibnamefont {Di~Valentino}},\ }\bibfield  {title} {\bibinfo {title} {{Teleparallel gravity: from theory to cosmology}},\ }\href {https://doi.org/10.1088/1361-6633/ac9cef} {\bibfield  {journal} {\bibinfo  {journal} {Rept. Prog. Phys.}\ }\textbf {\bibinfo {volume} {86}},\ \bibinfo {pages} {026901} (\bibinfo {year} {2023})},\ \Eprint {https://arxiv.org/abs/2106.13793}
  {arXiv:2106.13793 [gr-qc]} \BibitemShut {NoStop}%
\bibitem [{\citenamefont {Alves~Batista}\ \emph {et~al.}(2021)\citenamefont {Alves~Batista} \emph {et~al.}}]{AlvesBatista:2021gzc}%
  \BibitemOpen
  \bibfield  {author} {\bibinfo {author} {\bibfnamefont {R.}~\bibnamefont {Alves~Batista}} \emph {et~al.},\ }\bibfield  {title} {\bibinfo {title} {{EuCAPT White Paper: Opportunities and Challenges for Theoretical Astroparticle Physics in the Next Decade}},\ }\href@noop {} {\  (\bibinfo {year} {2021})},\ \Eprint {https://arxiv.org/abs/2110.10074} {arXiv:2110.10074 [astro-ph.HE]} \BibitemShut {NoStop}%
\bibitem [{\citenamefont {Addazi}\ \emph {et~al.}(2022)\citenamefont {Addazi} \emph {et~al.}}]{Addazi:2021xuf}%
  \BibitemOpen
  \bibfield  {author} {\bibinfo {author} {\bibfnamefont {A.}~\bibnamefont {Addazi}} \emph {et~al.},\ }\bibfield  {title} {\bibinfo {title} {{Quantum gravity phenomenology at the dawn of the multi-messenger era{\textemdash}A review}},\ }\href {https://doi.org/10.1016/j.ppnp.2022.103948} {\bibfield  {journal} {\bibinfo  {journal} {Prog. Part. Nucl. Phys.}\ }\textbf {\bibinfo {volume} {125}},\ \bibinfo {pages} {103948} (\bibinfo {year} {2022})},\ \Eprint {https://arxiv.org/abs/2111.05659} {arXiv:2111.05659 [hep-ph]} \BibitemShut {NoStop}%
\bibitem [{\citenamefont {Deffayet}\ \emph {et~al.}(2009{\natexlab{a}})\citenamefont {Deffayet}, \citenamefont {Deser},\ and\ \citenamefont {Esposito-Farese}}]{Deffayet:2009mn}%
  \BibitemOpen
  \bibfield  {author} {\bibinfo {author} {\bibfnamefont {C.}~\bibnamefont {Deffayet}}, \bibinfo {author} {\bibfnamefont {S.}~\bibnamefont {Deser}},\ and\ \bibinfo {author} {\bibfnamefont {G.}~\bibnamefont {Esposito-Farese}},\ }\bibfield  {title} {\bibinfo {title} {{Generalized Galileons: All scalar models whose curved background extensions maintain second-order field equations and stress-tensors}},\ }\href {https://doi.org/10.1103/PhysRevD.80.064015} {\bibfield  {journal} {\bibinfo  {journal} {Phys. Rev. D}\ }\textbf {\bibinfo {volume} {80}},\ \bibinfo {pages} {064015} (\bibinfo {year} {2009}{\natexlab{a}})},\ \Eprint {https://arxiv.org/abs/0906.1967} {arXiv:0906.1967 [gr-qc]} \BibitemShut {NoStop}%
\bibitem [{\citenamefont {Deffayet}\ \emph {et~al.}(2009{\natexlab{b}})\citenamefont {Deffayet}, \citenamefont {Esposito-Farese},\ and\ \citenamefont {Vikman}}]{Deffayet:2009wt}%
  \BibitemOpen
  \bibfield  {author} {\bibinfo {author} {\bibfnamefont {C.}~\bibnamefont {Deffayet}}, \bibinfo {author} {\bibfnamefont {G.}~\bibnamefont {Esposito-Farese}},\ and\ \bibinfo {author} {\bibfnamefont {A.}~\bibnamefont {Vikman}},\ }\bibfield  {title} {\bibinfo {title} {{Covariant Galileon}},\ }\href {https://doi.org/10.1103/PhysRevD.79.084003} {\bibfield  {journal} {\bibinfo  {journal} {Phys. Rev. D}\ }\textbf {\bibinfo {volume} {79}},\ \bibinfo {pages} {084003} (\bibinfo {year} {2009}{\natexlab{b}})},\ \Eprint {https://arxiv.org/abs/0901.1314} {arXiv:0901.1314 [hep-th]} \BibitemShut {NoStop}%
\bibitem [{\citenamefont {Horndeski}(1974)}]{Horndeski:1974wa}%
  \BibitemOpen
  \bibfield  {author} {\bibinfo {author} {\bibfnamefont {G.~W.}\ \bibnamefont {Horndeski}},\ }\bibfield  {title} {\bibinfo {title} {{Second-order scalar-tensor field equations in a four-dimensional space}},\ }\href {https://doi.org/10.1007/BF01807638} {\bibfield  {journal} {\bibinfo  {journal} {Int. J. Theor. Phys.}\ }\textbf {\bibinfo {volume} {10}},\ \bibinfo {pages} {363} (\bibinfo {year} {1974})}\BibitemShut {NoStop}%
\bibitem [{\citenamefont {Kobayashi}(2019)}]{Kobayashi:2019hrl}%
  \BibitemOpen
  \bibfield  {author} {\bibinfo {author} {\bibfnamefont {T.}~\bibnamefont {Kobayashi}},\ }\bibfield  {title} {\bibinfo {title} {{Horndeski theory and beyond: a review}},\ }\href {https://doi.org/10.1088/1361-6633/ab2429} {\bibfield  {journal} {\bibinfo  {journal} {Rept. Prog. Phys.}\ }\textbf {\bibinfo {volume} {82}},\ \bibinfo {pages} {086901} (\bibinfo {year} {2019})},\ \Eprint {https://arxiv.org/abs/1901.07183} {arXiv:1901.07183 [gr-qc]} \BibitemShut {NoStop}%
\bibitem [{\citenamefont {Linde}(1982)}]{1982PhLB..108..389L}%
  \BibitemOpen
  \bibfield  {author} {\bibinfo {author} {\bibfnamefont {A.~D.}\ \bibnamefont {Linde}},\ }\bibfield  {title} {\bibinfo {title} {{A New Inflationary Universe Scenario: A Possible Solution of the Horizon, Flatness, Homogeneity, Isotropy and Primordial Monopole Problems}},\ }\href {https://doi.org/10.1016/0370-2693(82)91219-9} {\bibfield  {journal} {\bibinfo  {journal} {Phys. Lett. B}\ }\textbf {\bibinfo {volume} {108}},\ \bibinfo {pages} {389} (\bibinfo {year} {1982})}\BibitemShut {NoStop}%
\bibitem [{\citenamefont {Guth}\ and\ \citenamefont {Pi}(1982)}]{1982PhRvL..49.1110G}%
  \BibitemOpen
  \bibfield  {author} {\bibinfo {author} {\bibfnamefont {A.~H.}\ \bibnamefont {Guth}}\ and\ \bibinfo {author} {\bibfnamefont {S.~Y.}\ \bibnamefont {Pi}},\ }\bibfield  {title} {\bibinfo {title} {{Fluctuations in the New Inflationary Universe}},\ }\href {https://doi.org/10.1103/PhysRevLett.49.1110} {\bibfield  {journal} {\bibinfo  {journal} {Phys. Rev. Lett.}\ }\textbf {\bibinfo {volume} {49}},\ \bibinfo {pages} {1110} (\bibinfo {year} {1982})}\BibitemShut {NoStop}%
\bibitem [{\citenamefont {Poulin}\ \emph {et~al.}(2023)\citenamefont {Poulin}, \citenamefont {Smith},\ and\ \citenamefont {Karwal}}]{Poulin:2023lkg}%
  \BibitemOpen
  \bibfield  {author} {\bibinfo {author} {\bibfnamefont {V.}~\bibnamefont {Poulin}}, \bibinfo {author} {\bibfnamefont {T.~L.}\ \bibnamefont {Smith}},\ and\ \bibinfo {author} {\bibfnamefont {T.}~\bibnamefont {Karwal}},\ }\bibfield  {title} {\bibinfo {title} {{The Ups and Downs of Early Dark Energy solutions to the Hubble tension: A review of models, hints and constraints circa 2023}},\ }\href {https://doi.org/10.1016/j.dark.2023.101348} {\bibfield  {journal} {\bibinfo  {journal} {Phys. Dark Univ.}\ }\textbf {\bibinfo {volume} {42}},\ \bibinfo {pages} {101348} (\bibinfo {year} {2023})},\ \Eprint {https://arxiv.org/abs/2302.09032} {arXiv:2302.09032 [astro-ph.CO]} \BibitemShut {NoStop}%
\bibitem [{\citenamefont {Bahamonde}\ \emph {et~al.}(2019)\citenamefont {Bahamonde}, \citenamefont {Dialektopoulos},\ and\ \citenamefont {Levi~Said}}]{Bahamonde:2019shr}%
  \BibitemOpen
  \bibfield  {author} {\bibinfo {author} {\bibfnamefont {S.}~\bibnamefont {Bahamonde}}, \bibinfo {author} {\bibfnamefont {K.~F.}\ \bibnamefont {Dialektopoulos}},\ and\ \bibinfo {author} {\bibfnamefont {J.}~\bibnamefont {Levi~Said}},\ }\bibfield  {title} {\bibinfo {title} {{Can Horndeski Theory be recast using Teleparallel Gravity?}},\ }\href {https://doi.org/10.1103/PhysRevD.100.064018} {\bibfield  {journal} {\bibinfo  {journal} {Phys. Rev. D}\ }\textbf {\bibinfo {volume} {100}},\ \bibinfo {pages} {064018} (\bibinfo {year} {2019})},\ \Eprint {https://arxiv.org/abs/1904.10791} {arXiv:1904.10791 [gr-qc]} \BibitemShut {NoStop}%
\bibitem [{\citenamefont {Bahamonde}\ \emph {et~al.}(2020)\citenamefont {Bahamonde}, \citenamefont {Dialektopoulos}, \citenamefont {Gakis},\ and\ \citenamefont {Levi~Said}}]{Bahamonde:2019ipm}%
  \BibitemOpen
  \bibfield  {author} {\bibinfo {author} {\bibfnamefont {S.}~\bibnamefont {Bahamonde}}, \bibinfo {author} {\bibfnamefont {K.~F.}\ \bibnamefont {Dialektopoulos}}, \bibinfo {author} {\bibfnamefont {V.}~\bibnamefont {Gakis}},\ and\ \bibinfo {author} {\bibfnamefont {J.}~\bibnamefont {Levi~Said}},\ }\bibfield  {title} {\bibinfo {title} {{Reviving Horndeski theory using teleparallel gravity after GW170817}},\ }\href {https://doi.org/10.1103/PhysRevD.101.084060} {\bibfield  {journal} {\bibinfo  {journal} {Phys. Rev. D}\ }\textbf {\bibinfo {volume} {101}},\ \bibinfo {pages} {084060} (\bibinfo {year} {2020})},\ \Eprint {https://arxiv.org/abs/1907.10057} {arXiv:1907.10057 [gr-qc]} \BibitemShut {NoStop}%
\bibitem [{\citenamefont {Gleyzes}\ \emph {et~al.}(2015)\citenamefont {Gleyzes}, \citenamefont {Langlois}, \citenamefont {Piazza},\ and\ \citenamefont {Vernizzi}}]{Gleyzes:2014qga}%
  \BibitemOpen
  \bibfield  {author} {\bibinfo {author} {\bibfnamefont {J.}~\bibnamefont {Gleyzes}}, \bibinfo {author} {\bibfnamefont {D.}~\bibnamefont {Langlois}}, \bibinfo {author} {\bibfnamefont {F.}~\bibnamefont {Piazza}},\ and\ \bibinfo {author} {\bibfnamefont {F.}~\bibnamefont {Vernizzi}},\ }\bibfield  {title} {\bibinfo {title} {{Exploring gravitational theories beyond Horndeski}},\ }\href {https://doi.org/10.1088/1475-7516/2015/02/018} {\bibfield  {journal} {\bibinfo  {journal} {JCAP}\ }\textbf {\bibinfo {volume} {02}},\ \bibinfo {pages} {018}},\ \Eprint {https://arxiv.org/abs/1408.1952} {arXiv:1408.1952 [astro-ph.CO]} \BibitemShut {NoStop}%
\bibitem [{\citenamefont {Ohashi}\ \emph {et~al.}(2015)\citenamefont {Ohashi}, \citenamefont {Tanahashi}, \citenamefont {Kobayashi},\ and\ \citenamefont {Yamaguchi}}]{Ohashi:2015fma}%
  \BibitemOpen
  \bibfield  {author} {\bibinfo {author} {\bibfnamefont {S.}~\bibnamefont {Ohashi}}, \bibinfo {author} {\bibfnamefont {N.}~\bibnamefont {Tanahashi}}, \bibinfo {author} {\bibfnamefont {T.}~\bibnamefont {Kobayashi}},\ and\ \bibinfo {author} {\bibfnamefont {M.}~\bibnamefont {Yamaguchi}},\ }\bibfield  {title} {\bibinfo {title} {{The most general second-order field equations of bi-scalar-tensor theory in four dimensions}},\ }\href {https://doi.org/10.1007/JHEP07(2015)008} {\bibfield  {journal} {\bibinfo  {journal} {JHEP}\ }\textbf {\bibinfo {volume} {07}},\ \bibinfo {pages} {008}},\ \Eprint {https://arxiv.org/abs/1505.06029} {arXiv:1505.06029 [gr-qc]} \BibitemShut {NoStop}%
\bibitem [{\citenamefont {Abbott}\ \emph {et~al.}(2017)\citenamefont {Abbott} \emph {et~al.}}]{LIGOScientific:2017vwq}%
  \BibitemOpen
  \bibfield  {author} {\bibinfo {author} {\bibfnamefont {B.~P.}\ \bibnamefont {Abbott}} \emph {et~al.} (\bibinfo {collaboration} {LIGO Scientific, Virgo}),\ }\bibfield  {title} {\bibinfo {title} {{GW170817: Observation of Gravitational Waves from a Binary Neutron Star Inspiral}},\ }\href {https://doi.org/10.1103/PhysRevLett.119.161101} {\bibfield  {journal} {\bibinfo  {journal} {Phys. Rev. Lett.}\ }\textbf {\bibinfo {volume} {119}},\ \bibinfo {pages} {161101} (\bibinfo {year} {2017})},\ \Eprint {https://arxiv.org/abs/1710.05832} {arXiv:1710.05832 [gr-qc]} \BibitemShut {NoStop}%
\bibitem [{\citenamefont {Goldstein}\ \emph {et~al.}(2017)\citenamefont {Goldstein} \emph {et~al.}}]{Goldstein:2017mmi}%
  \BibitemOpen
  \bibfield  {author} {\bibinfo {author} {\bibfnamefont {A.}~\bibnamefont {Goldstein}} \emph {et~al.},\ }\bibfield  {title} {\bibinfo {title} {{An Ordinary Short Gamma-Ray Burst with Extraordinary Implications: Fermi-GBM Detection of GRB 170817A}},\ }\href {https://doi.org/10.3847/2041-8213/aa8f41} {\bibfield  {journal} {\bibinfo  {journal} {Astrophys. J. Lett.}\ }\textbf {\bibinfo {volume} {848}},\ \bibinfo {pages} {L14} (\bibinfo {year} {2017})},\ \Eprint {https://arxiv.org/abs/1710.05446} {arXiv:1710.05446 [astro-ph.HE]} \BibitemShut {NoStop}%
\bibitem [{\citenamefont {Capozziello}\ \emph {et~al.}(2018)\citenamefont {Capozziello}, \citenamefont {Dialektopoulos},\ and\ \citenamefont {Sushkov}}]{Capozziello:2018gms}%
  \BibitemOpen
  \bibfield  {author} {\bibinfo {author} {\bibfnamefont {S.}~\bibnamefont {Capozziello}}, \bibinfo {author} {\bibfnamefont {K.~F.}\ \bibnamefont {Dialektopoulos}},\ and\ \bibinfo {author} {\bibfnamefont {S.~V.}\ \bibnamefont {Sushkov}},\ }\bibfield  {title} {\bibinfo {title} {{Classification of the Horndeski cosmologies via Noether Symmetries}},\ }\href {https://doi.org/10.1140/epjc/s10052-018-5939-1} {\bibfield  {journal} {\bibinfo  {journal} {Eur. Phys. J. C}\ }\textbf {\bibinfo {volume} {78}},\ \bibinfo {pages} {447} (\bibinfo {year} {2018})},\ \Eprint {https://arxiv.org/abs/1803.01429} {arXiv:1803.01429 [gr-qc]} \BibitemShut {NoStop}%
\bibitem [{\citenamefont {Dialektopoulos}\ \emph {et~al.}(2022)\citenamefont {Dialektopoulos}, \citenamefont {Said},\ and\ \citenamefont {Oikonomopoulou}}]{Dialektopoulos:2021ryi}%
  \BibitemOpen
  \bibfield  {author} {\bibinfo {author} {\bibfnamefont {K.~F.}\ \bibnamefont {Dialektopoulos}}, \bibinfo {author} {\bibfnamefont {J.~L.}\ \bibnamefont {Said}},\ and\ \bibinfo {author} {\bibfnamefont {Z.}~\bibnamefont {Oikonomopoulou}},\ }\bibfield  {title} {\bibinfo {title} {{Classification of teleparallel Horndeski cosmology via Noether symmetries}},\ }\href {https://doi.org/10.1140/epjc/s10052-022-10201-7} {\bibfield  {journal} {\bibinfo  {journal} {Eur. Phys. J. C}\ }\textbf {\bibinfo {volume} {82}},\ \bibinfo {pages} {259} (\bibinfo {year} {2022})},\ \Eprint {https://arxiv.org/abs/2112.15045} {arXiv:2112.15045 [gr-qc]} \BibitemShut {NoStop}%
\bibitem [{\citenamefont {Miranda}\ \emph {et~al.}(2024)\citenamefont {Miranda}, \citenamefont {Capozziello},\ and\ \citenamefont {Vernieri}}]{Miranda:2024hhe}%
  \BibitemOpen
  \bibfield  {author} {\bibinfo {author} {\bibfnamefont {M.}~\bibnamefont {Miranda}}, \bibinfo {author} {\bibfnamefont {S.}~\bibnamefont {Capozziello}},\ and\ \bibinfo {author} {\bibfnamefont {D.}~\bibnamefont {Vernieri}},\ }\bibfield  {title} {\bibinfo {title} {{General analysis of Noether symmetries in Horndeski gravity}},\ }\href {https://doi.org/10.1140/epjc/s10052-024-13088-8} {\bibfield  {journal} {\bibinfo  {journal} {Eur. Phys. J. C}\ }\textbf {\bibinfo {volume} {84}},\ \bibinfo {pages} {771} (\bibinfo {year} {2024})},\ \Eprint {https://arxiv.org/abs/2408.09018} {arXiv:2408.09018 [gr-qc]} \BibitemShut {NoStop}%
\bibitem [{\citenamefont {Ratra}\ and\ \citenamefont {Peebles}(1988)}]{RatraPeebles_1988}%
  \BibitemOpen
  \bibfield  {author} {\bibinfo {author} {\bibfnamefont {B.}~\bibnamefont {Ratra}}\ and\ \bibinfo {author} {\bibfnamefont {P.~J.~E.}\ \bibnamefont {Peebles}},\ }\bibfield  {title} {\bibinfo {title} {Cosmological consequences of a rolling homogeneous scalar field},\ }\href {https://doi.org/10.1103/PhysRevD.37.3406} {\bibfield  {journal} {\bibinfo  {journal} {Phys. Rev. D}\ }\textbf {\bibinfo {volume} {37}},\ \bibinfo {pages} {3406} (\bibinfo {year} {1988})}\BibitemShut {NoStop}%
\bibitem [{\citenamefont {Roy}\ and\ \citenamefont {Bhadra}(2018)}]{Roy_2018}%
  \BibitemOpen
  \bibfield  {author} {\bibinfo {author} {\bibfnamefont {N.}~\bibnamefont {Roy}}\ and\ \bibinfo {author} {\bibfnamefont {N.}~\bibnamefont {Bhadra}},\ }\bibfield  {title} {\bibinfo {title} {Dynamical systems analysis of phantom dark energy models},\ }\href {https://doi.org/10.1088/1475-7516/2018/06/002} {\bibfield  {journal} {\bibinfo  {journal} {Journal of Cosmology and Astroparticle Physics}\ }\textbf {\bibinfo {volume} {2018}}\bibinfo  {number} { (06)},\ \bibinfo {pages} {002–002}}\BibitemShut {NoStop}%
\bibitem [{\citenamefont {Hrycyna}\ and\ \citenamefont {Szydłowski}(2013{\natexlab{a}})}]{Hrycyna_2013a}%
  \BibitemOpen
\bibfield  {number} {  }\bibfield  {author} {\bibinfo {author} {\bibfnamefont {O.}~\bibnamefont {Hrycyna}}\ and\ \bibinfo {author} {\bibfnamefont {M.}~\bibnamefont {Szydłowski}},\ }\bibfield  {title} {\bibinfo {title} {Brans-dicke theory and the emergence of $\lambda$cdm model},\ }\bibfield  {journal} {\bibinfo  {journal} {Physical Review D}\ }\textbf {\bibinfo {volume} {88}},\ \href {https://doi.org/10.1103/physrevd.88.064018} {10.1103/physrevd.88.064018} (\bibinfo {year} {2013}{\natexlab{a}})\BibitemShut {NoStop}%
\bibitem [{\citenamefont {Avilez}\ and\ \citenamefont {Skordis}(2014)}]{Avilez:2013dxa}%
  \BibitemOpen
  \bibfield  {author} {\bibinfo {author} {\bibfnamefont {A.}~\bibnamefont {Avilez}}\ and\ \bibinfo {author} {\bibfnamefont {C.}~\bibnamefont {Skordis}},\ }\bibfield  {title} {\bibinfo {title} {{Cosmological constraints on Brans-Dicke theory}},\ }\href {https://doi.org/10.1103/PhysRevLett.113.011101} {\bibfield  {journal} {\bibinfo  {journal} {Phys. Rev. Lett.}\ }\textbf {\bibinfo {volume} {113}},\ \bibinfo {pages} {011101} (\bibinfo {year} {2014})},\ \Eprint {https://arxiv.org/abs/1303.4330} {arXiv:1303.4330 [astro-ph.CO]} \BibitemShut {NoStop}%
\bibitem [{\citenamefont {Faraoni}\ \emph {et~al.}(2020)\citenamefont {Faraoni}, \citenamefont {C{\^o}t{\'e}},\ and\ \citenamefont {Giusti}}]{Faraoni:2019sxw}%
  \BibitemOpen
  \bibfield  {author} {\bibinfo {author} {\bibfnamefont {V.}~\bibnamefont {Faraoni}}, \bibinfo {author} {\bibfnamefont {J.}~\bibnamefont {C{\^o}t{\'e}}},\ and\ \bibinfo {author} {\bibfnamefont {A.}~\bibnamefont {Giusti}},\ }\bibfield  {title} {\bibinfo {title} {{Do solar system experiments constrain scalar{\textendash}tensor gravity?}},\ }\href {https://doi.org/10.1140/epjc/s10052-020-7721-4} {\bibfield  {journal} {\bibinfo  {journal} {Eur. Phys. J. C}\ }\textbf {\bibinfo {volume} {80}},\ \bibinfo {pages} {132} (\bibinfo {year} {2020})},\ \Eprint {https://arxiv.org/abs/1906.05957} {arXiv:1906.05957 [gr-qc]} \BibitemShut {NoStop}%
\bibitem [{\citenamefont {Amirhashchi}\ and\ \citenamefont {Yadav}(2020)}]{Amirhashchi:2019jpf}%
  \BibitemOpen
  \bibfield  {author} {\bibinfo {author} {\bibfnamefont {H.}~\bibnamefont {Amirhashchi}}\ and\ \bibinfo {author} {\bibfnamefont {A.~K.}\ \bibnamefont {Yadav}},\ }\bibfield  {title} {\bibinfo {title} {{Constraining an exact Brans{\textendash}Dicke gravity theory with recent observations}},\ }\href {https://doi.org/10.1016/j.dark.2020.100711} {\bibfield  {journal} {\bibinfo  {journal} {Phys. Dark Univ.}\ }\textbf {\bibinfo {volume} {30}},\ \bibinfo {pages} {100711} (\bibinfo {year} {2020})},\ \Eprint {https://arxiv.org/abs/1908.04735} {arXiv:1908.04735 [gr-qc]} \BibitemShut {NoStop}%
\bibitem [{\citenamefont {De~Felice}\ \emph {et~al.}(2011)\citenamefont {De~Felice}, \citenamefont {Kobayashi},\ and\ \citenamefont {Tsujikawa}}]{DeFelice:2011hq}%
  \BibitemOpen
  \bibfield  {author} {\bibinfo {author} {\bibfnamefont {A.}~\bibnamefont {De~Felice}}, \bibinfo {author} {\bibfnamefont {T.}~\bibnamefont {Kobayashi}},\ and\ \bibinfo {author} {\bibfnamefont {S.}~\bibnamefont {Tsujikawa}},\ }\bibfield  {title} {\bibinfo {title} {{Effective gravitational couplings for cosmological perturbations in the most general scalar-tensor theories with second-order field equations}},\ }\href {https://doi.org/10.1016/j.physletb.2011.11.028} {\bibfield  {journal} {\bibinfo  {journal} {Phys. Lett. B}\ }\textbf {\bibinfo {volume} {706}},\ \bibinfo {pages} {123} (\bibinfo {year} {2011})},\ \Eprint {https://arxiv.org/abs/1108.4242} {arXiv:1108.4242 [gr-qc]} \BibitemShut {NoStop}%
\bibitem [{\citenamefont {Garriga}\ and\ \citenamefont {Mukhanov}(1999)}]{Garriga:1999vw}%
  \BibitemOpen
  \bibfield  {author} {\bibinfo {author} {\bibfnamefont {J.}~\bibnamefont {Garriga}}\ and\ \bibinfo {author} {\bibfnamefont {V.~F.}\ \bibnamefont {Mukhanov}},\ }\bibfield  {title} {\bibinfo {title} {{Perturbations in k-inflation}},\ }\href {https://doi.org/10.1016/S0370-2693(99)00602-4} {\bibfield  {journal} {\bibinfo  {journal} {Phys. Lett. B}\ }\textbf {\bibinfo {volume} {458}},\ \bibinfo {pages} {219} (\bibinfo {year} {1999})},\ \Eprint {https://arxiv.org/abs/hep-th/9904176} {arXiv:hep-th/9904176} \BibitemShut {NoStop}%
\bibitem [{\citenamefont {Kobayashi}\ \emph {et~al.}(2011)\citenamefont {Kobayashi}, \citenamefont {Yamaguchi},\ and\ \citenamefont {Yokoyama}}]{Kobayashi:2011nu}%
  \BibitemOpen
  \bibfield  {author} {\bibinfo {author} {\bibfnamefont {T.}~\bibnamefont {Kobayashi}}, \bibinfo {author} {\bibfnamefont {M.}~\bibnamefont {Yamaguchi}},\ and\ \bibinfo {author} {\bibfnamefont {J.}~\bibnamefont {Yokoyama}},\ }\bibfield  {title} {\bibinfo {title} {{Generalized G-inflation: Inflation with the most general second-order field equations}},\ }\href {https://doi.org/10.1143/PTP.126.511} {\bibfield  {journal} {\bibinfo  {journal} {Prog. Theor. Phys.}\ }\textbf {\bibinfo {volume} {126}},\ \bibinfo {pages} {511} (\bibinfo {year} {2011})},\ \Eprint {https://arxiv.org/abs/1105.5723} {arXiv:1105.5723 [hep-th]} \BibitemShut {NoStop}%
\bibitem [{\citenamefont {Nojiri}\ \emph {et~al.}(2019)\citenamefont {Nojiri}, \citenamefont {Odintsov},\ and\ \citenamefont {Oikonomou}}]{Nojiri:2019dqc}%
  \BibitemOpen
  \bibfield  {author} {\bibinfo {author} {\bibfnamefont {S.}~\bibnamefont {Nojiri}}, \bibinfo {author} {\bibfnamefont {S.~D.}\ \bibnamefont {Odintsov}},\ and\ \bibinfo {author} {\bibfnamefont {V.~K.}\ \bibnamefont {Oikonomou}},\ }\bibfield  {title} {\bibinfo {title} {{$k$-essence $f(R)$ gravity inflation}},\ }\href {https://doi.org/10.1016/j.nuclphysb.2019.02.008} {\bibfield  {journal} {\bibinfo  {journal} {Nucl. Phys. B}\ }\textbf {\bibinfo {volume} {941}},\ \bibinfo {pages} {11} (\bibinfo {year} {2019})},\ \Eprint {https://arxiv.org/abs/1902.03669} {arXiv:1902.03669 [gr-qc]} \BibitemShut {NoStop}%
\bibitem [{\citenamefont {Hrycyna}\ and\ \citenamefont {Szydłowski}(2013{\natexlab{b}})}]{Hrycyna_2013b}%
  \BibitemOpen
  \bibfield  {author} {\bibinfo {author} {\bibfnamefont {O.}~\bibnamefont {Hrycyna}}\ and\ \bibinfo {author} {\bibfnamefont {M.}~\bibnamefont {Szydłowski}},\ }\bibfield  {title} {\bibinfo {title} {Dynamical complexity of the brans-dicke cosmology},\ }\href {https://doi.org/10.1088/1475-7516/2013/12/016} {\bibfield  {journal} {\bibinfo  {journal} {Journal of Cosmology and Astroparticle Physics}\ }\textbf {\bibinfo {volume} {2013}}\bibinfo  {number} { (12)},\ \bibinfo {pages} {016–016}}\BibitemShut {NoStop}%
\bibitem [{\citenamefont {Hrycyna}\ \emph {et~al.}(2014)\citenamefont {Hrycyna}, \citenamefont {Szydłowski},\ and\ \citenamefont {Kamionka}}]{Hrycyna_2014}%
  \BibitemOpen
\bibfield  {number} {  }\bibfield  {author} {\bibinfo {author} {\bibfnamefont {O.}~\bibnamefont {Hrycyna}}, \bibinfo {author} {\bibfnamefont {M.}~\bibnamefont {Szydłowski}},\ and\ \bibinfo {author} {\bibfnamefont {M.}~\bibnamefont {Kamionka}},\ }\bibfield  {title} {\bibinfo {title} {Dynamics and cosmological constraints on brans-dicke cosmology},\ }\bibfield  {journal} {\bibinfo  {journal} {Physical Review D}\ }\textbf {\bibinfo {volume} {90}},\ \href {https://doi.org/10.1103/physrevd.90.124040} {10.1103/physrevd.90.124040} (\bibinfo {year} {2014})\BibitemShut {NoStop}%
\bibitem [{\citenamefont {Hrycyna}\ and\ \citenamefont {Szydłowski}(2015)}]{Hrycyna_2015}%
  \BibitemOpen
  \bibfield  {author} {\bibinfo {author} {\bibfnamefont {O.}~\bibnamefont {Hrycyna}}\ and\ \bibinfo {author} {\bibfnamefont {M.}~\bibnamefont {Szydłowski}},\ }\bibfield  {title} {\bibinfo {title} {Cosmological dynamics with non-minimally coupled scalar field and a constant potential function},\ }\href {https://doi.org/10.1088/1475-7516/2015/11/013} {\bibfield  {journal} {\bibinfo  {journal} {Journal of Cosmology and Astroparticle Physics}\ }\textbf {\bibinfo {volume} {2015}}\bibinfo  {number} { (11)},\ \bibinfo {pages} {013–013}}\BibitemShut {NoStop}%
\bibitem [{\citenamefont {Ahmedov}\ \emph {et~al.}(2025)\citenamefont {Ahmedov}, \citenamefont {Caruana}, \citenamefont {Dialektopoulos}, \citenamefont {Levi~Said}, \citenamefont {Nosirov}, \citenamefont {Oikonomopoulou},\ and\ \citenamefont {Yunusov}}]{Ahmedov:2024aez}%
  \BibitemOpen
\bibfield  {number} {  }\bibfield  {author} {\bibinfo {author} {\bibfnamefont {B.}~\bibnamefont {Ahmedov}}, \bibinfo {author} {\bibfnamefont {M.}~\bibnamefont {Caruana}}, \bibinfo {author} {\bibfnamefont {K.~F.}\ \bibnamefont {Dialektopoulos}}, \bibinfo {author} {\bibfnamefont {J.}~\bibnamefont {Levi~Said}}, \bibinfo {author} {\bibfnamefont {A.}~\bibnamefont {Nosirov}}, \bibinfo {author} {\bibfnamefont {Z.}~\bibnamefont {Oikonomopoulou}},\ and\ \bibinfo {author} {\bibfnamefont {O.}~\bibnamefont {Yunusov}},\ }\bibfield  {title} {\bibinfo {title} {{Gauge invariant perturbations in teleparallel Horndeski gravity}},\ }\href {https://doi.org/10.1016/j.dark.2025.101846} {\bibfield  {journal} {\bibinfo  {journal} {Phys. Dark Univ.}\ }\textbf {\bibinfo {volume} {48}},\ \bibinfo {pages} {101846} (\bibinfo {year} {2025})},\ \Eprint {https://arxiv.org/abs/2412.01349} {arXiv:2412.01349 [gr-qc]} \BibitemShut {NoStop}%
\bibitem [{\citenamefont {Aurrekoetxea}\ \emph {et~al.}(2025)\citenamefont {Aurrekoetxea}, \citenamefont {Clough},\ and\ \citenamefont {Lim}}]{Aurrekoetxea:2024ypv}%
  \BibitemOpen
  \bibfield  {author} {\bibinfo {author} {\bibfnamefont {J.~C.}\ \bibnamefont {Aurrekoetxea}}, \bibinfo {author} {\bibfnamefont {K.}~\bibnamefont {Clough}},\ and\ \bibinfo {author} {\bibfnamefont {E.~A.}\ \bibnamefont {Lim}},\ }\bibfield  {title} {\bibinfo {title} {{Cosmology using numerical relativity}},\ }\href {https://doi.org/10.1007/s41114-025-00058-z} {\bibfield  {journal} {\bibinfo  {journal} {Living Rev. Rel.}\ }\textbf {\bibinfo {volume} {28}},\ \bibinfo {pages} {5} (\bibinfo {year} {2025})},\ \Eprint {https://arxiv.org/abs/2409.01939} {arXiv:2409.01939 [gr-qc]} \BibitemShut {NoStop}%
\bibitem [{\citenamefont {Copeland}\ \emph {et~al.}(1998)\citenamefont {Copeland}, \citenamefont {Liddle},\ and\ \citenamefont {Wands}}]{Copeland_1998}%
  \BibitemOpen
  \bibfield  {author} {\bibinfo {author} {\bibfnamefont {E.~J.}\ \bibnamefont {Copeland}}, \bibinfo {author} {\bibfnamefont {A.~R.}\ \bibnamefont {Liddle}},\ and\ \bibinfo {author} {\bibfnamefont {D.}~\bibnamefont {Wands}},\ }\bibfield  {title} {\bibinfo {title} {Exponential potentials and cosmological scaling solutions},\ }\href {https://doi.org/10.1103/physrevd.57.4686} {\bibfield  {journal} {\bibinfo  {journal} {Physical Review D}\ }\textbf {\bibinfo {volume} {57}},\ \bibinfo {pages} {4686–4690} (\bibinfo {year} {1998})}\BibitemShut {NoStop}%
\bibitem [{\citenamefont {Steinhardt}\ \emph {et~al.}(1999)\citenamefont {Steinhardt}, \citenamefont {Wang},\ and\ \citenamefont {Zlatev}}]{Steinhardt_1999}%
  \BibitemOpen
  \bibfield  {author} {\bibinfo {author} {\bibfnamefont {P.~J.}\ \bibnamefont {Steinhardt}}, \bibinfo {author} {\bibfnamefont {L.}~\bibnamefont {Wang}},\ and\ \bibinfo {author} {\bibfnamefont {I.}~\bibnamefont {Zlatev}},\ }\bibfield  {title} {\bibinfo {title} {Cosmological tracking solutions},\ }\bibfield  {journal} {\bibinfo  {journal} {Physical Review D}\ }\textbf {\bibinfo {volume} {59}},\ \href {https://doi.org/10.1103/physrevd.59.123504} {10.1103/physrevd.59.123504} (\bibinfo {year} {1999})\BibitemShut {NoStop}%
\bibitem [{\citenamefont {Macorra}\ and\ \citenamefont {Piccinelli}(2000)}]{Macorra_2000}%
  \BibitemOpen
  \bibfield  {author} {\bibinfo {author} {\bibfnamefont {A.~d.~l.}\ \bibnamefont {Macorra}}\ and\ \bibinfo {author} {\bibfnamefont {G.}~\bibnamefont {Piccinelli}},\ }\bibfield  {title} {\bibinfo {title} {Cosmological evolution of general scalar fields and quintessence},\ }\bibfield  {journal} {\bibinfo  {journal} {Physical Review D}\ }\textbf {\bibinfo {volume} {61}},\ \href {https://doi.org/10.1103/physrevd.61.123503} {10.1103/physrevd.61.123503} (\bibinfo {year} {2000})\BibitemShut {NoStop}%
\bibitem [{\citenamefont {Bertotti}\ \emph {et~al.}(2003)\citenamefont {Bertotti}, \citenamefont {Iess},\ and\ \citenamefont {Tortora}}]{Bertotti:2003rm}%
  \BibitemOpen
  \bibfield  {author} {\bibinfo {author} {\bibfnamefont {B.}~\bibnamefont {Bertotti}}, \bibinfo {author} {\bibfnamefont {L.}~\bibnamefont {Iess}},\ and\ \bibinfo {author} {\bibfnamefont {P.}~\bibnamefont {Tortora}},\ }\bibfield  {title} {\bibinfo {title} {{A test of general relativity using radio links with the Cassini spacecraft}},\ }\href {https://doi.org/10.1038/nature01997} {\bibfield  {journal} {\bibinfo  {journal} {Nature}\ }\textbf {\bibinfo {volume} {425}},\ \bibinfo {pages} {374} (\bibinfo {year} {2003})}\BibitemShut {NoStop}%
\bibitem [{\citenamefont {Sch{\"o}neberg}\ \emph {et~al.}(2022)\citenamefont {Sch{\"o}neberg}, \citenamefont {Franco~Abell{\'a}n}, \citenamefont {P{\'e}rez~S{\'a}nchez}, \citenamefont {Witte}, \citenamefont {Poulin},\ and\ \citenamefont {Lesgourgues}}]{Schoneberg:2021qvd}%
  \BibitemOpen
  \bibfield  {author} {\bibinfo {author} {\bibfnamefont {N.}~\bibnamefont {Sch{\"o}neberg}}, \bibinfo {author} {\bibfnamefont {G.}~\bibnamefont {Franco~Abell{\'a}n}}, \bibinfo {author} {\bibfnamefont {A.}~\bibnamefont {P{\'e}rez~S{\'a}nchez}}, \bibinfo {author} {\bibfnamefont {S.~J.}\ \bibnamefont {Witte}}, \bibinfo {author} {\bibfnamefont {V.}~\bibnamefont {Poulin}},\ and\ \bibinfo {author} {\bibfnamefont {J.}~\bibnamefont {Lesgourgues}},\ }\bibfield  {title} {\bibinfo {title} {{The H0 Olympics: A fair ranking of proposed models}},\ }\href {https://doi.org/10.1016/j.physrep.2022.07.001} {\bibfield  {journal} {\bibinfo  {journal} {Phys. Rept.}\ }\textbf {\bibinfo {volume} {984}},\ \bibinfo {pages} {1} (\bibinfo {year} {2022})},\ \Eprint {https://arxiv.org/abs/2107.10291} {arXiv:2107.10291 [astro-ph.CO]} \BibitemShut {NoStop}%
\bibitem [{\citenamefont {G{\'o}mez-Valent}\ \emph {et~al.}(2022)\citenamefont {G{\'o}mez-Valent}, \citenamefont {Zheng}, \citenamefont {Amendola}, \citenamefont {Wetterich},\ and\ \citenamefont {Pettorino}}]{Gomez-Valent:2022bku}%
  \BibitemOpen
  \bibfield  {author} {\bibinfo {author} {\bibfnamefont {A.}~\bibnamefont {G{\'o}mez-Valent}}, \bibinfo {author} {\bibfnamefont {Z.}~\bibnamefont {Zheng}}, \bibinfo {author} {\bibfnamefont {L.}~\bibnamefont {Amendola}}, \bibinfo {author} {\bibfnamefont {C.}~\bibnamefont {Wetterich}},\ and\ \bibinfo {author} {\bibfnamefont {V.}~\bibnamefont {Pettorino}},\ }\bibfield  {title} {\bibinfo {title} {{Coupled and uncoupled early dark energy, massive neutrinos, and the cosmological tensions}},\ }\href {https://doi.org/10.1103/PhysRevD.106.103522} {\bibfield  {journal} {\bibinfo  {journal} {Phys. Rev. D}\ }\textbf {\bibinfo {volume} {106}},\ \bibinfo {pages} {103522} (\bibinfo {year} {2022})},\ \Eprint {https://arxiv.org/abs/2207.14487} {arXiv:2207.14487 [astro-ph.CO]} \BibitemShut {NoStop}%
\bibitem [{\citenamefont {D'Agostino}\ \emph {et~al.}(2022)\citenamefont {D'Agostino}, \citenamefont {Luongo},\ and\ \citenamefont {Muccino}}]{DAgostino:2022fcx}%
  \BibitemOpen
  \bibfield  {author} {\bibinfo {author} {\bibfnamefont {R.}~\bibnamefont {D'Agostino}}, \bibinfo {author} {\bibfnamefont {O.}~\bibnamefont {Luongo}},\ and\ \bibinfo {author} {\bibfnamefont {M.}~\bibnamefont {Muccino}},\ }\bibfield  {title} {\bibinfo {title} {{Healing the cosmological constant problem during inflation through a unified quasi-quintessence matter field}},\ }\href {https://doi.org/10.1088/1361-6382/ac8af2} {\bibfield  {journal} {\bibinfo  {journal} {Class. Quant. Grav.}\ }\textbf {\bibinfo {volume} {39}},\ \bibinfo {pages} {195014} (\bibinfo {year} {2022})},\ \Eprint {https://arxiv.org/abs/2204.02190} {arXiv:2204.02190 [gr-qc]} \BibitemShut {NoStop}%
\bibitem [{\citenamefont {Bhattacharya}\ \emph {et~al.}(2024)\citenamefont {Bhattacharya}, \citenamefont {Borghetto}, \citenamefont {Malhotra}, \citenamefont {Parameswaran}, \citenamefont {Tasinato},\ and\ \citenamefont {Zavala}}]{Bhattacharya:2024hep}%
  \BibitemOpen
  \bibfield  {author} {\bibinfo {author} {\bibfnamefont {S.}~\bibnamefont {Bhattacharya}}, \bibinfo {author} {\bibfnamefont {G.}~\bibnamefont {Borghetto}}, \bibinfo {author} {\bibfnamefont {A.}~\bibnamefont {Malhotra}}, \bibinfo {author} {\bibfnamefont {S.}~\bibnamefont {Parameswaran}}, \bibinfo {author} {\bibfnamefont {G.}~\bibnamefont {Tasinato}},\ and\ \bibinfo {author} {\bibfnamefont {I.}~\bibnamefont {Zavala}},\ }\bibfield  {title} {\bibinfo {title} {{Cosmological constraints on curved quintessence}},\ }\href {https://doi.org/10.1088/1475-7516/2024/09/073} {\bibfield  {journal} {\bibinfo  {journal} {JCAP}\ }\textbf {\bibinfo {volume} {09}},\ \bibinfo {pages} {073}},\ \Eprint {https://arxiv.org/abs/2405.17396} {arXiv:2405.17396 [astro-ph.CO]} \BibitemShut {NoStop}%
\bibitem [{\citenamefont {Arbey}\ and\ \citenamefont {Mahmoudi}(2021)}]{Arbey:2021gdg}%
  \BibitemOpen
  \bibfield  {author} {\bibinfo {author} {\bibfnamefont {A.}~\bibnamefont {Arbey}}\ and\ \bibinfo {author} {\bibfnamefont {F.}~\bibnamefont {Mahmoudi}},\ }\bibfield  {title} {\bibinfo {title} {{Dark matter and the early Universe: a review}},\ }\href {https://doi.org/10.1016/j.ppnp.2021.103865} {\bibfield  {journal} {\bibinfo  {journal} {Prog. Part. Nucl. Phys.}\ }\textbf {\bibinfo {volume} {119}},\ \bibinfo {pages} {103865} (\bibinfo {year} {2021})},\ \Eprint {https://arxiv.org/abs/2104.11488} {arXiv:2104.11488 [hep-ph]} \BibitemShut {NoStop}%
\bibitem [{\citenamefont {Strogatz}(2018)}]{Strogatz_2018}%
  \BibitemOpen
  \bibfield  {author} {\bibinfo {author} {\bibfnamefont {S.}~\bibnamefont {Strogatz}},\ }\href@noop {} {\emph {\bibinfo {title} {Nonlinear Dynamics and Chaos: With Applications to Physics, Biology, Chemistry and Engineering}}}\ (\bibinfo  {publisher} {CRC Press},\ \bibinfo {year} {2018})\BibitemShut {NoStop}%
\bibitem [{\citenamefont {Boehmer}\ \emph {et~al.}(2010)\citenamefont {Boehmer}, \citenamefont {Harko},\ and\ \citenamefont {Sabau}}]{boehmer2010}%
  \BibitemOpen
  \bibfield  {author} {\bibinfo {author} {\bibfnamefont {C.~G.}\ \bibnamefont {Boehmer}}, \bibinfo {author} {\bibfnamefont {T.}~\bibnamefont {Harko}},\ and\ \bibinfo {author} {\bibfnamefont {S.~V.}\ \bibnamefont {Sabau}},\ }\href {https://arxiv.org/abs/1010.5464} {\bibinfo {title} {Jacobi stability analysis of dynamical systems -- applications in gravitation and cosmology}} (\bibinfo {year} {2010}),\ \Eprint {https://arxiv.org/abs/1010.5464} {arXiv:1010.5464 [math-ph]} \BibitemShut {NoStop}%
\bibitem [{\citenamefont {Charters}\ \emph {et~al.}(2001)\citenamefont {Charters}, \citenamefont {Nunes},\ and\ \citenamefont {Mimoso}}]{Charters_2001}%
  \BibitemOpen
  \bibfield  {author} {\bibinfo {author} {\bibfnamefont {T.~C.}\ \bibnamefont {Charters}}, \bibinfo {author} {\bibfnamefont {A.}~\bibnamefont {Nunes}},\ and\ \bibinfo {author} {\bibfnamefont {J.~P.}\ \bibnamefont {Mimoso}},\ }\bibfield  {title} {\bibinfo {title} {Stability analysis of cosmological models through lyapunov’s method},\ }\href {https://doi.org/10.1088/0264-9381/18/9/307} {\bibfield  {journal} {\bibinfo  {journal} {Classical and Quantum Gravity}\ }\textbf {\bibinfo {volume} {18}},\ \bibinfo {pages} {1703–1713} (\bibinfo {year} {2001})}\BibitemShut {NoStop}%
\bibitem [{\citenamefont {Leon}\ \emph {et~al.}(2013)\citenamefont {Leon}, \citenamefont {Leyva},\ and\ \citenamefont {Socorro}}]{leon2013}%
  \BibitemOpen
  \bibfield  {author} {\bibinfo {author} {\bibfnamefont {G.}~\bibnamefont {Leon}}, \bibinfo {author} {\bibfnamefont {Y.}~\bibnamefont {Leyva}},\ and\ \bibinfo {author} {\bibfnamefont {J.}~\bibnamefont {Socorro}},\ }\href {https://arxiv.org/abs/1208.0061} {\bibinfo {title} {Quintom phase-space: beyond the exponential potential}} (\bibinfo {year} {2013}),\ \Eprint {https://arxiv.org/abs/1208.0061} {arXiv:1208.0061 [gr-qc]} \BibitemShut {NoStop}%
\bibitem [{\citenamefont {Rendall}(2002)}]{Rendall_2002}%
  \BibitemOpen
  \bibfield  {author} {\bibinfo {author} {\bibfnamefont {A.~D.}\ \bibnamefont {Rendall}},\ }\bibfield  {title} {\bibinfo {title} {Cosmological models and centre manifold theory},\ }\href {https://doi.org/10.1023/a:1019734703162} {\bibfield  {journal} {\bibinfo  {journal} {General Relativity and Gravitation}\ }\textbf {\bibinfo {volume} {34}},\ \bibinfo {pages} {1277–1294} (\bibinfo {year} {2002})}\BibitemShut {NoStop}%
\bibitem [{\citenamefont {Wiggins}(1990)}]{Wiggins_1990}%
  \BibitemOpen
  \bibfield  {author} {\bibinfo {author} {\bibfnamefont {S.}~\bibnamefont {Wiggins}},\ }\href@noop {} {\emph {\bibinfo {title} {Introduction to applied nonlinear dynamical systems and chaos}}},\ \bibinfo {edition} {1st}\ ed.\ (\bibinfo  {publisher} {Springer},\ \bibinfo {year} {1990})\BibitemShut {NoStop}%
\bibitem [{\citenamefont {Bahamonde}\ \emph {et~al.}(2018{\natexlab{b}})\citenamefont {Bahamonde}, \citenamefont {Böhmer}, \citenamefont {Carloni}, \citenamefont {Copeland}, \citenamefont {Fang},\ and\ \citenamefont {Tamanini}}]{Bahamonde_2018}%
  \BibitemOpen
  \bibfield  {author} {\bibinfo {author} {\bibfnamefont {S.}~\bibnamefont {Bahamonde}}, \bibinfo {author} {\bibfnamefont {C.~G.}\ \bibnamefont {Böhmer}}, \bibinfo {author} {\bibfnamefont {S.}~\bibnamefont {Carloni}}, \bibinfo {author} {\bibfnamefont {E.~J.}\ \bibnamefont {Copeland}}, \bibinfo {author} {\bibfnamefont {W.}~\bibnamefont {Fang}},\ and\ \bibinfo {author} {\bibfnamefont {N.}~\bibnamefont {Tamanini}},\ }\bibfield  {title} {\bibinfo {title} {Dynamical systems applied to cosmology: Dark energy and modified gravity},\ }\href {https://doi.org/10.1016/j.physrep.2018.09.001} {\bibfield  {journal} {\bibinfo  {journal} {Physics Reports}\ }\textbf {\bibinfo {volume} {775–777}},\ \bibinfo {pages} {1–122} (\bibinfo {year} {2018}{\natexlab{b}})}\BibitemShut {NoStop}%
\bibitem [{\citenamefont {Aghanim}\ \emph {et~al.}(2020)\citenamefont {Aghanim} \emph {et~al.}}]{Planck:2018vyg}%
  \BibitemOpen
  \bibfield  {author} {\bibinfo {author} {\bibfnamefont {N.}~\bibnamefont {Aghanim}} \emph {et~al.} (\bibinfo {collaboration} {Planck}),\ }\bibfield  {title} {\bibinfo {title} {{Planck 2018 results. VI. Cosmological parameters}},\ }\href {https://doi.org/10.1051/0004-6361/201833910} {\bibfield  {journal} {\bibinfo  {journal} {Astron. Astrophys.}\ }\textbf {\bibinfo {volume} {641}},\ \bibinfo {pages} {A6} (\bibinfo {year} {2020})},\ \bibinfo {note} {[Erratum: Astron.Astrophys. 652, C4 (2021)]},\ \Eprint {https://arxiv.org/abs/1807.06209} {arXiv:1807.06209 [astro-ph.CO]} \BibitemShut {NoStop}%
\bibitem [{\citenamefont {Zlatev}\ \emph {et~al.}(1999)\citenamefont {Zlatev}, \citenamefont {Wang},\ and\ \citenamefont {Steinhardt}}]{Zlatev_1999}%
  \BibitemOpen
  \bibfield  {author} {\bibinfo {author} {\bibfnamefont {I.}~\bibnamefont {Zlatev}}, \bibinfo {author} {\bibfnamefont {L.}~\bibnamefont {Wang}},\ and\ \bibinfo {author} {\bibfnamefont {P.~J.}\ \bibnamefont {Steinhardt}},\ }\bibfield  {title} {\bibinfo {title} {Quintessence, cosmic coincidence, and the cosmological constant},\ }\href {https://doi.org/10.1103/physrevlett.82.896} {\bibfield  {journal} {\bibinfo  {journal} {Physical Review Letters}\ }\textbf {\bibinfo {volume} {82}},\ \bibinfo {pages} {896–899} (\bibinfo {year} {1999})}\BibitemShut {NoStop}%
\bibitem [{\citenamefont {Baumann}\ and\ \citenamefont {McAllister}(2014)}]{baumann2014inflationstringtheory}%
  \BibitemOpen
  \bibfield  {author} {\bibinfo {author} {\bibfnamefont {D.}~\bibnamefont {Baumann}}\ and\ \bibinfo {author} {\bibfnamefont {L.}~\bibnamefont {McAllister}},\ }\href {https://arxiv.org/abs/1404.2601} {\bibinfo {title} {Inflation and string theory}} (\bibinfo {year} {2014}),\ \Eprint {https://arxiv.org/abs/1404.2601} {arXiv:1404.2601 [hep-th]} \BibitemShut {NoStop}%
\bibitem [{\citenamefont {Ureña-López}(2012)}]{Urena_Lopez_2012}%
  \BibitemOpen
  \bibfield  {author} {\bibinfo {author} {\bibfnamefont {L.~A.}\ \bibnamefont {Ureña-López}},\ }\bibfield  {title} {\bibinfo {title} {Unified description of the dynamics of quintessential scalar fields},\ }\href {https://doi.org/10.1088/1475-7516/2012/03/035} {\bibfield  {journal} {\bibinfo  {journal} {Journal of Cosmology and Astroparticle Physics}\ }\textbf {\bibinfo {volume} {2012}}\bibinfo  {number} { (03)},\ \bibinfo {pages} {035–035}}\BibitemShut {NoStop}%
\bibitem [{\citenamefont {Tamanini}(2014)}]{Tamanini_2014}%
  \BibitemOpen
\bibfield  {number} {  }\bibfield  {author} {\bibinfo {author} {\bibfnamefont {N.}~\bibnamefont {Tamanini}},\ }\bibfield  {title} {\bibinfo {title} {Dynamics of cosmological scalar fields},\ }\bibfield  {journal} {\bibinfo  {journal} {Physical Review D}\ }\textbf {\bibinfo {volume} {89}},\ \href {https://doi.org/10.1103/physrevd.89.083521} {10.1103/physrevd.89.083521} (\bibinfo {year} {2014})\BibitemShut {NoStop}%
\end{thebibliography}
\end{document}